%% file: paper.tex
\documentclass[10pt,journal,compsoc]{IEEEtran}

\ifCLASSOPTIONcompsoc
  \usepackage[nocompress]{cite}
\else
  \usepackage{cite}
\fi
\ifCLASSINFOpdf
\else
\fi

\usepackage{mathtools}

\usepackage{booktabs} 
\usepackage[rightcaption]{sidecap}
\usepackage{bold-extra}
\usepackage{amsmath,amsfonts,amssymb}

\usepackage{multirow}
\usepackage{xcolor}

\usepackage{xspace}

\usepackage{paralist}
\usepackage{enumitem}

\usepackage{graphicx}
\usepackage{grffile}
\usepackage{siunitx}

\usepackage{epstopdf}

\usepackage{boxedminipage}

\usepackage{tabularx}
\usepackage{array}
\usepackage{subfig}
\usepackage{tabularx}
\usepackage{array}
\usepackage[hidelinks]{hyperref}
\usepackage{textcase}
\usepackage{caption}
\usepackage{stfloats}
\usepackage[noabbrev, capitalise]{cleveref}

\newcommand{\ffrac}[2]{\ensuremath{\frac{\displaystyle #1}{\displaystyle #2}}}

\newcommand{\Sec}[1]{\hyperref[sec:#1]{\S\ref*{sec:#1}}} 
\newcommand{\Eqn}[1]{\hyperref[eqn:#1]{(\ref*{eqn:#1})}} 
\newcommand{\Fig}[1]{\hyperref[fig:#1]{Fig.\,\ref*{fig:#1}}} 
\newcommand{\Tab}[1]{\hyperref[tab:#1]{Tab.\,\ref*{tab:#1}}} 
\newcommand{\Thm}[1]{\hyperref[thm:#1]{Thm.\,\ref*{thm:#1}}} 
\newcommand{\Lem}[1]{\hyperref[lem:#1]{Lemma\,\ref*{lem:#1}}} 
\newcommand{\Prop}[1]{\hyperref[prop:#1]{Prop.~\ref*{prop:#1}}} 
\newcommand{\Cor}[1]{\hyperref[cor:#1]{Cor.~\ref*{cor:#1}}} 
\newcommand{\Def}[1]{\hyperref[def:#1]{Defn.~\ref*{def:#1}}} 
\newcommand{\Alg}[1]{\hyperref[alg:#1]{Alg.\,\ref*{alg:#1}}} 
\newcommand{\Ex}[1]{\hyperref[ex:#1]{Ex.~\ref*{ex:#1}}} 
\newcommand{\Clm}[1]{\hyperref[clm:#1]{Claim~\ref*{clm:#1}}} 
\newcommand{\Step}[1]{\hyperref[step:#1]{Step~\ref*{step:#1}}}

\definecolor{applegreen}{rgb}{0.55, 0.71, 0.0}
\definecolor{asparagus}{rgb}{0.53, 0.66, 0.42}
\definecolor{brightgreen}{rgb}{0.4, 1.0, 0.0}
\definecolor{caribbeangreen}{rgb}{0.0, 0.8, 0.6}
\definecolor{chromeyellow}{rgb}{1.0, 0.65, 0.0}
\definecolor{darkolivegreen}{rgb}{0.33, 0.42, 0.18}
\definecolor{darkpastelgreen}{rgb}{0.01, 0.75, 0.24}

\definecolor{myblue}{RGB}{0,186,255}
\definecolor{myred}{RGB}{233,3,7}
\definecolor{mygreen}{RGB}{93,199,70}
\definecolor{myyellow}{RGB}{255,158,23}

\newcommand{\newstuff}[1]{{#1}}

\usepackage{soul}

\let\oldnl\nl
\newcommand{\nonl}{\renewcommand{\nl}{\let\nl\oldnl}}

\newcommand{\ti}[1]{\texttt{#1}}

\newcommand{\dw}{\textit{only-}$\Delta_W$\xspace}
\newcommand{\dc}{\textit{only-}$\Delta_C$\xspace}

\newcommand{\ignore}[1]{\iffalse{#1}\fi}

\usepackage{pdfrender}

\usepackage{pifont}
\newcommand{\cmark}{\textcolor{blue}{\ding{52}}} 
\newcommand{\xmark}{\textcolor{red}{\ding{56}}}

\newcommand{\com}[1]{{\iffalse#1\fi}}

\begin{document}
\title{Temporal Network Motifs:\\ Models, Limitations, Evaluation}

\author{Penghang~Liu,
        Valerio~Guarrasi,
        and~Ahmet~Erdem~Sar{\i}y\"{u}ce%
\IEEEcompsocitemizethanks{\IEEEcompsocthanksitem Penghang Liu and Ahmet~Erdem~Sar{\i}y\"{u}ce are with the Department
of Computer Science and Engineering, University at Buffalo.\protect\\
E-mail: \{penghang, erdem\}@buffalo.edu
\IEEEcompsocthanksitem Valerio~Guarrasi is with the Sapienza University of Rome.\protect\\
E-mail: valerio.guarrasi@uniroma1.it}%
}

\markboth{Journal of \LaTeX\ Class Files,~Vol.~14, No.~8, August~2015}%
{Shell \MakeLowercase{\textit{et al.}}: Bare Demo of IEEEtran.cls for Computer Society Journals}

\IEEEtitleabstractindextext{%
\begin{abstract}
Investigating the frequency and distribution of small subgraphs with a few nodes/edges, i.e., motifs, is an effective analysis method for static networks.
Motif-driven analysis is also useful for temporal networks where the spectrum of motifs is significantly larger due to the additional temporal information on edges.
This variety makes it challenging to design a temporal motif model that can consider all aspects of temporality.
In the literature, previous works have introduced various models that handle different characteristics.
In this work, we compare the existing temporal motif models and evaluate the facets of temporal networks that are overlooked in the literature.
We first survey four temporal motif models and highlight their differences.
Then, we evaluate the advantages and limitations of these models with respect to the temporal inducedness and  timing constraints.
In addition, we suggest a new lens, event pairs, to investigate temporal correlations.
We believe that our comparative survey and extensive evaluation will catalyze the research on temporal network motif models.

\end{abstract}

}

\maketitle

\IEEEdisplaynontitleabstractindextext
\IEEEpeerreviewmaketitle

\IEEEraisesectionheading{\section{Introduction}\label{sec:introduction}}

\IEEEPARstart An important local property of networks is the network motifs, which are defined as the limited size, recurrent and statistically significant patterns~\cite{Milo02}. Motifs are shown to be more effective when considered as connected, non-isomorphic, and induced subgraphs~\cite{Przulj07}.
Motifs are used to model and examine interactions among small sets of vertices in networks. Finding frequent patterns of interactions can reveal functions of participating entities~\cite{SePiKo14, Jha15, Pinar17, Ahmed16, Bressan17, Jain17} and help characterize the network. Also known as higher-order structures, motifs are regarded as basic building blocks of complex networks in domains such as social networks, food webs, and neural networks~\cite{Milo02}.
The triangle, for example, is the most basic motif in simple undirected networks and plays an important role in defining the global network characteristics such as clustering coefficients.

\newstuff{In temporal networks, edges are associated with temporal information ( timestamps), which indicate the occurrence time. We call the temporal edges as events. For example, in an email network each node is an email address and an event represents the email sent from one address to another at a specific time, or an event in the call detail records is a phone call from a person to another at a certain time.}

Temporality brings new challenges for network analysis~\cite{holme2012temporal}.
Motif-driven techniques, for instance, should consider the temporal information on edges which significantly increases the spectrum of motifs with respect to static networks.
The event order, inter-event time intervals, and durations are some of the aspects that need to be incorporated~\cite{Masuda16}.
Thus, it is beyond non-trivial to design a model for temporal network motifs that considers all those characteristics while being practical.
There are several studies~\cite{K11, S14, H15, P17} and each proposes a temporal motif model in a different way.
A motif can be valid in some models but not in the others, due to the specific constraints required by the different models.
Also, the prior studies are introduced in various subfields of computer and network science and mostly unaware of each other.
Consequently, there does not exist a unified approach that can address the limitations of those models while leveraging their novelties.
A comparative evaluation on these model is essential in that respect.
\newstuff{Note that the motif-driven approach is different from the temporal subgraph isomorphism or periodic subgraph mining~\cite{lahiri2008mining}. The motifs have limited sizes, when compared to the subgraphs, and it enables the exploration of all the combinations of edges among the nodes and all the permutations of edges in temporal order.}

In this work, we introduce a comparative survey for the four models: by Kovanen et al.~\cite{K11, kovanen2013book}, Song et al.~\cite{S14}, Hulovatyy et al.~\cite{H15}, and Paranjape et al.~\cite{P17}.
\newstuff{These four models are the first works that introduced new approaches while handling various aspects of temporality. They are also used heavily in various applications (\cref{sec:related}).}
\newstuff{We aim to address three questions:
\begin{enumerate}
\item What are the differences between these models in terms of the aspects of temporality and the constraints incorporated?
\item How do these differences affect the frequency and spectrum of the resulting temporal motifs?
\item What are the implications of these effects and how do they benefit real-world applications?
\end{enumerate}
}
\newstuff{
\noindent For the first, we give a comparison with respect to the different aspects of temporality, and highlight the advantages and limitations of each (\cref{sec:survey}). For the second, we evaluate two key aspects of temporality: temporal inducedness and timing constraints (\cref{sec:exps}). We focus on the temporality features that are overlooked in previous studies, such as the behavior of intermediate events.
We also suggest a new lens, event pairs, to analyze the characteristics of the sequences in temporal motifs and investigate the temporal correlations.
Lastly, we interpret the results in each experiment and discuss what domains can be affected by the assumptions in those four models (\cref{sec:exps}).}
We explore all three-event two-/three-nodes (36 in total) and four-event two-/three-/four-nodes (696 in total) motifs.
We believe that our work will steer the temporal motif research in a healthy direction.

\vspace{-3ex}
\section{Background}\label{sec:background}

$G(V, E)$ is a temporal network where $V$ is the set of nodes and $E$ is the set of events.
Each event $e_i \in E$ is a 4-tuple $(u_i, v_i, t_i, \Delta t_i)$~\cite{Masuda16}.
$u_i$ and $v_i$ are the (source and target) node pair where the $i$-th event occurs, $t_i$ is the starting time of $i$-th event, and  $\Delta t_i$ is its duration.
$E$ is a time-ordered list of $m$ events where the starting times are $t_1 \le t_2 \le t_3 \le \cdots \le t_m$, and $V$ is the set of nodes that appear in $E$.
In real-world temporal networks, it is very common that the inter-event time, $t_{i+1} - t_{i}$, is significantly larger than $\Delta t_{i}$, thus event durations can be ignored.
{\bf For simplicity, we also follow this convention in our work and consider each event in set $E$ as a 3-tuple $(u_i, v_i, t_i)$.}
Here we distinguish edges and events, where the edge $(u, v)$ is the static projection of an event $(u, v, t)$.
We also refer the set of motifs with the number of nodes and events; e.g., 3n3e motifs have three-nodes and three-events.

\vspace{-2ex}
\section{Related Work}\label{sec:related}

In this work, we focus on four temporal motif models~\cite{K11, S14, H15, P17} and give a detailed overview and comparison in~\cref{sec:survey} (Note that the works by Nicosia et al.~\cite{nicosia2013graph} and~\cite{holme2015modern} also provided short surveys covering a few works available).
Here we summarize the previous work on (1) applications of temporal motifs, (2) algorithmic improvements for temporal motif finding, and (3) temporal subgraph isomorphism, a related topic to our subject.

\noindent {\bf Applications.} Motif has been a versatile tool for several application domains that are engaged with the temporal networks.
Building on the notion of static motifs~\cite{Milo02,Przulj07}, there have been several works that considered snapshot-based adaptations to incorporate the temporality, listed below in chronological order.
Jin et al.~\cite{jin2007trend} investigated the temporal node-weighted networks and devised trend motifs based on the weight changes on the nodes over a specified period of time.
They counted the trend motifs in financial and protein-interaction networks.
Chechik et al. analyzed a yeast metabolic network by activity motifs to understand timings of transcriptions~\cite{chechik2008activity}.
Activity motif is defined as a partially (or totally) ordered combination of chains, forks, and joins to understand the interactions among gene activations and repressions.
Zhao et al. proposed communication motifs to study synchronous and asynchronous human communication networks, such as call detail records (CDR) and Facebook wall post interactions, and understand the information propagation~\cite{zhao2010communication}.
Communication motif is defined as a static network motif where each connected edge pair satisfies a timing constraint and there is no particular order defined among the edges.
\newstuff{Lastly, Sarkar et al.~\cite{sarkar2019using} investigated the undirected static motifs in the subsequences of microblog network snapshots to study the information diffusion process. They created motif based features to predict the network at the stage of inhibition.}

Bajardi et al. worked on the cattle trade movements among farms in Italy and used the dynamic motifs, which are defined as the chain motifs ordered in time, to model cause-effect relations~\cite{bajardi2011dynamical}.
Faisal and Milenkovic introduced static-temporal motifs to study human aging~\cite{Faisal14}.
Kovanen et al. introduced the first holistic temporal motif model that is independent of a particular topology and explicitly considers the temporal adjacency~\cite{K11} (more details in~\cref{sec:survey}).
Most of the following studies embraced this model.
Jurgens and Lu investigated the editor interactions in Wikipedia with temporal motifs~\cite{ jurgens2012temporal}.
Kovanen et al. adapted their model~\cite{K11} for colored networks, where the colors denote categorical node attributes, and studied a CDR with respect to sex, age group, and subscription type attributes~\cite{kovanen2013temporal}.
Zhang et al. analyzed two bipartite networks, ship-chartering and ship order-to-build networks, to see how the 2,2-bicliques are formed in time~\cite{zhang2014dynamic}.
Their analysis does not rely on Kovanen et al.'s model and can be considered as a snapshot based approach.
Li et al. performed a study on mobile communication networks to analyze the ordering among edges in three-node motifs~\cite{li2014statistically}.
Zhang et al. introduced an extensive analysis for online and offline human interactions~\cite{zhang2015human} by using the temporal motif model by Kovanen et al.~\cite{K11}.
They analyzed phone messages as well as face-to-face interactions (by RFID) and sexual contacts by using 3-event and 4-event temporal motifs.
Xuan et al. studied the online task-oriented networks, where people collaborate to work on tasks, and used temporal motifs to understand the collaboration patterns~\cite{xuan2015temporal}.
Li et al. considered temporal motifs for heterogeneous networks where there are multiple types of nodes and edges~\cite{li2018temporal}.
They analyzed the DBLP network (of papers, authors, terms, venues, and years), meme-tracker data, and news articles (of documents, locations, and topics).
More recently, Kosyfaki et al. proposed flow motifs to analyze the dynamics of evolution in edge-weighted temporal networks, such as bitcoin transactions, passenger flows, and facebook interactions (aggregated over certain time intervals)~\cite{Kos19}.
    
\noindent {\bf Algorithmic improvements.} Algorithmic improvements are studied for faster temporal motif finding and counting.
Gurukar et al. proposed a fast technique called COMMIT~\cite{gurukar2015commit} to find the communication motifs proposed by~\cite{zhao2010communication}.
Building on Paranjape etl al.'s model~\cite{P17}, Kumar et al. introduced an efficient algorithm to enumerate temporal cycles of any length~\cite{Kumar18} by extending the seminal cycle counting algorithms of Johnson~\cite{Johnson75}. 
Sun et al. introduced fast algorithms to find temporal motifs (by~\cite{K11}'s model) by using a time-first search approach~\cite{sun2019tm,sun2019new}.
Liu et al. presented sampling algorithms to find the approximate frequency of temporal motifs up to two orders of magnitude faster~\cite{Liu19}.
 
\noindent {\bf Temporal subgraph isomorphism.} A related subject is the problem of temporal subgraph isomorphism.
Redmond and Cunningham first introduced the problem to find a query subgraph with temporal constraints in a given temporal network~\cite{Ursula13, redmond2016subgraph}.
Unlike temporal motifs, there is no size constraint in the isomorphism problem.
Mackey et al. proposed another variation where there is a total ordering among the edges~\cite{Mackey18}.
Franzke et al. introduced another definition where there is more flexibility in defining the temporal order of edges; e.g., relative time intervals can be specified for each edge in the temporal subgraph~\cite{franzke2018pattern}.

\section{Overview of Temporal Motif Models}\label{sec:survey}

To the best of our knowledge, there are four models: %

\begin{asparaitem}
\setlength\itemsep{0.1ex}
\setlength{\itemindent}{0.5ex}
\item \textbf{Kovanen et al.~\cite{K11}} proposed the first model that has the notion of temporal adjacency to relate the events in a motif.
\item \textbf{Song et al.~\cite{S14}} introduced another model for streaming workloads where the motifs are found on-the-fly and the events in a motif can be partially ordered. %
\item \textbf{Hulovatyy et al.~\cite{H15}} considered new relaxations and restrictions to improve Kovanen et al.'s model and also discussed the events with durations. %
\item \textbf{Paranjape et al.~\cite{P17}} proposed a practical model with a specified time window to bound all events in a motif. %
\end{asparaitem}

\newstuff{\noindent Note that we focus on motif models in this work and refrain from discussing the temporal subgraphs, which are larger in size and often only considered with respect to a template.}
We first briefly explain the main idea in each paper and then compare them with respect to different aspects.

The first temporal network motif model is introduced by Kovanen et al.~\cite{K11}.
The idea is to use edge timestamps to build more expressive motifs than the classical network motifs in static networks~\cite{Milo02}.
Kovanen et al. define the temporal motif as an ordered set of events with two features: (1) time difference between each pair of consecutive events (in the whole set) is less than the threshold $\Delta_C$, an input parameter, (2) for each node in the motif, its adjacent events in the motif are consecutive, i.e., the node does not participate in any other event between its events in the motif.
This is aimed to consider causality among the events.

\begin{table*}[!t]
\centering
\caption{\it \\Aspects of Temporal Motif Models}
\vspace{-2ex}
\small
\begin{tabular}{|l|c|c|c|c|}
\hline
\textbf{Article}                & \textbf{Kovanen et al.\cite{K11}}   & \textbf{Song et al.\cite{S14}}   & \textbf{Hulovatyy et al.\cite{H15}}   & \textbf{Paranjape et al.\cite{P17}}   \\ \hline
Induced subgraph {\it (Sec.~\ref{sec:ind})} & {\it Node-based temporal} & \xmark & {\it Static only} & {\it Static only} \\  \hline
Event durations {\it (Sec.~\ref{sec:dur})} & \xmark \com{omitted for simplicity} & \xmark \com{treated as label} & \cmark & \xmark \com{claimed to be handled easily}  \\ \hline
Partial ordering {\it (Sec.~\ref{sec:part})}& \cmark & \cmark & \xmark  & \xmark  \\ \hline
Directed edges {\it (Sec.~\ref{sec:att})} & \cmark & \cmark & \xmark  & \cmark \\ \hline
Node/Edge labels {\it (Sec.~\ref{sec:att})} & \xmark  & \cmark & \xmark  & \xmark  \\ \hline
Adjacent events in $\Delta_C$ {\it (Sec.~\ref{sec:deltas})} & \cmark & \xmark  & \cmark & \xmark  \\ \hline
Entire motif in $\Delta_W$ {\it (Sec.~\ref{sec:deltas})} & \xmark  & \cmark & \xmark  & \cmark \\ \hline
\end{tabular}
\vspace{-4ex}
\label{tab:models}
\end{table*}

\begin{figure}[b!]
\vspace{-3ex}
\centerline{\includegraphics[width=\linewidth]{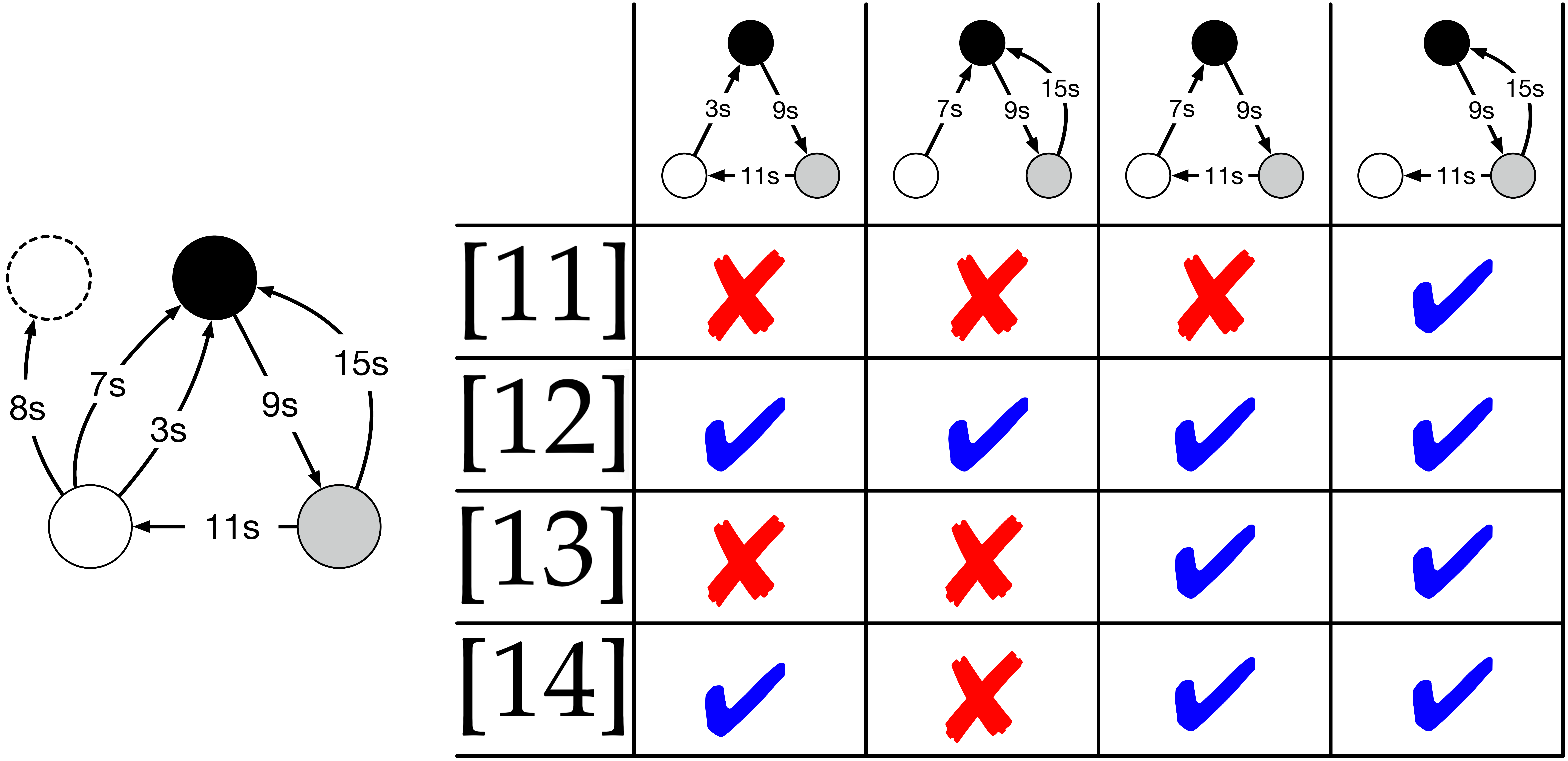}}
\vspace{-1ex}
\caption{\it Comparison of four temporal motif models~\cite{K11, S14, H15, P17}. A temporal network with six events is shown on the left. The table on the right has four motifs and shows whether they are valid motifs according to the four models, where $\Delta_C$=5s (inter-event timing constraint) and $\Delta_W$=10s (entire motif timing constraint). The first motif is not valid according to~\cite{K11, H15} because it breaks the $\Delta_C$ constraint; the second motif is invalid in~\cite{P17} since it is not an induced subgraph, and it also violates the $\Delta_C$ constraint in~\cite{K11, H15}; the third motif violates the consecutive event restriction, thus not a valid motif in~\cite{K11}; and the last motif is valid according to all the models.}
\label{fig:intro}
\end{figure}

Song et al.~\cite{S14} proposed the event pattern matching problem for the real-time graph streams.
By considering the graph structure, event pattern matching improves the traditional complex event processing~\cite{CEP}. 
They approached the problem from time series and stream processing perspective and used the graph structure as a new feature.
Event pattern is simply a temporal motif model that considers node/edge labels, partial orderings among events, and an input parameter, $\Delta_W$, which is an upper bound for the time difference between the first and the last events in the motif.

Hulovatyy et al.~\cite{H15} came up with another temporal motif model based on the notion of graphlets (induced motifs) in static networks~\cite{Przulj07}.
Hulovatyy et al. improved the model by Kovanen et al. by (1) only considering the induced subgraphs, i.e., all interactions among a given set of nodes are taken into account, and (2) relaxing the constraint that adjacent events of a node should be consecutive.
They also discussed the use of events with durations in temporal network motifs, for the first time.
Furthermore, they introduced an additional restriction (constrained dynamic graphlets) to reduce the computational complexity while obtaining approximate results.
Authors have shown that their new model captures various temporal motifs from each node's perspective and gives more effective results than the prior techniques for predicting aging-related genes in humans.

The last model, by Paranjape et al.~\cite{P17}, also considers a relaxation of the first model proposed by~\cite{K11}. The constraint that the adjacent events of a node should be consecutive (in~\cite{K11}) is relaxed so that the motifs that occur in a short burst can be caught. 
Time window, $\Delta_W$, to bound the time difference between the last and the first events in a motif.

\newstuff{A given motif can be valid in some models but not in the others, due to the specific constraints required by the different models.~\cref{fig:intro} gives four example motifs in a temporal network and describes the validity of each motif with respect to the four models.}

There are several aspects of temporal networks and motifs that are handled differently in each of those four models.
Table~\ref{tab:models} presents an overview.
Those aspects are crucial across the diverse application space of temporal networks.
For instance, fraudsters in financial transaction networks camouflage their identities by getting involved in repetitive legal transactions. A strictly induced temporal motif is helpless in this context since it considers all the transactions among a set of entities in which the few fraudulent transactions can be overlooked.
In a communication network, on the other hand, it might make more sense to use induced motifs to fully understand the information dissemination.
Here we discuss each aspect and highlight the advantages and limitations of the four models accordingly.

\subsection{Motif as induced subgraph}\label{sec:ind}

In static networks, considering all the edges among a given set of nodes (rather than selecting a subset) has been shown to be more effective in motif-based analysis~\cite{Przulj04}.
Because the non-induced motifs become artificially recurrent and shadow the importance of larger induced structures.
For instance, an induced square motif ($1$$\rightarrow$$2$, $2$$\rightarrow$$3$, $3$$\rightarrow$$4$,$1$$\rightarrow$$4$) implies that no diagonal edges exist (i.e., $1$$\rightarrow$$3$ and $2$$\rightarrow$$4$ are missing) whereas a non-induced square motif has no such restriction (i.e., every $4$-clique is also a square).
In temporal networks, ``inducedness'' is more involved.
In order to capture all the interactions among a given set of nodes, one should consider the time interval in which those events occur and also watch for the number of events in that period.
Formally, a temporal subgraph induced by the node set $V' \in G$ for a time interval $[t_s, t_f]$ includes all the events $(u, v, t)$ in $G$ such that $u, v \in V'$ and $t_s\leq t\leq t_f$.
If there is an additional restriction for the number of events, say $k$ (as in the motif definition), then one can consider only the consecutive $k$ events to form a motif. This means that there should not be any other event $(u,v,t) \in G$ such that $u, v \in V'$ and  $t_i\leq t\leq t_{i+1}$ for any consecutive event pair in the motif with timestamps $t_i$ and $t_{i+1}$. 

The first model in~\cite{K11} does not require a motif to be induced in the static sense, e.g., a diagonal edge in the square example above can be allowed.
However, one important condition in Kovanen et al's model~\cite{K11} is that a node's adjacent events in a motif should be consecutive in time, i.e., there cannot be any adjacent event outside the motif which occurs between two events in the motif.
For example, if there is a temporal motif with events $(u, v, 5)$, $(v, w, 8)$, and $(u, v, 12)$, then there cannot be any other event in the graph that contains $u$ and occurs in $\lbrack 5,12 \rbrack$ interval. Likewise, no event adjacent to $v$ can exist in $[8,12]$ interval.
We call this {\bf consecutive events} restriction.
This can be seen as {\it node-based temporal inducedness} since all the events that are adjacent to a node should be part of the motif for a given time interval.
This also avoids the exponentially many motifs in certain scenarios, like when a node has a burst of events in the form of a star.
Note that, Hulovatty et al.~\cite{H15} and Paranjape et al.~\cite{P17} argued that this constraint is too restrictive and shadows important motifs occurring in short bursts.
Song et al.'s model~\cite{S14} does not rely on such a restriction either.

Song et al.~\cite{S14} approach the inducedness from a different perspective.
They argue that if the temporal data is observed in a streaming setting, it is often desirable to catch certain events in a specific structure, e.g., some temporal and non-induced motifs (like squares) in financial transaction networks are a strong indicator of fraud~\cite{Hoffman14, Kumar18}.
Therefore, unlike static networks, non-induced motifs in temporal networks are valuable, especially when the network is being streamed and time-sensitive decision making is demanded. Hence,~\cite{S14} does not consider induced subgraphs.

Hulovatyy et al.~\cite{H15} is the first to discuss the lack of induced subgraphs in the initial model by Kovanen et al.~\cite{K11}. Authors argue that the temporal motif must be induced by relying on Przulj et al.'s work where they proposed to use induced motifs for static networks since the non-induced counts would be artificially amplified~\cite{Przulj04}.
However, the proposed model in~\cite{H15} only captures the static inducedness, i.e., considering all the edges among the given set of nodes in a static projection (e.g., a diagonal edge in the square example above is not allowed). 
There is no restriction for the temporal behavior, even in the node-level since the {\it consecutive events} restriction (of~\cite{K11}) does not exist.
For example, given four events $(a,b,2), (b,c,4), (c,a,5), (c,a,6)$; the triangle formed by the first, second, and fourth events is a valid motif in~\cite{H15}'s model (third event can be ignored).
There is another approach in~\cite{H15}, named {\it constrained dynamic graphlet}, that is somehow related to the inducedness.
Considering a snapshot-based representation where consecutive snapshots have a similar dense structure for some set of nodes, the events that are being repeated in the new snapshots do not give any new information and also cause redundancy in computation. Hence, watching for the new events that are not observed in the prior snapshots is more interesting and also more efficient.
Hulovatyy et al. incorporates this observation in the {\bf constrained dynamic graphlet} model such that if two events $(u_1, v_1, t_1)$ and $(u_2, v_2, t_2)$ are consecutive in a temporal motif (where $u_1, v_1 \neq u_2, v_2$), then there is no event $(u_2, v_2, t')$ in the temporal graph for which $t_1 \leq t' \leq t_2$.
In a sense, this enables the discovery of temporal motifs where there are causal relationships.

The model by Paranjape et al.~\cite{P17} requires the motifs to be induced, but only in the static projections, as in~\cite{H15}. 

\vspace{-2ex}
\subsection{Event durations}\label{sec:dur}
 
As defined in~\cref{sec:background}, each event in a temporal network can have a duration that denotes how long the event exists, e.g., each phone call in CDR has its duration.
Temporal motifs should be modeled to take the durations into account in such cases.
This is acknowledged in~\cite{K11} but omitted from the model for simplicity.
Likewise,~\cite{P17} mentions that their model can be generalized for the events with duration but it is not clear how the timing constraints would be adjusted.
In~\cite{S14}, authors consider the event duration as an edge label where the motifs can be defined accordingly, e.g., a motif where the duration of each edge is less than 30 secs.
The only work that incorporated the duration into the model definition is~\cite{H15} where
they propose the dynamic graphlet model; the time difference between two consecutive events are determined with respect to the end time of the first and the start time of the second event.

\vspace{-2ex}
\subsection{Partial ordering}\label{sec:part}

Temporal networks have timestamps associated with the edges and a temporal motif is often defined with respect to the ordering among its edges.
However, the ordering among the events of a motif can be total or partial.
In the former, each event pair is ordered and the order of all events are unique.
In the partial ordering, there can be some event pairs for which the ordering is undefined.
For instance, an acyclic-triangle motif with nodes $A, B, C$ and edges $A$$\rightarrow$$B$, $A$$\rightarrow$$ C$, $B$$\rightarrow$$C$ can be defined in a way that $B$$\rightarrow$$C$ precedes both $A$$\rightarrow$$B$ and $A$$\rightarrow$$C$.
A temporal motif with a partial ordering can always be expressed as a set of multiple motifs where each covers a different total ordering for the nodes who are partially ordered, i.e., the example above is union of $(B$$\rightarrow$$C)$$\prec$$(A$$\rightarrow$$C)$$\prec$$(A$$\rightarrow$$B)$ and $(B$$\rightarrow$$C)$$\prec$$(A$$\rightarrow$$B)$$\prec$$(A$$\rightarrow$$C)$.
However, finding all possibilities is not practical and also redundant.
~\cite{K11} and~\cite{S14} consider this and define temporal motif models with a partial ordering.
In both models, there is a strict partial ordering among the events, which means that the ordering is irreflexive ($e_i \nprec e_i$), transitive ($(e_i\prec e_j \land e_j\prec e_k) \rightarrow e_i\prec e_k$), and asymmetric ($e_i\prec e_j \rightarrow e_j \nprec e_i$).
Note that, this ordering assumes that each event in the given temporal network has a unique timestamp (timestamps are strictly increasing), which is not realistic: many real-world temporal networks have events (on different edges) with the same timestamp (timestamps are non-strictly increasing), e.g., an email network where a person can send an email to multiple people at the same time.
Furthermore, if the temporal network's timespan is too large, it is often desirable to reduce the resolution by creating snapshots.
Thus, the models in~\cite{K11} and~\cite{S14} do not handle such networks.
~\cite{P17} only mentions that their model can be extended to handle partial ordering but does not provide further details.
In~\cite{H15}, partial ordering is not taken into account at all and the motif model is defined to have a total ordering among its events.
Note that the models that assume a total ordering also fail to handle the networks where event timestamps are not unique.

\vspace{-2ex}
\subsection{Directed edges, node/edge labels}\label{sec:att}

All models, except~\cite{H15}, considered directed edges in their definitions.
Hulovatty et al. only mention that their model is extendible for directed edges~\cite{H15}.
For the node/edge labels, only~\cite{S14} established their model accordingly.
We believe that other models can also handle node/edge labels.
Note that none of the models consider edge weights, which is also challenging and mostly overlooked for static networks.

\vspace{-2ex}
\subsection{Timing Constraints}\label{sec:deltas}

Connectivity in the temporal dimension is a key feature for temporal motif models.
There have been different approaches to formalize temporal connectedness in the previous works~\cite{K11, S14, H15, P17}.
Kovanen et al. defined the temporal motif as a connected temporal subgraph such that for any pair of consecutive events that share a node, the time difference should be less than $\Delta_C$~\cite{K11}.
The same approach is also used by Hulovatyy et al.~\cite{H15}.
Note that both models require graph connectivity to consider consecutive events.
On the other hand,~\cite{S14} and~\cite{P17} consider a window-based temporal connectivity where all the events in a temporal motif must occur within a given time interval, denoted as $\Delta_W$.
Namely, the time difference between the last and the first events is limited by $\Delta_W$.

Those two approaches yield temporal motifs with different semantics.
$\Delta_C$ is useful to detect temporal correlations since each pair of consecutive events should occur in a certain time period, but it fails to bound all the events in a motif, i.e., can only give a loose limit, $(|E'|-1)*\Delta_C$, for the entire motif where $E'$ is the set of events.
$\Delta_W$, on the other hand, presents a holistic temporal view for the entire motif but cannot consider the temporal correlations between consecutive events.
Consider a connected temporal motif with three ordered events where we set $\Delta_C$=$5$ for the models in~\cite{K11, H15} and $\Delta_W$=$10$ for the models in~\cite{S14, P17}.
If the timestamps of those three events are 1, 9, and 10, $\Delta_W$ based models consider this motif valid, but $\Delta_C$ based models do not since the first two events are not close enough for $\Delta_C$=$5$. 
One can say that the first two events are too far apart, thus there is no temporal correlations.
However, it is also possible that the third event occurred right after the second only because of some important information that is initially conveyed by the first event; thus there is a temporal correlation between the first and third events.
Either interpretation can make sense depending on the type of temporal network being considered.
\newstuff{For instance, a temporal cycle can be formed this way where person $A$ tells person $B$ that she started a new job. Then $B$ sees $C$ like a week later and tells about $A$'s new job, which immediately results in $C$ calling $A$ for congratulations. Here the delayed convey of the information can be captured with $\Delta_W$ parameter.}

One can consider to use both parameters to have a trade-off between the two extremes of $\Delta_W$ and $\Delta_C$.
{\bf Depending on the number of events in the temporal motif, one of those two timing constraints can be useless for certain values of $\Delta_W$ and $\Delta_C$.}
Given a motif with $m$ events and ${\Delta_C}/{\Delta_W}$ ratio, we have:

$\textrm{Constraints} =
  \begin{cases}
    \Delta_C  & \quad \text{if } 0 \leq \ffrac{\Delta_C}{\Delta_W} \leq \ffrac{1}{m -1}\\
    \Delta_C, \Delta_W  & \quad \text{if } \ffrac{1}{m -1} < \ffrac{\Delta_C}{\Delta_W} < 1\\
    \Delta_W  & \quad \text{if } \ffrac{\Delta_C}{\Delta_W} \geq 1
  \end{cases}$
\vspace{2ex}

\noindent There are $m-1$ time intervals among $m$ events.
The loose bound defined by $\Delta_C$ for the entire motif window is $\Delta_C*(m-1)$.
In the first case, any $\Delta_W$ larger than that bound is meaningless; so satisfying $\Delta_C$ constraint is sufficient.
It is also meaningless to consider a $\Delta_C$ value (third case) that is not smaller than $\Delta_W$; just considering $\Delta_W$ would also satisfy the other.
The only case where both constraints make sense is when $\Delta_W$ is smaller than the loose bound $\Delta_C*(m-1)$ (second case).
Exploring the parameter space in that case may enable to consider temporal motifs with respect to both inter-event timings ($\Delta_C$) and entire motif timing ($\Delta_W$).

\input{exps}

\vspace{-4ex}
\section{Discussion}

In this work, we introduced a comparative survey for the existing temporal motif models. We evaluated both advantages and limitations of these models with respect to two key aspects: temporal inducedness and timing constraints.
In addition, we use the event pairs to analyze the sequences in the observed motifs.
Our experimental evaluation shows that; (1) The temporal inducedness restrictions (consecutive events restriction and constrained dynamic graphlets) exhibit a bias towards certain types of motifs, consistently in most datasets, (2) Timing constraints, ${\Delta_C}$ and ${\Delta_W}$, have complementary strengths, where the former fails to bound timespans whereas the latter introduces bias for the occurrence of intermediate events; hence combining both parameters by choosing a ratio for ${\Delta_C}/{\Delta_W}$ in $(\frac{1}{m -1}, 1)$ interval for $m$ event motifs can yield a trade-off; (3) Sequences of event pairs suggest interesting findings about the interplays among different types of pairs, such as asymmetrical trends.

There are several directions worth to explore as a future work.
Besides temporal inducedness and timing constraints, for instance, temporal motifs with event durations is a promising avenue.
We believe that devising an ultimate unifying model would be too ambitious due to the diverse characteristics of temporal networks and application-driven models (e.g., for phone calls) are likely to yield more effective and practical results.
Also, understanding how the observed motif counts are related to the intrinsic network characteristics is important and recently introduced random reference models can be helpful~\cite{gauvin2018randomized}.
\newstuff{We also intend to utilize the sequence of event pairs for the event prediction.}

\ifCLASSOPTIONcaptionsoff
  \newpage
\fi

\vspace{-3ex}
\bibliographystyle{IEEEtran}
\bibliography{paper,afosr}

\vspace{-10ex}
\begin{IEEEbiography}[{\includegraphics[width=1in,clip,keepaspectratio]{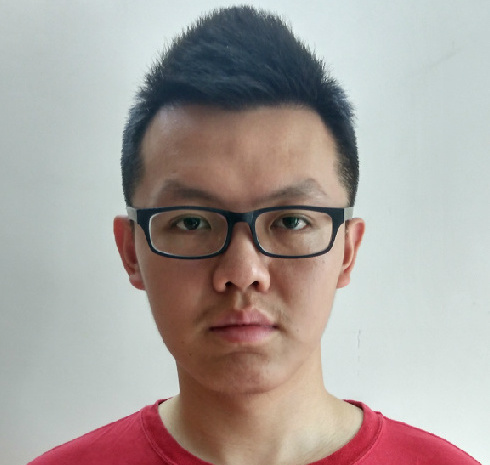}}]{Penghang Liu}
received his B.S. degree in geographic information science from Wuhan University, Wuhan, China, and the M.S. degree from University at Buffalo. He is working toward the PhD degree in computer science at University at Buffalo. His research interests include temporal network motifs and dense subgraph discovery.
\end{IEEEbiography}

\vspace{-15ex}
\begin{IEEEbiography}[{\includegraphics[width=1in,clip,keepaspectratio]{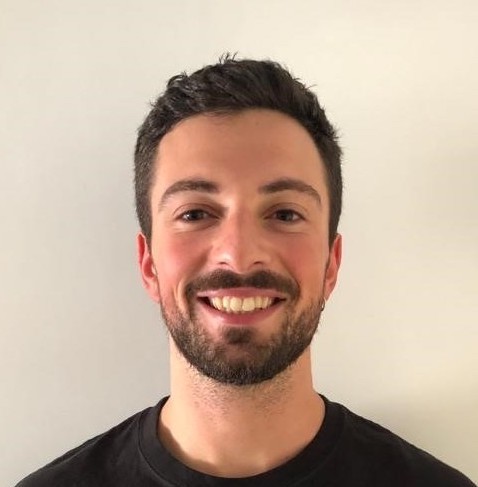}}]{Valerio Guarrasi}
received the B.C degree in Management Engineering, and the M.S. degree in Data Science from Sapienza University of Rome, Italy. He is persuing the PhD degree in Data Science at Sapienza University of Rome. His research interests include machine learning, especially with bio-medical applications.
\end{IEEEbiography}
\vspace{-15ex}

\begin{IEEEbiography}[{\includegraphics[width=1in,clip,keepaspectratio]{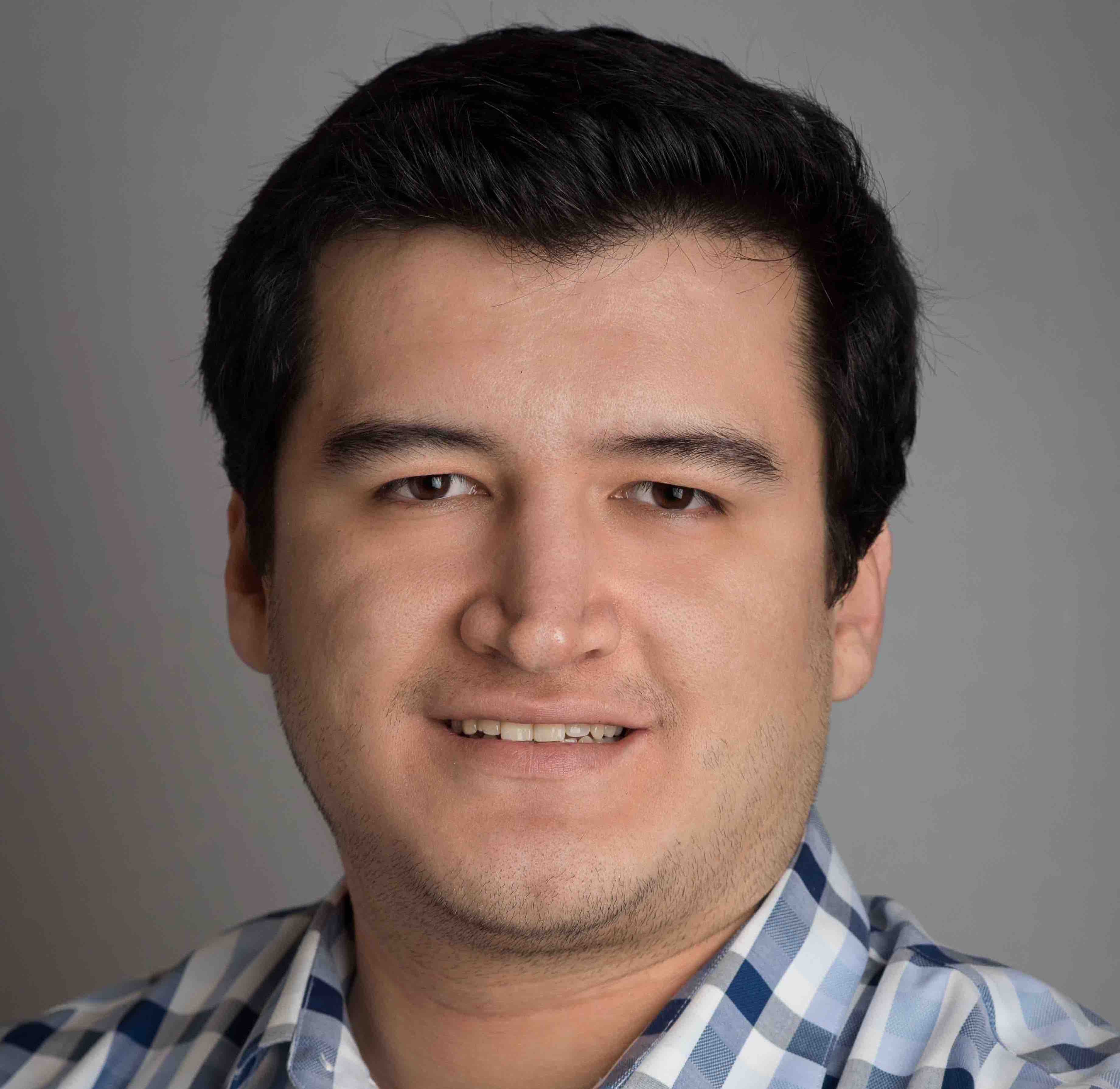}}]{Ahmet Erdem Sar{\i}y\"{u}ce} received the B.S. degree in computer engineering from Middle East Technical University and Ph.D. degree in computer science and engineering from the Ohio State University. He is currently an assistant professor at University at Buffalo. His research is in graph mining and management.
\end{IEEEbiography}
\vspace{-5ex}

\input{appendix}

\end{document}

%% file: exps.tex
\vspace{-2ex}
\section{Experimental Evaluation} \label{sec:exps}

\begin{table}[!t]
\captionsetup{justification=centering}
\caption{\it \\Important statistics for the temporal network datasets we use in our experiments. For each network, we provide the number of nodes, events (interactions), edges (unique pairs of nodes in events), timestamps (\#T, unique timestamps across the entire timespan), the percentage of events with unique timestamp ($|E_{u}|/|E|$),
 and the median interevent-time (m($\Delta_t$), in seconds). 
}
\vspace{-1ex}
\small
\setlength\tabcolsep{1.5pt}
\centering
\begin{tabular}{|l|r|r|r|r|r|r|r|}
\hline
Name     & Nodes & Events & Edges & \#T & $|E_{u}|/|E|$ &m($\Delta_t$) \\ \hline
\texttt{Bitcoin-otc}   & 5.88K & 35.6K  & 35.6K & 35.4K &99.2\%     & 707                    \\
\texttt{CollegeMsg}      & 1.90K & 59.8K  & 20.3K & 58.9K & 97.2\%      & 37                     \\
\texttt{Calls-Copen.}    & 536   & 3.60K  & 924   & 3.59K  &99.7\%    &194                    \\
\texttt{SMS-Copen.}       & 568   & 24.3K  & 1.30K & 24.0K &97.6\%     &32                     \\
\texttt{Email}               & 986   & 332K   & 24.9K & 208K   &50.5\%     &15                     \\
\texttt{FBWall}             & 47.0K & 877K   & 274K  & 868K   &98.0\%   &42                     \\
\texttt{SMS-A}          & 44.4K & 548K   & 69.0K    & 470K  &73.1\%     &3                      \\
\texttt{StackOver.} & 260K  & 6.35M  & 4.15M & 5.97M    &88.2\%  &6                      \\
\texttt{SuperUser} & 194K  & 1.44M  & 925K  & 1.44M    &99.2\%  &83                    \\ \hline
\end{tabular}
\label{tab:data}
\vspace{-2ex}
\end{table}

In this section, we present an extensive evaluation for the various aspects considered in the four temporal network motifs~\cite{K11, S14, H15, P17}. We are particularly interested in the impacts of temporal inducedness and timing constraints. {\bf Our code (including plot scripts) is available for reproducibility: \url{https://shorturl.at/erBP1}}.

\noindent {\bf Datasets.} We select various directed temporal network datasets from several domains, including phone messages, emails, Facebook wall interactions, posts in Q/A websites, and call detail records (CDR).
\cref{tab:data} gives several statistics about our datasets.
The time resolution of all the networks is one second.
In addition to the number of events, edges,
 and the percentage of events with unique timestamp, we give the median interevent-time for each dataset, which is the median of time intervals between all pairs of consecutive events in the network.
This gives us an idea about how to choose the timing parameters ($\Delta_C$ and $\Delta_W$) in order to address the trade-off between discovering more motifs and reducing the computational costs. 
In the phone message networks, an event $(u, v, t)$ represents a message sent to person $v$ by person $u$ at time $t$; we have \texttt{SMS-A}~\cite{Wu18803}, \texttt{SMS(Copenhagen)}~\cite{sapiezynski2019interaction}, and \texttt{College-messages}~\cite{snap}.
\texttt{Email} presents the emails between members of European research institution~\cite{snap}, where an event $(u, v, t)$ denotes an email sent from person $u$ to person $v$ at time $t$.
Among the online social networks; \texttt{FBWall} is the collection of posts between users on Facebook in the New Orleans region~\cite{viswanath2009}, where an event $(u, v, t)$ indicates user $u$ posted on the user $v$'s wall at time $t$; \texttt{StackOverflow} and \texttt{SuperUser} are the interaction networks on two stack exchange websites~\cite{snap}, where event $(u, v, t)$ stands for the answer/comment user $u$ posted on user $v$'s question/answer at time $t$.
Note that we slice out the earliest 10 percents of the events from the original \texttt{StackOverflow} datasets for the efficiency purposes. %
We also study the \texttt{Bitcoin-otc}, a trust network where users rate each other to reflect their trust regarding the bitcoin transactions~\cite{snap}. 
We also use \texttt{Calls(Copenhagen)}, which is the collection of phone calls between university students over a period of four weeks~\cite{sapiezynski2019interaction}.

\begin{figure}[b!]
\vspace{-2ex}
\centering
\includegraphics[width=0.9\linewidth]{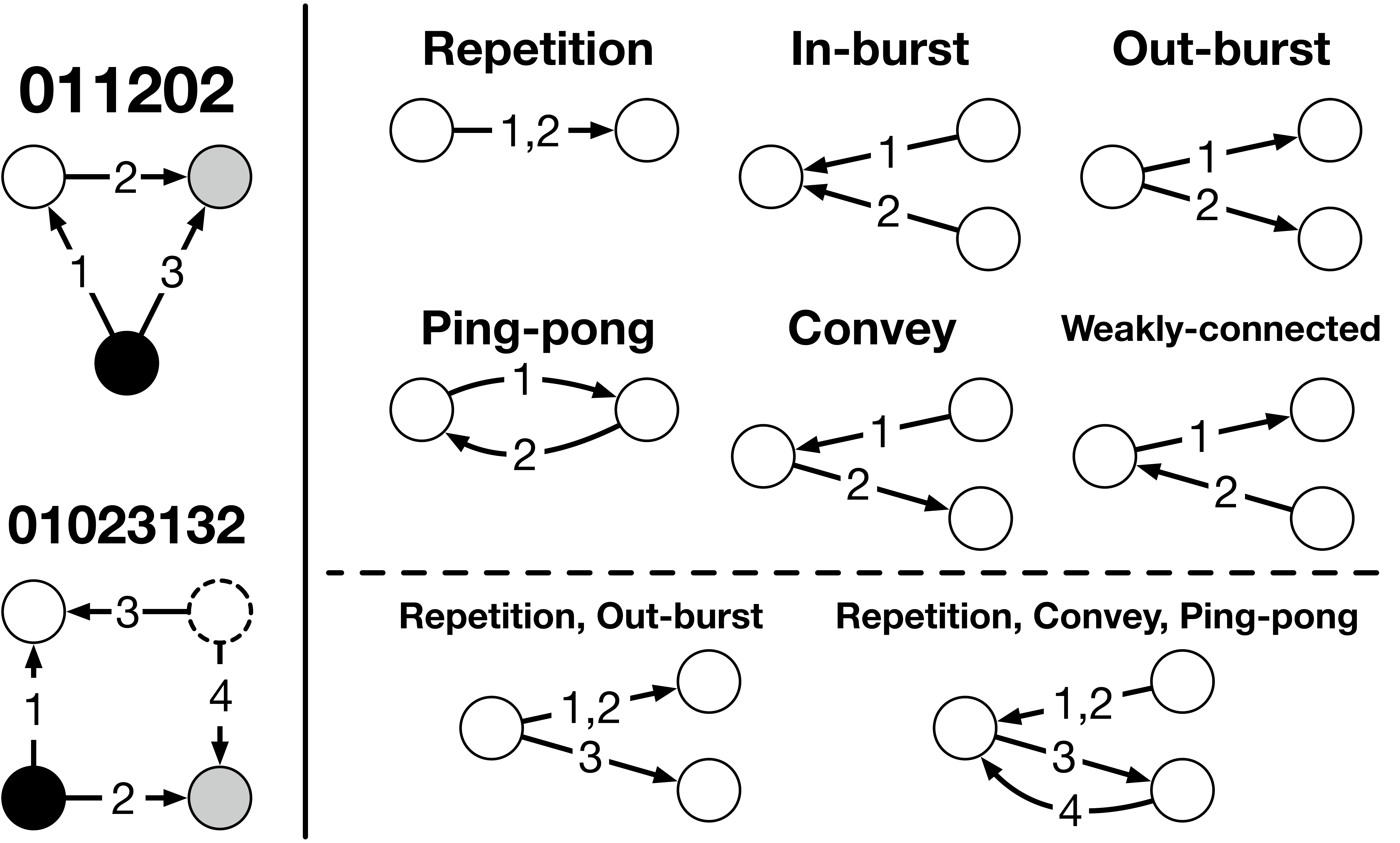}
\caption{\it [Left] We use $2n$ digits to denote a temporal motif with $n$ events.
Each event is given by a pair of digits, where the source node is the first and the target node is the second digit.
The first two digits are always $01$, to denote that the first event occurred from node $0$ to node $1$.
The sequence of pairs and the digit number of each node follow the chronological order of events and nodes.
[Right] Event pair representations, all six types are listed. At the bottom, a three-node three-event motif is denoted as a sequence of repetition and out-burst, and a four-event motif is denoted as a sequence of repetition, convey, and ping-pong.
}
\label{fig:notation}
\end{figure}

\noindent {\bf Motif notation. }We consider three-event and four-event motifs in this work and it is not feasible to visualize all. Instead, we introduce a notation to refer to the temporal motifs in our experiments.
We use $2n$ digits to denote a temporal motif with $n$ events.
Each pair of digits is an event from the node represented by the first digit to the node denoted by the second digit.
The first two digits are always $01$, to denote that the first event occurred from node $0$ to node $1$.
The sequence of pairs and the digit number of each node follow the chronological order of events and nodes.
~\cref{fig:notation} presents some examples.
For instance, 011202 (top-left) corresponds to a triangle temporal motif where the first event is from the black node (0) to the white node (1), the second one is from the white node (1) to the gray node (2), and the last event is from the black node (0) to the gray node (2).
Note that we only consider the motifs that grow as a single component, by adding one event at a time.

\noindent {\bf A new lens: Event pairs.}
In this work, we consider an alternative way to interpret the temporal motifs~\cite{zhang2015human}.
We simply look at the sequence of two events that share a node (i.e., event pairs).
Event pairs can be seen as the building blocks of the larger motifs (with $\ge 2$ events).
Given a pair of consecutive events that share a node in the motif, $(u_1, v_1, t_1)$ and $(u_2, v_2, t_2)$, where $t_1 < t_2$, there are six types of event pairs (also shown in~\cref{fig:notation}):

\begin{asparaitem}
\setlength\itemsep{0.1ex}
\setlength{\itemindent}{0.5ex}
\item \textbf{Repetition.} Two events occur on the same edge ($u_1$=$u_2$, $v_1$=$v_2$).
\item \textbf{Ping-pong.} Second event is reverse of the first one ($u_1$=$v_2$, $v_1$=$u_2$). %
\item \textbf{In-burst.} Two events share the same target ($u_1$$\neq$$u_2$, $v_1$=$v_2$). %
\item \textbf{Out-burst.} Two events share the same source ($u_1$=$u_2$, $v_1$$\neq$$v_2$). %
\item \textbf{Convey.} The source of the second event is the target of the first event ($v_1$=$u_2$, $u_1$$\neq$$v_2$). %
\item \textbf{Weakly-connected.} The target of the second event is the source of the first event ($u_1$=$v_2$, $v_1$$\neq$$u_2$). %
\end{asparaitem}

A motif with $m$ events can be represented as a sequence of $m-1$ event pairs, as long as each consecutive event pair shares a node (see~\cref{fig:notation} for examples).
Considering event pairs as the building blocks is useful for three reasons:
\begin{asparaitem}
\setlength\itemsep{0.1ex}
\setlength{\itemindent}{0.5ex}
\item {\bf It is a 6-letter alphabet}, thus a simple and succinct way to deal with the temporal and structural complexity.
\item {\bf It is expressive}; can exactly represent all 2n3e or 3n3e motifs (36 in total, $6^2$) and all 3n4e or 2n4e motifs (216 in total, $6^3$). It also gives 216 ($6^3$) broad descriptions (not exact) for the 480 4n4e motifs that grow as a single component.
\item {\bf It is well-suited to capture the temporal correlations}: event pairs within larger motifs can reveal the mesoscale characteristics while providing an easy interpretation.
\end{asparaitem}

\noindent{\bf Questions.} {In our experimental evaluation, we are looking to answer the following questions about temporal network motifs by using the various real-world temporal networks in our dataset:

\noindent{\bf Q1: What are the implications of temporal inducedness constraints discussed in~\cref{sec:ind}?} In particular, is there any bias in the spectrum or counts of motifs obtained by consecutive event restriction in~\cite{K11} and constrained dynamic graphlets in~\cite{H15}? We answer those in~\cref{sec:tempinduced}. 

\noindent{\bf Q2: How the timing constraints, $\Delta_C$ and $\Delta_W$, impact the spectrum or counts of motifs?} What are the biases implied by each constraint? How would the combination of those two constraints compare? We consider those in~\cref{sec:timing}. Note that we avoid the experiments in prior studies~\cite{P17} about the choice of parameters.

\noindent{\bf Q3: What new insights can we capture by using the lens of event pairs?} What kind of event pairs observed in each dataset? Is there any commonality for all datasets or for the ones from the same domain? We answer all in~\cref{sec:eventpairs}.

\newstuff{\noindent{\bf Comparison criteria.} We are only interested in the motif counts and behavior of the motif spectrum in our evaluation; runtime performance is not in our scope in this work (but is a promising future direction).
Note that, in static motif analysis~\cite{Milo02} a null model (randomized graph) is often used as reference to measure the statistical significance of motifs. However, it is difficult to choose a proper null model for temporal motifs: the intertwined temporal and structural complexity in temporal networks makes it challenging to consider a null model that can mimic the realistic features. This issue is also highlighted by Kovanen et al.~\cite{K11}. Although there has been several new random reference models for temporal networks~\cite{gauvin2018randomized}, our preliminary analysis showed that there is no reliable random graph model which can mimic both the structural and temporal features in real-world networks. We tried several link-shuffling and time-shuffling models from~\cite{gauvin2018randomized}, some are too restrictive where the motif counts barely change, and some others are too loose where all the motifs are reported as significant. Therefore, as also done in~\cite{K11}, we consider the motif count as the main indicator of motif significance, which gives a relative comparison.
}
We consider three-event and four-event motifs in our experiments.

\vspace{-2ex}
\subsection{Temporal inducedness constraints}\label{sec:tempinduced}

As discussed in~\cref{sec:ind}, temporal network motif models have different approaches to define temporal inducedness. Here we evaluate the impact of two restrictions by using 3n3e (three-nodes, three-events) motifs: (1) Consecutive events, proposed in~\cite{K11} to ensure node-based temporal inducedness, (2) Constrained dynamic graphlets, proposed in~\cite{H15} to filter out the stale information.

\subsubsection{Consecutive events restriction}

\begin{table}[!t]
\small
\captionsetup{justification=centering}
\caption{\it \\The impact of consecutive event restriction. We compare the total counts of 3n3e motifs without or with consecutive event restriction for $\Delta_C$=$1500s$ (2nd and 3rd columns). We also give the ranking changes of the four motifs among all 32 3n3e motifs after the consecutive restriction is added (last four columns). Overall, the given motifs are amplified when the consecutive event restriction is applied. 
}
\setlength\tabcolsep{1.6pt}
\centering
\begin{tabular}{|l|r|r|r|r|r|r|}
\hline
Network            & Non-cons. & Cons. & 010210 & 011210 & 012010 & 012110 \\ \hline
\texttt{Calls-Cop.} & 1.66K            & 77          & +7   & -9        & 0       & 0       \\ \hline
\texttt{CollegeMsg}       & 1.59M        & 2.55K         & +18    & +23    & +10    & +16    \\ \hline
\texttt{SMS-Copen.}   & 198K          & 389         & +16    & +18    & +14    & +17   \\ \hline
\texttt{SMS-A}            & 1.02M         & 35          &+11    & +11       &  +2      &  +6      \\ \hline
\texttt{Email}            & 1.84M         & 1.82K        &+1        & +4    &  +4      & +4      \\ \hline
\texttt{FBWall}           & 268K          & 904         &+14   & +10    & +13    &  +2     \\ \hline
\texttt{Bitcoin-otc}      & 6.42K            & 1.93K        &0        &0        & +2       &  +3     \\ \hline
\texttt{StackOver.}    & 3.78M         & 1.02K        &-11        & +8       &  -5      &    0    \\ \hline
\texttt{SuperUser}        & 481K          & 10.9K       &+4        & +4       &  -4      & +5   \\ \hline
\end{tabular}
\label{tab:consecutive}
\vspace{-3ex}
\end{table}

In this part, we evaluate the advantages and limitations of adding the consecutive event restriction from~\cite{K11}. As explained in~\cref{sec:ind}, consecutive event restriction is a node-based constraint.
If a node is part of a motif, then the adjacent events of the node in the motif should be consecutive, i.e., the node cannot have an adjacent event outside the motif while it is engaged in the motif. The third column in~\cref{fig:intro} also demonstrates this restriction, where white node interacts with the dashed node at $t=8$ thus the motif is not valid according to~\cite{K11}.
Consecutive events restriction is useful when handling a star node for instance, which is adjacent to many events. The continuity requirement ensures that the node can only be engaged with a linear number of motifs, avoiding quadratically many.
However this is a double-edged sword: it permits fast counting and analysis, but can miss important patterns.
Here we count the motifs with and without consecutive event restriction on all 32 3n3e motifs and use $\Delta_C=1500s$ timing configuration (i.e., no $\Delta_W$ is specified).
~\cref{tab:consecutive} presents the results (all results are available in the supplementary materials).

The motif counts show that in all datasets except \texttt{Bitcoin-otc}, over 95\% of the motifs are removed if the restriction is applied.
{\bf We observe that four motifs, 010210, 011210, 012010, and 012110, are often amplified by the consecutive event restriction.} These motifs are not frequently observed in the non-consecutive scenario but rise to a significantly higher ranking after the consecutive restriction is applied.
The amplification of these motifs is most commonly observed in message networks, such as \texttt{CollegeMsg} and \texttt{SMS(Copenhagen)}
All of these motifs follow an ask-reply pattern, where the last event replies the first event. It is not an immediate response and another node is involved in the second event, which appears to be another conversation. 
Note that if the consecutive events restriction is not considered, adjacent events of a node are not required to be consecutive. Then, the second event would be allowed to form a different motif, by skipping the third event (reply) and adding another adjacent event from another conversation. \newstuff{This would shadow the original ask-reply pattern between the first and the third events and, in turn, overstates the motifs where the first and second events form a star. When it is important to distinguish ask-reply patterns from the big star structures, consecutive events restriction can be advantageous. For instance, using consecutive events restriction can be useful for temporal networks of text messages and emails. A burst of messages or emails from one individual to others may be spam (or the information value may not be high), while two reciprocal messages or emails in short time are likely to indicate a real conversation. On the other hand, considering this restriction in other temporal networks, like online interactions in social networks, may suggest biased results by artificially amplifying the ask-reply patterns.}
\textbf{All in all, while the consecutive event restriction filters out the discontinuous engagements from the node's perspective, it yields biased motif counts by consistently amplifying ask-reply motifs where the first and last events are reciprocal.} \newstuff{This can be advantageous in the analysis of message and emails (where bursts of messages/emails are too common) but may introduce bias in the analysis of other temporal networks.}

\subsubsection{Constrained dynamic graphlets}
As defined in~\cref{sec:ind}, constrained dynamic graphlets are defined in~\cite{H15} to exclude stale information in the temporal motifs. Constrained dynamic graphlet only includes the new events that are not observed in the prior timestamps; formally, if two events $(u_1, v_1, t_1)$ and $(u_2, v_2, t_2)$ are consecutive in the graphlet (where $u_1, v_1 \neq u_2, v_2$), then there must be no event $(u_2, v_2, t')$ in the temporal graph for which $t_1 \leq t' \leq t_2$.

Here we investigate how the constrained dynamic graphlets impact the spectrum of observed motifs and give a comparison against the temporal motifs without such restriction.
Note that this restriction is mainly motivated by the snapshot-based representation, which implies that there are multiple events occurring at the same timestamp.
For most real-world temporal networks, the raw data has a fine resolution (e.g., 1 second) and it is unlikely to observe multiple events occurring at the same timestamp.
We also show this in~\cref{tab:data} in the second last column: more than 97\% of the events have unique timestamps in most datasets.
Hence, nearly all motifs would be able to escape from the constrained dynamic graphlet restriction since almost always only one event occurs in a give timestamp.
Thus we degrade the resolution of our datasets to 300s in order to highlight the difference between constrained dynamic graphlet counting and vanilla temporal motif counting (i.e., without constraints). 
Note that this will have an additional impact on both types of the motif counts: each will have less motifs when compared to 1s resolution (This is because we assume the events in a motif have a total ordering, so events within the same timestamp are not included in the same temporal motif). 
In both scenarios, we only consider $\Delta_C = 1500s$ as the timing constraint and investigate the counts of 3n3e motifs.

\begin{table}[!t]
\captionsetup{justification=centering}
\caption{\it \\The motif proportion changes (percentage) when going from vanilla temporal motifs to the constrained dynamic graphlets. The resolution of all datasets is degraded to 300s, and the variance of proportion changes for all 3n3e motifs is shown. \texttt{Email} has the greatest variance in motif proportion changes. 010102, 010202, 012020, and 010201 are the four motifs that show the most significant changes. For \texttt{Bitcoin-otc}, the two methods show no difference since no repetitions allowed in the dataset.
}
\setlength\tabcolsep{1.5pt}
\small
\centering
\begin{tabular}{|l|r|r|r|r|r|}
\hline
Network             & Variance & 010102    & 010202   & 012020 & 010201 \\ \hline
\texttt{Calls-Cop.} & 1.70     & +0.22\%  & -3.45\%  & +1.39\% & -5.60\%   \\ \hline
\texttt{CollegeMsg}        & 3.36     & +3.31\%  & +4.36\%   & +3.76\% & -2.12\%  \\ \hline
\texttt{SMS-Copen.}   & 3.49     & +2.37\%   & +3.23\%   & +3.26\% & -0.99\% \\ \hline
\texttt{SMS-A}            & 2.34     & +4.09\%  & +4.98\%   & +1.54\% & -1.93\%  \\ \hline
\texttt{Email}            & 18.98    & -9.63\% & -10.05\% & +2.88\% & -18.00\% \\ \hline
\texttt{FBWall}           & 1.03     & +1.06\%  & +1.09\%   & +0.75\% & -0.78\%  \\ \hline
\texttt{Bitcoin-otc}      & 0.00     & 0.00\%  & 0.00\%   & 0.00\%   & 0.00\% \\  \hline
\texttt{StackOver.}    & 0.04     & +0.26\%  & +0.27\%   & +0.26\% & -0.09\%  \\  \hline
\texttt{SuperUser}        & 0.06     & +0.63\%  & +0.65\%   & +0.24\% & -0.14\%  \\ \hline
\end{tabular}
\vspace{-3ex}
\label{tab:dgc}
\end{table}

We first check the impact of degrading the resolution on vanilla temporal motif counts.
Message networks are affected most; we observe 80\% decrease in counts from 1s to 300s. Degrading the resolution has less impacts on the stack exchange networks, where the inter-event time intervals are larger than the other datasets.
Comparing the ratios of each motif at 1s and 300s resolutions shows that the change of proportion is always less than 1\%.
Hence degrading the resolution affects all motifs equally, so does not introduce any bias into our evaluation. 

Now we compare the two methods after the resolution is degraded to 300s. ~\cref{tab:dgc} presents the results (complete results are available in the supplementary materials). We look at the changes in the ratios for all 3n3e motifs (ratio of a particular motif count to the sum). We observe the largest variance in \texttt{Email}, which indicates that some motifs change significantly than others. Message networks also show strong variance while the variances are less than 0.1 in the stack exchange networks.
There are some commonalities across all datasets: {\bf 010102, 010202, 012020, and 010201 are the motifs that show the most significant proportion changes when going from vanilla counts to restricted counts.}
The decrease in 010201 is due to the fact that the constrained dynamic graphlet does not favor the {\it delayed repetitions}: if there exists many 01s after the 02, only the first one is considered and the rest is ignored.
Note that it is often likely to see such delayed repetitions in communication networks such as messaging and email.
The decrease in 010201 translates to increases in the motifs with {\it immediate repetitions}, such as 010102, 010202, and 012020. 
Note that \texttt{Email} shows a different behavior with respect to these motifs where most ratios decrease drastically.
\newstuff{The common feature of the motifs with decreases (010102, 010202, 010201) is that there is a repetition and out-burst from the initial sender node (0), either one after the other or interleaved. We believe that the prevalence of out-bursts is due to the the carbon copies (cc) in emails.
It is also likely that the carbon copies occur at the both timestamps of the repetition, thus those motifs are removed by the constrained dynamic graphlet restriction.
{\bf Using the constrained dynamic graphlet restriction can adversely impact the analysis of communication networks (such as text messages and emails) where two parties are engaged in a conversation.} Delayed repetitions are likely to occur when the sender is engaged in another conversation and ignoring those motifs will amplify the motifs with immediate repetitions.}

\noindent{\bf Summary.} Overall, we observe that both the consecutive events restriction and constrained dynamic graphlets exhibit a bias towards certain types of motifs, consistently in most datasets. This can have an adverse affect in the analysis of certain types of temporal networks.

\subsection{Impact of timing constraints}\label{sec:timing}

Both timing constraints ($\Delta_C$ and $\Delta_W$) have their specialties in capturing temporal motifs: $\Delta_C$ emphasizes the continuity, while $\Delta_W$ provides a holistic view and brings a strict bound to the motif timespan. Here we use both parameters and compare the patterns we observed in the two extremes (\dc and \dw). 
\textbf{For all experiments, we set the $\Delta_W$ as 3000 seconds and change the $\Delta_C$ to obtain different ${\Delta_C}/{\Delta_W}$ ratios.}
Our choice relies on the same principle used in~\cite{P17}; we use inter-event times as proxy.
As we discussed in~\cref{sec:deltas}, the ${\Delta_C}/{\Delta_W}$ ratio must be in $(\frac{1}{m -1}, 1)$ interval to make the both constraints useful, i.e., we are considering \dc if the ${\Delta_C}/{\Delta_W}$ ratio is too small and \dw if the ratio is too large.
For three-event motifs we select three configurations: ${\Delta_C}/{\Delta_W}$ = 0.5 (\dc), 0.66 ($\Delta_W$-and-$\Delta_C$), and 1.0 (\dw); for four-event motifs there are four configurations: ${\Delta_C}/{\Delta_W}$ = 0.33 (\dc), 0.5, 0.66, and 1.0 (\dw). 
Since we only change the $\Delta_C$ here, some motifs in larger ${\Delta_C}/{\Delta_W}$ configurations will not satisfy the $\Delta_C$ constraint when the ratio becomes smaller, thus the set of motifs observed under a smaller ${\Delta_C}/{\Delta_W}$ ratio is a subset of a larger ${\Delta_C}/{\Delta_W}$ configuration.

\begin{table}[!t]
\captionsetup{justification=centering}
\caption{\it \\Counts of event pairs and their reduction rates when going from \dw to $\Delta_W$-and-$\Delta_C$ and \dc configurations. The \textit{R, P, I, O, C, W} are the types of event-pairs. All counts are decreasing when going from \dw to \dc configuration. However, the reduction in \textit{R, P, I, O} is more significant in all datasets. Overall, \dw amplifies \textit{R, P, I, O} occurrences. Using both constraints (${\Delta_C}/{\Delta_w}=0.66$) reaches a balance between the two extremes, where the \textit{R, P, I, O} motifs are significantly reduced while the \textit{C, W} motifs are mostly preserved.
}
\small
\centering
\renewcommand{\tabcolsep}{0.5ex}
\centering
\begin{tabular}{|l|c|r|r|r|r|r|}
\hline
\multirow{2}{*}{Network} & \multirow{2}{*}{Motif Type} & \dw & \multicolumn{2}{c|}{$\Delta_W$-and-$\Delta_C$} & \multicolumn{2}{c|}{\dc} \\
& & \centering Count & Count & Ratio & Count & Ratio \\ \hline
\multirow{2}{*}{\texttt{College.}}  & \textit{R, P, I, O}     & 514K   & 421K  & 81.9\% & 292K  & 56.8\% \\
            & \textit{C, W}    & 68.3K  & 56.4K & 82.6\% & 40.2K & 58.9\% \\ \hline
\multirow{2}{*}{\texttt{FBWall}}     & \textit{R, P, I, O}     & 395K   & 315K  & 79.7\% & 242K  & 61.3\% \\
            & \textit{C, W}   & 45.9K  & 40.6K & 88.4\% & 32.7K & 71.2\% \\ \hline
\multirow{2}{*}{\texttt{Bitcoin.}} & \textit{R, P, I, O}    & 8.91K  & 7.21K & 80.9\% & 5.94K & 66.6\% \\
            & \textit{C, W}    & 338    & 316   & 93.5\% & 282   & 83.4\% \\ \hline
\multirow{2}{*}{\texttt{SMS-Cop.}}   & \textit{R, P, I, O}     & 293K   & 241K  & 82.1\% & 177K  & 60.3\% \\
            & \textit{C, W}    & 31.1K  & 27.0K & 86.8\% & 21.0K & 67.5\% \\ \hline
\multirow{2}{*}{\texttt{SMS-A}}       & \textit{R, P, I, O}    & 894K   & 745K  & 83.4\% & 561K  & 62.8\% \\
            & \textit{C, W}    & 66.0K  & 58.6K & 88.8\% & 43.6K & 66.1\% \\ \hline
\end{tabular}
\label{tab:reduction}
\vspace{-2ex}
\end{table}

We first discuss the motif counts and spectrum under different timing constraints by using the event pairs (\cref{subsec:count}). Then, we look at the behavior of intermediate events (non-first, non-last) (\cref{subsec:inter}). Last, we check how the motif timespans are shaped by the different timing parameters (\cref{subsec:span}).

\subsubsection{Motifs counts and event pairs}\label{subsec:count}
We compare the motif counts obtained with different timing constraint configurations and analyze the frequency and ratio of the event pairs (R, P, I, O, C, W) in 3n3e motifs.

\begin{figure}[b!]
\vspace{-4ex}
\centering
\captionsetup[subfigure]{captionskip=0.5ex}

\subfloat[{3e motifs (\texttt{Stack.})}]
{\label{fig:stack-pie-3}
\hspace{-2ex}
\includegraphics[width=0.24\linewidth]{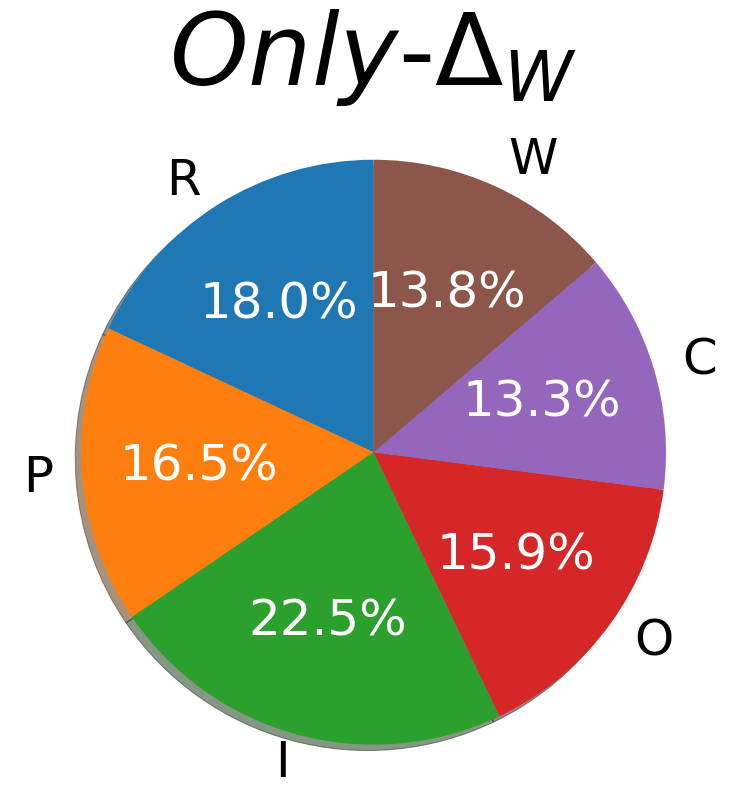}
\hspace{-2ex}
\includegraphics[width=0.24\linewidth]{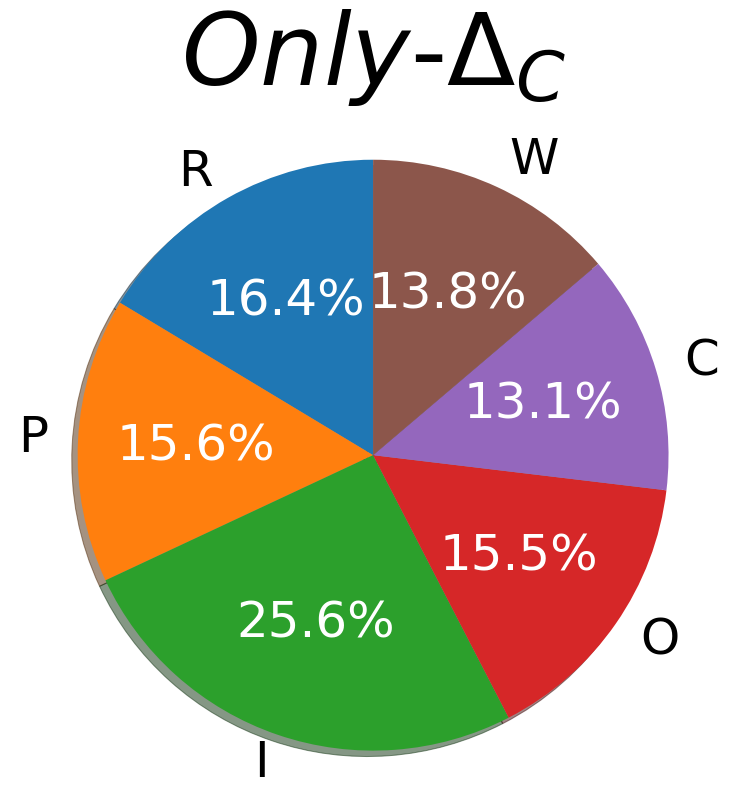}
}
\hspace{-1ex}
\includegraphics[width=0.07\linewidth]{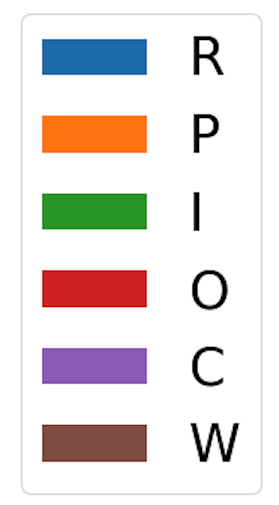}
\hspace{-1ex}
\subfloat[{4e motifs (\texttt{Calls-Cop.})}]
{\label{fig:calls-pie-4}\includegraphics[width=0.24\linewidth]{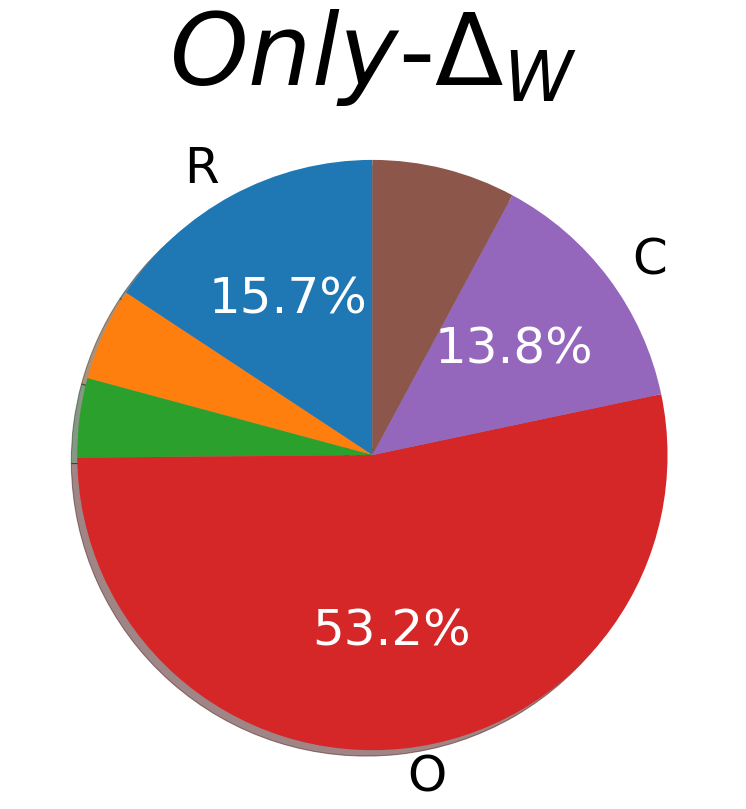}
\hspace{-2ex}
\includegraphics[width=0.24\linewidth]{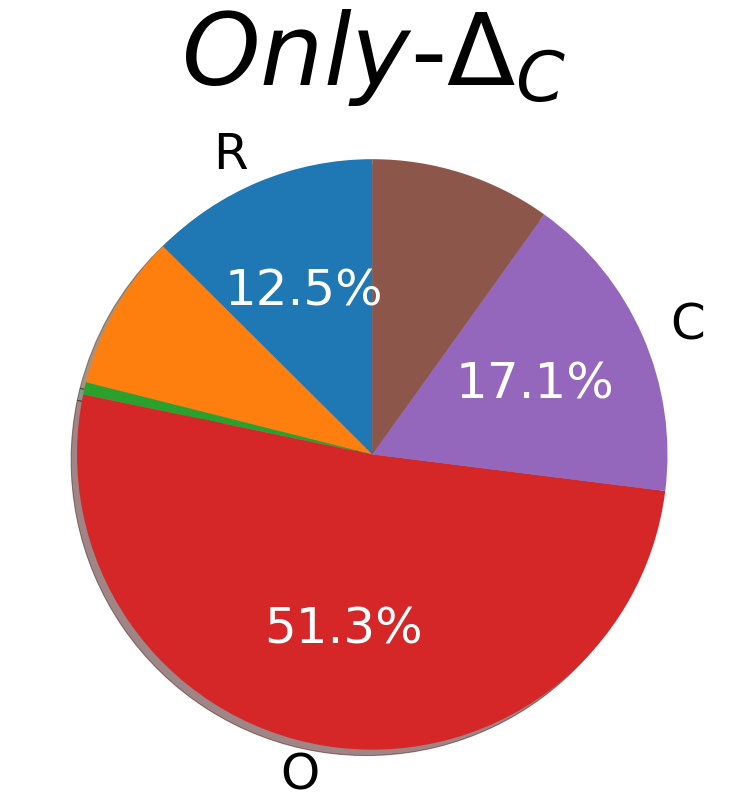}
\hspace{-1ex}
}
\caption{\it Ratios of event pairs in \dw and \dc configurations. Each figure shows a pair of pie charts for the ratio of event pairs in six categories.~\cref{fig:stack-pie-3} has the distributions for 3e motifs in \texttt{StackOverflow};~\cref{fig:calls-pie-4} gives the ratios for 4e motifs in \texttt{Calls-Copenhagen}. The proportion of repetitions decreases in almost all datasets when going from \dw to \dc while the ones with increasing ratios are different. 
}
\label{fig:ratiosep}
\end{figure}

\textbf{We observe that considering \dw amplifies the number of \textit{R, P, I, O} motifs.}
Enforcing the $\Delta_C$ constraint helps to find less \textit{R, P, I, O} motifs since consecutive events should be close to each other.
\cref{tab:reduction} shows that \textit{R, P, I, O} motifs are overrepresented in the situation where $\Delta_C$ is ignored.
The number of \textit{R, P, I, O} motifs is 10 times greater than \textit{C, W} motifs in all datasets.
The number of \textit{R, P, I, O} motifs observed in \dc configuration is 40 percent less than the \dw configuration. 
Among all datasets message networks are affected most, where the number of \textit{R, P, I, O} motifs is reduced to 56.8\% in \texttt{CollegeMsg} and 60.3\% in \texttt{SMS-Copenhagen}.
The number of \textit{C, W} motifs is also reduced when switched to \dc configuration, but the reduction rates are smaller than the \textit{R, P, I, O} motifs.
Using both constraints can achieve a balance between the two extremes, where the number of \textit{R, P, I, O} motifs is reduced to nearly 80 percent and the \textit{C, W} motifs is reduced to 90 percent.

\begin{figure*}[t!]
\centering
\subfloat[\ti{010102} motif for \texttt{SMS-Copenhagen}] %
{
\label{fig:sms-intermediate}
\includegraphics[width=0.33\linewidth]{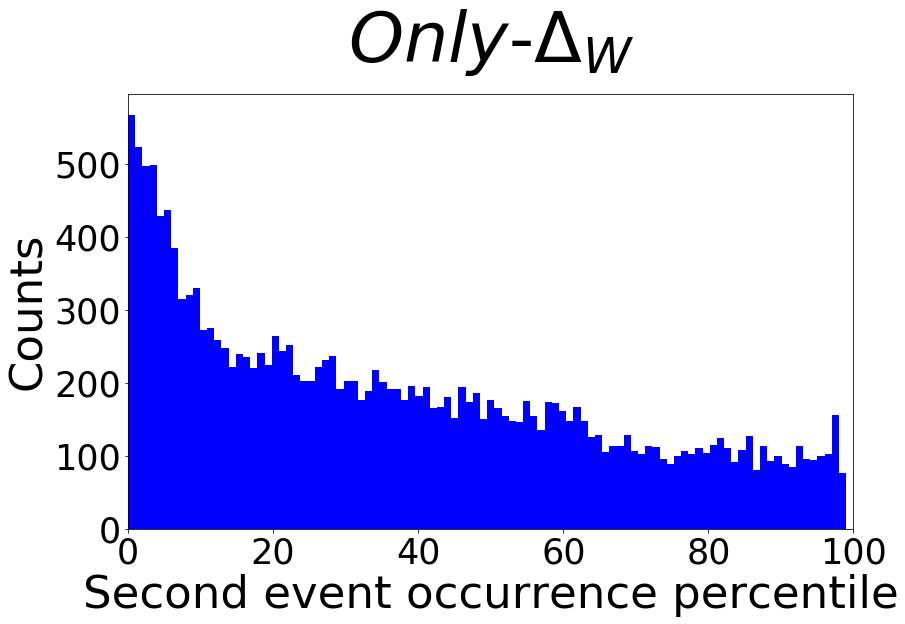}
\includegraphics[width=0.33\linewidth]{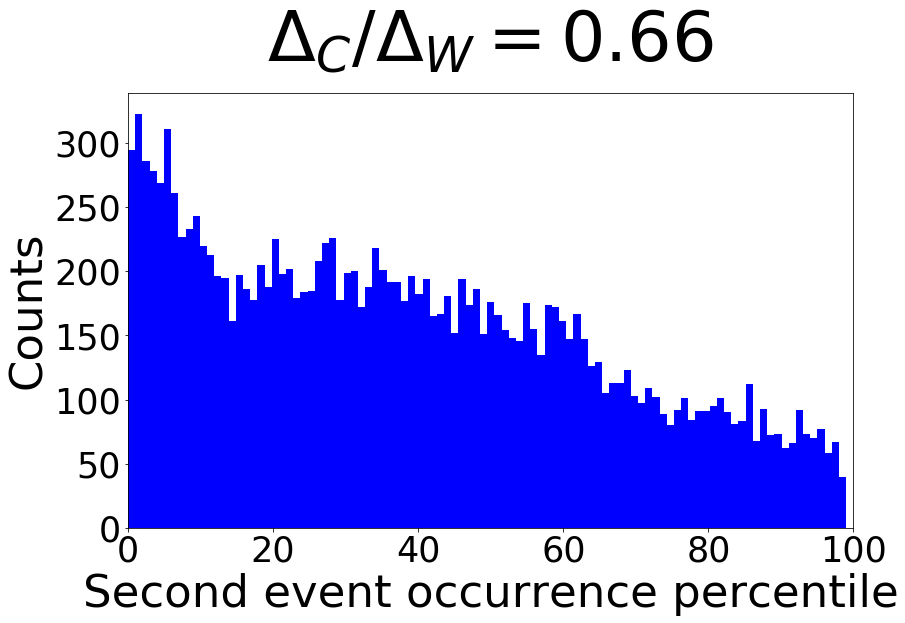}
\includegraphics[width=0.33\linewidth]{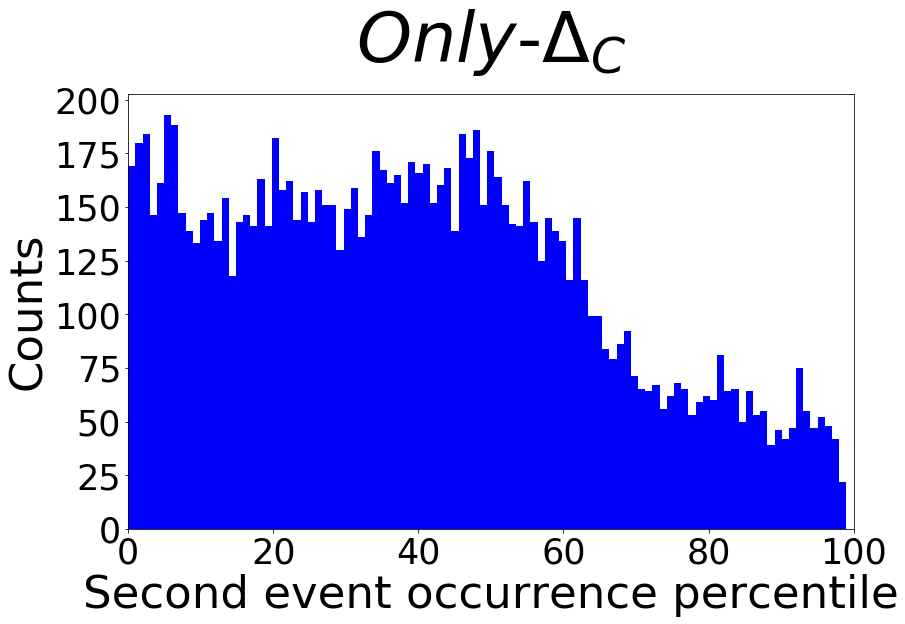}
}

\vspace{-2ex}

\subfloat[\ti{011221} motif for \texttt{FBWall}] %
{
\label{fig:fb-intermediate}
\includegraphics[width=0.33\linewidth]{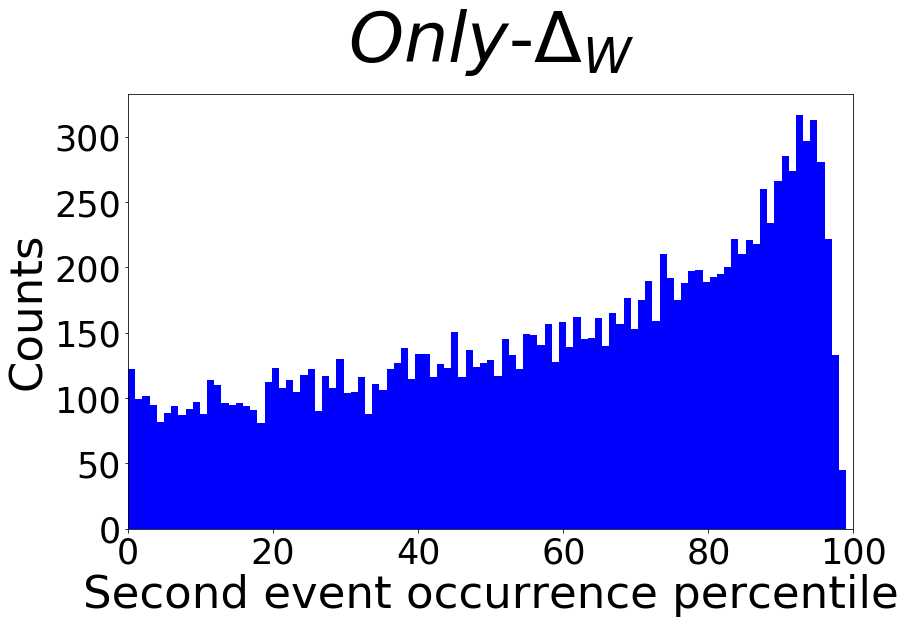}
\includegraphics[width=0.33\linewidth]{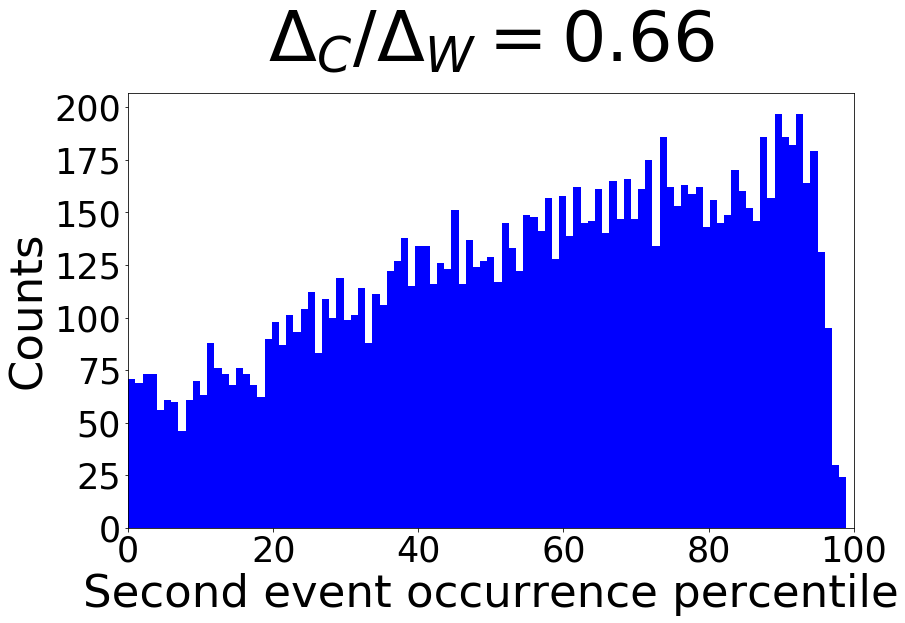}
\includegraphics[width=0.33\linewidth]{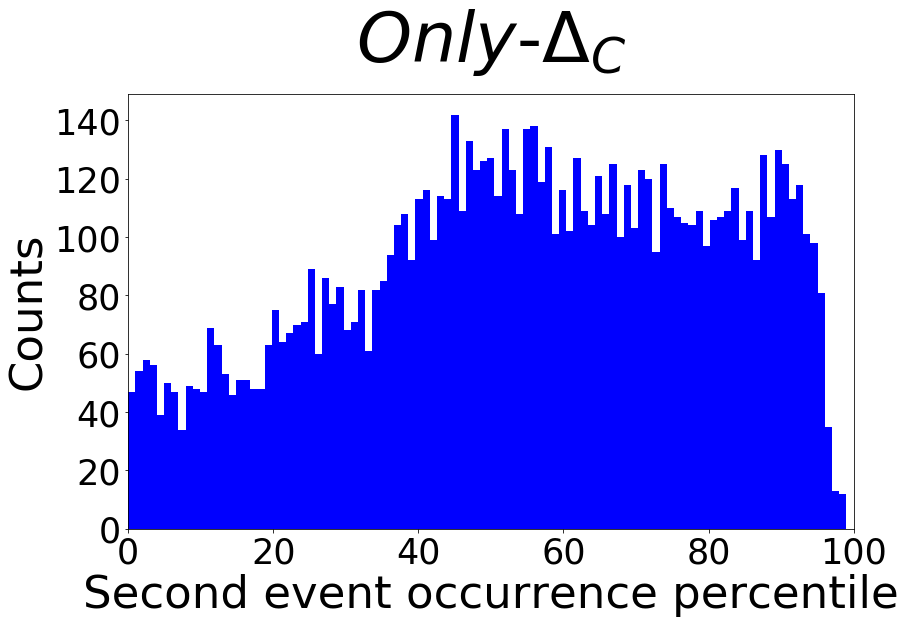}
}

\vspace{-2ex}

\subfloat[\ti{01212303} motif for \texttt{CollegeMsg}] %
{
\label{fig:CollegeMsg-intermediate}
\includegraphics[width=0.24\linewidth]{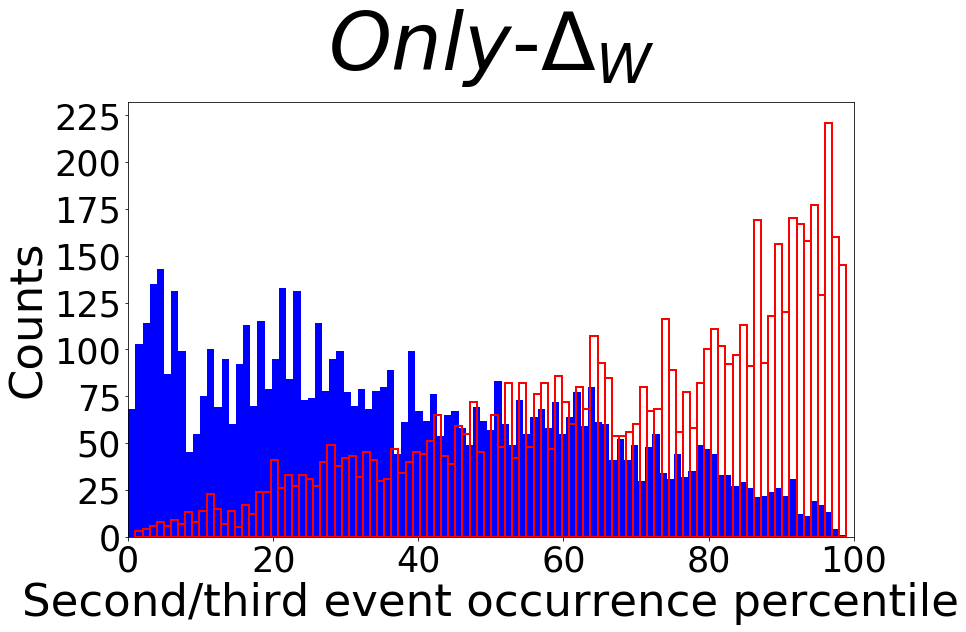}
\includegraphics[width=0.24\linewidth]{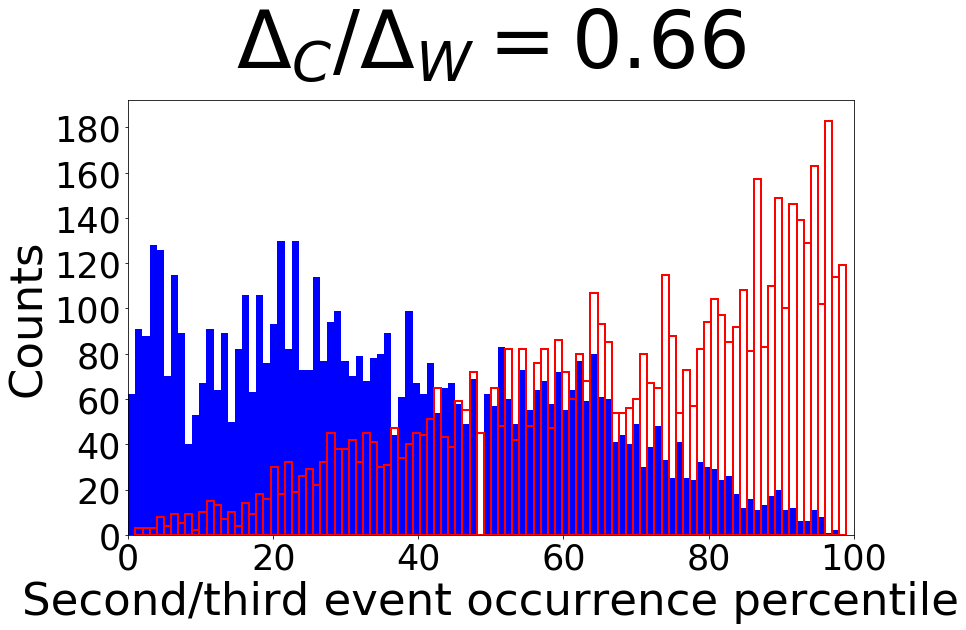}
\includegraphics[width=0.24\linewidth]{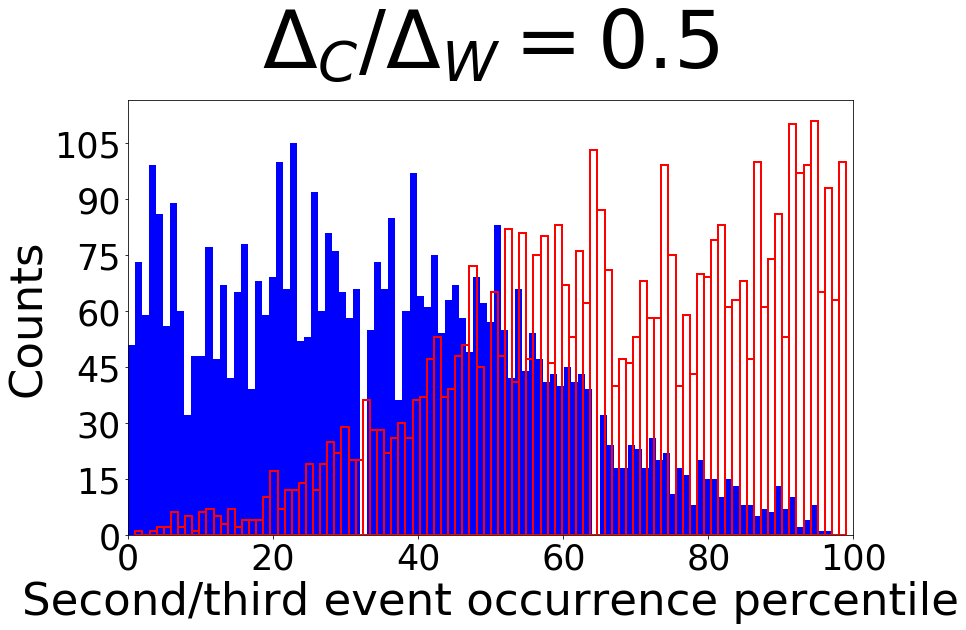}
\includegraphics[width=0.24\linewidth]{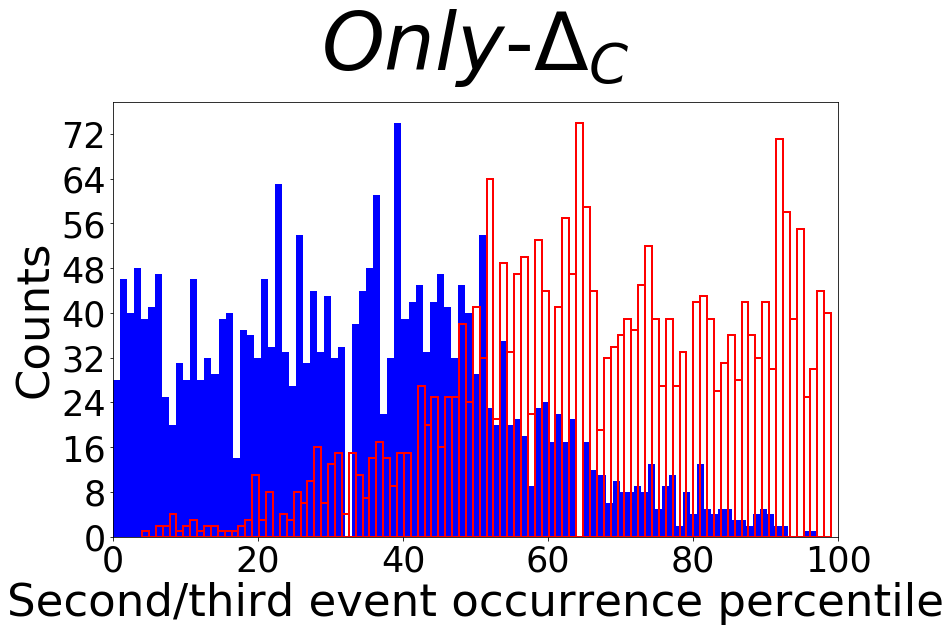}
}

\vspace{-2ex}

\caption{\it Behavior of intermediate event occurrences. In each figure, the x-axis denotes the occurrence time of intermediate events with respect to the first and last events. 0\% denotes the first event occurrence and 100\% represents the last event occurrence. The y-axis shows the frequency of the intermediate events; second events (blue) in three-event motifs and second \& third events (blue \& red) in four-event motifs. In all cases, enforcing the $\Delta_C$ constraint regularizes the skew in \dw case.
}
\label{fig:intermediate}
\vspace{-4ex}
\end{figure*}

We also compare the ratios of event pair types between \dc and \dw configurations for all three-events (3e) (two- or three-nodes) and four-events (4e) (two-, three-, or -four nodes) motifs.
~\cref{fig:ratiosep} presents representative results; for \texttt{StackOverflow} (3e motifs) and \texttt{Calls-Copenhagen} (4e motifs) (complete results are available in the supplementary materials).
\textbf{In most datasets, we observe that the ratio of repetitions decreases when going from \dw to \dc configuration whereas the increases show variety}.
The ratio of repetitions for 3e motifs in \texttt{StackOverflow} is reduced from 18.0\% in \dw to 16.4\% in \dc (\cref{fig:stack-pie-3}).
It is similar for 4e motifs; in \texttt{Calls-Copenhagen}, the repetitions decrease from 15.7\% in \dw to 12.5\% in \dc (\cref{fig:calls-pie-4}).
The decrease in repetitions also means an increase in other types of event pairs.
For stack exchange interactions, the ratio of in-bursts increases: for 3e motifs in \texttt{StackOverflow}, the in-bursts increases from 22.5\% in \dw to 25.6\% in \dc. As interactions in \texttt{StackOverflow} are answers and comments, this indicates that new posts are often answered and commented by many different persons in a short time period, and we can have better grasp of the in-bursts of answers and comments by bringing the $\Delta_C$ constraint. 
For 4e motifs in \texttt{Calls-Copenhagen}, on the other hand, the proportions of ping-pongs and conveys increase when going from \dw to \dc configuration. This implies that in CDR networks ask-reply and message delivering patterns tend to happen in a short time period, therefore become more prominent in the \dc configuration.

\subsubsection{Intermediate event behaviors}\label{subsec:inter}
Next, we investigate the intermediate event behaviors under different configurations.
Since $\Delta_W$ only limits the interval between the first and the last events, it has no control on the behavior of intermediate events, i.e., when the second event happens in three-event motifs or when the second/third events happen in four-event motifs.
Due to the bursty nature of temporal networks, the intermediate events in the motifs can be heavily skewed to the first event or the last event. 
Ideally, such bias is greatly reduced by $\Delta_C$ constraint, since it limits the time difference between consecutive events.
Here we show a few representative motifs in different datasets for 3n3e and 4n4e motifs in~\cref{fig:intermediate} (all results are available in the supplementary materials).
~\cref{fig:sms-intermediate} shows the intermediate event occurrence of motif 010102 in \texttt{SMS(Copenhagen)}. In \dw case, the occurrence of the second event is significantly skewed to the first event since the second event is a repetition of the first. The lapse between the second and the last event can cover the entire timespan of the motif, and it is hard to guarantee the relevance of the last event in this scenario. 
{\bf The skewness in the second event occurrence is reduced when ${\Delta_C}/{\Delta_W}=0.66$, and the distribution is further regularized in \dc configuration.}
Similar skewed patterns are also observed in the other 3n3e motifs. ~\cref{fig:fb-intermediate} displays the intermediate event occurrence of motif 011221 in \texttt{FBWall}. Due to the ping-pong behavior formed by the last two events, the second event occurrence is skewed to the last event for \dw, and the bias is reduced in ${\Delta_C}/{\Delta_W}=0.66$ and \dc configurations.
We also observed similar patterns in 4n4e motifs.~\cref{fig:CollegeMsg-intermediate} shows the intermediate event occurrence of motif 01212303 in \texttt{CollegeMsg}. In \dw case, the occurrence of the second event is skewed to the first event and the third event is skewed to the last event, as both first-second and third-fourth pairs are in-bursts.
{\bf Tuning ${\Delta_C}/{\Delta_W}$ from 1.0 to 0.33 regularizes the distribution of the occurrence }\newstuff{ and ensures that the timing of the intermediate event does not depend on the type of the motif.}

\begin{figure}[b!]
\vspace{-2ex}
\centering
\includegraphics[width=0.33\linewidth]{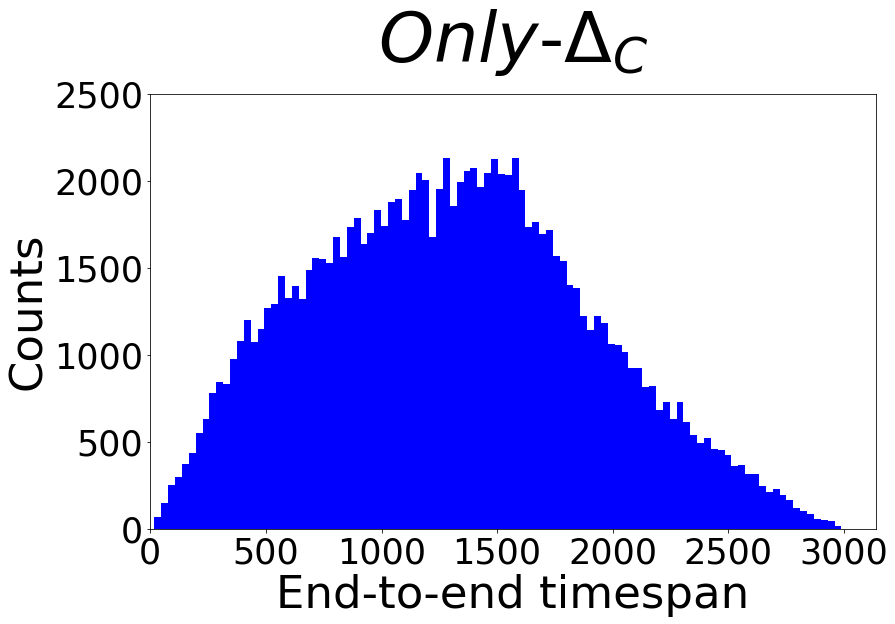}
\hspace{-1ex}
\includegraphics[width=0.33\linewidth]{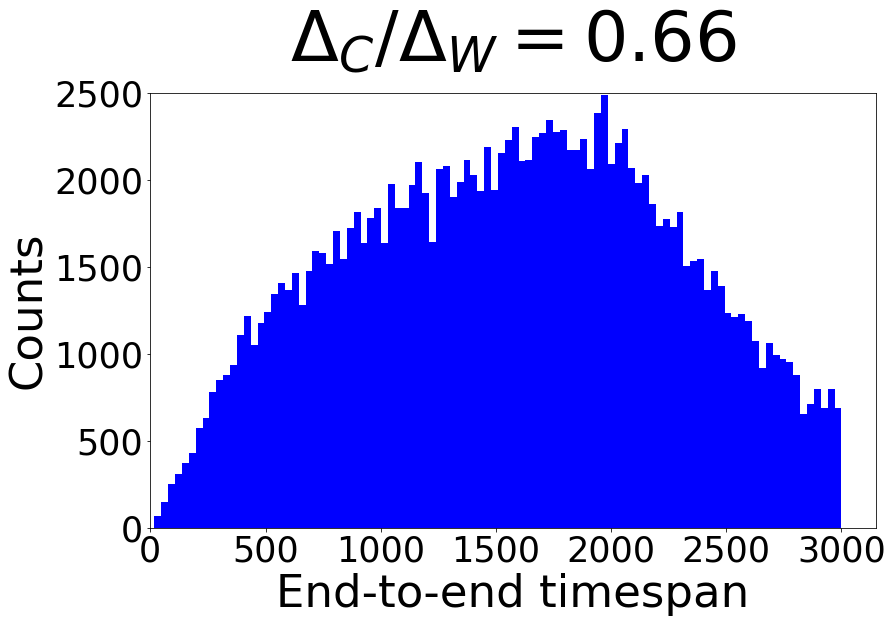}
\hspace{-1ex}
\includegraphics[width=0.33\linewidth]{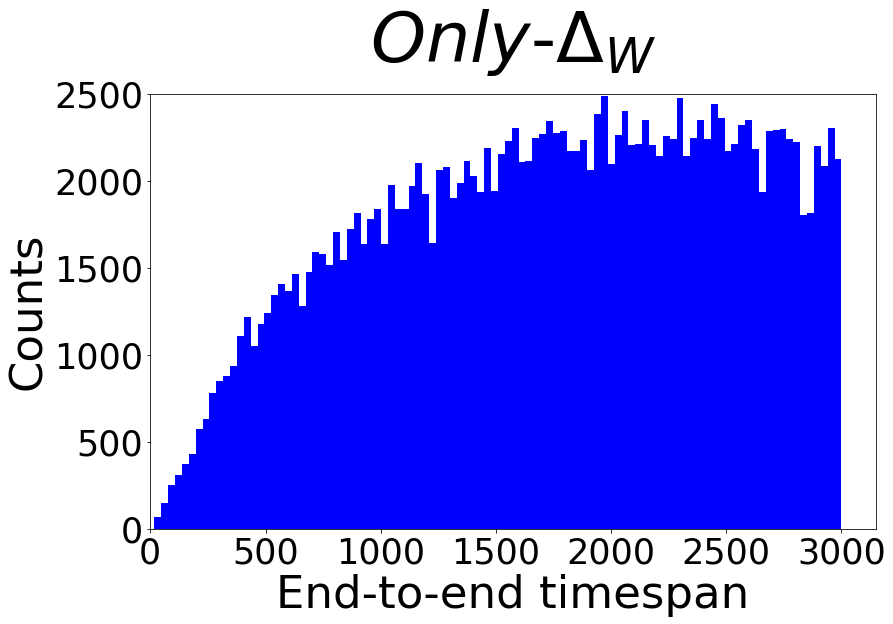}
\vspace{-4ex}
\caption{\it Distribution of the motif timespans for 010102 motifs in \texttt{CollegeMsg}. The x-axis denotes the timespan of the motif and the y-axis shows the count of such motifs. The distributions are more regularized when going from \dc to \dw.
}
\label{fig:span}
\end{figure}

\vspace{-1ex}
\subsubsection{The timespan of motifs}\label{subsec:span}

Lastly, we compare the distributions of motif timespans (the time difference between the last and first events) for \dc and \dw configurations. 
Note that the $\Delta_C$ only gives a loose bound to the timespan of motifs ($\Delta_C*(m-1)$ where $m$ is the number of events), while the $\Delta_W$ gives a hard limit.
\cref{fig:span} shows a representative result; the timespan of all 010102 motifs in \texttt{CollegeMsg} for different timing constraints (other results are in the supplementary materials).
In \dc, the distribution of motif timespans follows a normal distribution where the mean is approximately equal to the value of $\Delta_C$. This indicates that, although $\Delta_C$ provides a loose bound, using \dc fails to control the timespan of motifs. {\bf We observe that the motif timespans are more uniform in ${\Delta_C}/{\Delta_W}=0.66$ configuration, and the distribution is further regularized in \dw configuration.}
\newstuff{Obtaining a set of motifs with uniform timespan distribution can be important in the analysis of real-world events which shows a variety in the timespan. For example, people have different churn behaviors in subscription services (some leave after a week and some others unsubscribe after a month) and one may be interested to see the temporal motifs that lead to the attrition. In such cases, selecting the motifs with uniform time distribution can enable to see the patterns that are related to customer's timeline rather than the absolute period of subscription.}

\noindent{\bf Summary.} We observe that ${\Delta_C}$ and ${\Delta_W}$ have complementary features; the former fails to bound timespans whereas the latter introduces bias for the occurrence of intermediate events. Combining both parameters by choosing a ratio for ${\Delta_C}/{\Delta_W}$ in $(\frac{1}{m -1}, 1)$ interval can yield a trade-off.

\ignore{%
\begin{figure}[t!]
\centering
\captionsetup[subfigure]{captionskip=-1ex}
\includegraphics[width=\linewidth]{charts/fig11/facebook.png}
\caption{\bf The timespan distributions of motif 010202 and 011212 in \texttt{FBWall}. For a given type of motif, we count the number of motifs in 10 seconds time bins, and reveal the distribution of motif timespans. The solid lines denote the distribution in \dc configuration, and the dotted lines represent the distribution in  \dw configuration. Overrepresented motifs with large timespan are removed by the $\Delta_C$ constraint, while the regular motifs and motifs with small timespan are not affected.}
\label{fig11:FBWall}
\vspace{-3ex}
\end{figure}

In addition, we also investigate the difference between timespan distributions of different motifs. Generally we observe two patterns: 1) the motif timespan follows a normal distribution, 2) the motif timespan follows a uniform distribution, which means the motif count is invariant over time.
The distribution of timespan also demonstrates how the number of motifs is reduced by adding the $\Delta_C$ constraint. As we mentioned in \cref{subsec:count}, the $\Delta_C$ greatly reduces some overrepresented motifs, and ensures the causal relations between consecutive events.
\cref{fig11:FBWall} shows the pattern of motif 010202 and 011212 in \dw and \dc configurations. 
The motif 010202 is a overrepresented motif, which has the largest count among all three-node three-event motifs. The distribution of motif 010202 in \dw configuration indicates a large amount of motifs with timespan between 1000s and 2000s. This points out that if the $\Delta_W$ constraint is raised to a particular value, the number of motif 010202 explodes. Since the last two events in the motif correspond to a repetition, it implies that a irrelevant first event which occurred long time ago are taken into the motif when the $\Delta_W$ constraint is increased to the particular value. Such motifs are removed in \dc configuration, as the peak in the \dw configuration is eliminated. On the other hand,  motifs with timespan less than 1000s are not affected by the $\Delta_C$ constraint, as the distributions between two configurations are exactly the same. 
Note that those non-overrepresented motifs hardly decrease from \dw to \dc configuration. For example, the distributions of motif 011212 between two configurations show minor difference.
}

\begin{figure*}[!t]
\centering
\subfloat[\ti{\texttt{SMS-A}}]{\label{fig:heat-SMS-A}\includegraphics[width=0.22\linewidth]{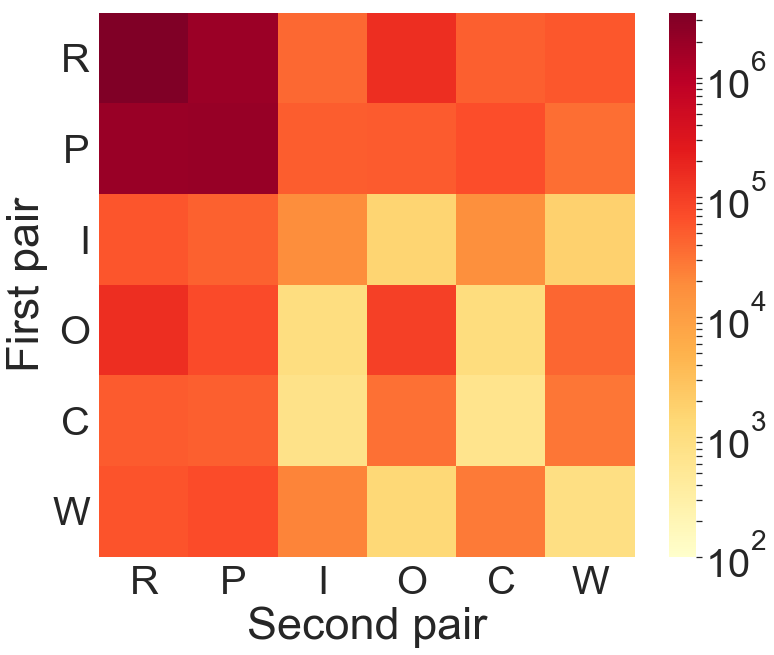}}
\subfloat[\ti{\texttt{SMS-Copenhagen}}]{\label{fig:heat-sms}\includegraphics[width=0.22\linewidth]{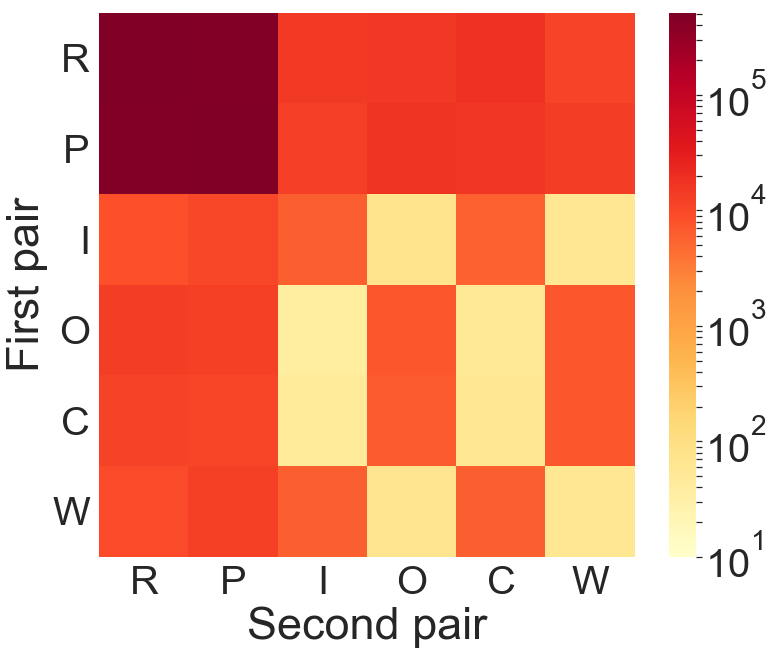}}
\subfloat[\ti{\texttt{Calls-Copenhagen}}]{\label{fig:heat-calls}\includegraphics[width=0.22\linewidth]{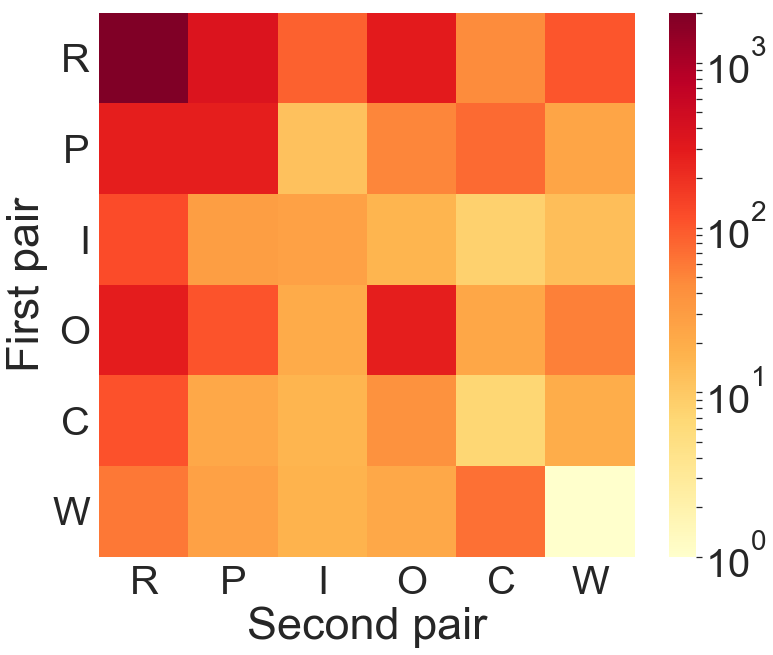}}
\subfloat[\ti{\texttt{Email}}]{\label{fig:heat-email}\includegraphics[width=0.22\linewidth]{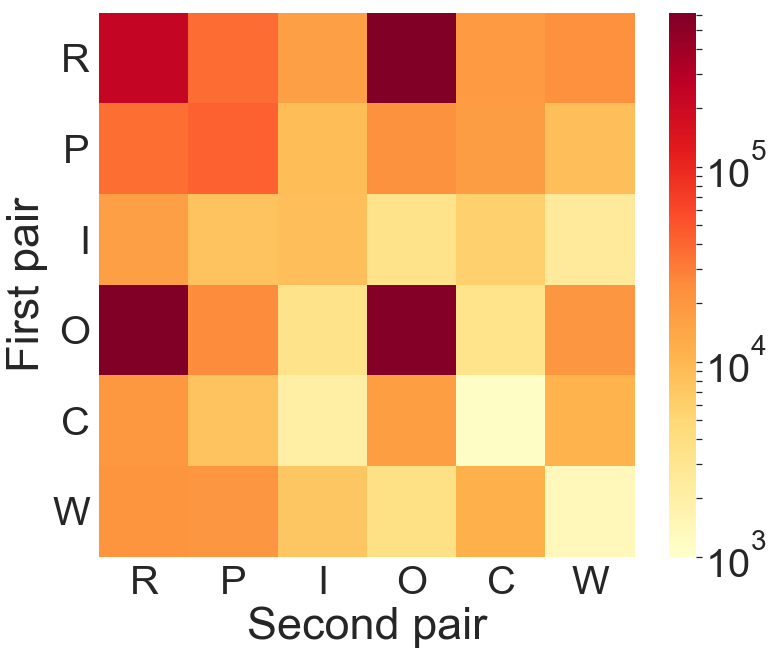}}
\vspace{-2ex}
\caption{\it Ordered sequences of event pairs for three-event motifs. Each block denotes a type of three-event motif, where y-axis shows the first pair of events (first and second event) and the x-axis shows the second pair of events (second and third event). The color indicates the motif counts in log scale and calculated with respect to the minimum and maximum counts in each dataset. Repetitions and ping-pongs are the most dominant in message networks, while repetitions and out-bursts are more common in calls and emails. Also, conveys, in-bursts, and out-bursts have asymmetrical trends in their compatibility.
}
\vspace{-4ex}
\label{fig:heat}
\end{figure*}

\vspace{-2ex}
\subsection{Motifs as sequence of event pairs}\label{sec:eventpairs}
In this section, we use our new lens, event pairs, to analyze the characteristics of the sequences.
In particular, we are interested in the
temporal correlations among different types of event pairs and how the ordered sequences of event pairs show variety in various datasets.
\cref{fig:heat} presents the heat maps for all three-event motifs (both two- and three-nodes) in \texttt{SMS-A}, \texttt{SMS-Copenhagen}, \texttt{Calls-Copenhagen}, and \texttt{Email} using both timing constraints ($\Delta_C=2000s$ and $\Delta_W=3000s$); densities are color coded with respect to the min. and max. values in each dataset (all results are available in the supplementary materials).
Since there are two pairs of events in the three-event motifs, we represent those as an ordered sequence of two pairs in a heat map.
Regarding the counts, \textbf{we often observe the majority of motifs are formed by the sequences involving repetitions, while only a few motifs are formed by the sequences including weakly-connected event pairs.} This is also observed in the previous studies~\cite{P17, zhao2010communication}.
\newstuff{The weakly-connected event pairs are rare since the two events take place in a non-transitive order, therefore are likely to be irrelevant}.

In terms of the ordered sequences, we observe highly similar patterns in all message networks. %
The sequences involving both the repetitions and ping-pongs are the majority.
Note that only two nodes are involved if there are only repetitions and ping-pongs; thus those interactions in message networks tend to be very local, and in most cases one-to-one conversations. 
Message networks also show a certain preference in the sequences involving in-bursts, out-bursts, and conveys. In-bursts and out-bursts seem to be incompatible as we do not observe many motifs that contain both.
{\bf Conveys are often followed by out-bursts but not by in-bursts or conveys, and in-bursts are often followed with conveys, while the opposites rarely happen.}
We believe it is due to nature of information. 
\newstuff{The information delivered by conveys is likely to be distributed to multiple nodes through out-bursts.
Meanwhile, the information gathered from multiple nodes through in-bursts is often further propagated through conveys. }
In \texttt{Calls-Copenhagen} and \texttt{Email}, there are less motifs formed by the sequences involving ping-pongs; however out-bursts are very dominant. Phone calls are not instantaneous events and happen for a duration; thus the information are mutually exchanged and repetitions are less likely to happen.
Similarly, email communications do not have information limit when compared to the messages.
Asymmetrical trends observed for message networks are also common for calls and emails.

\noindent{\bf Summary.} Sequences of event pairs suggest interesting findings about the interplays among different types of pairs. In particular, the commonalities observed for message networks and the asymmetrical trends are surprising thanks to event pairs based analysis.

%% file: appendix.tex
\appendices

\onecolumn
\section{Impact of timing constraints}
Here we present the full collection of figures that show the difference between \dw and \dc configurations.

\subsection{Ratio of event pairs with timing constraints}
\begin{figure*}[hp]
\centering
\captionsetup[subfigure]{captionskip=-1ex}

\subfloat[\ti{\texttt{Calls-Copenhagen}: three-event \dw vs. three-event \dc vs. four-event \dw vs. four-event \dc}]
{\label{fig:calls-pie}\includegraphics[width=0.25\linewidth]{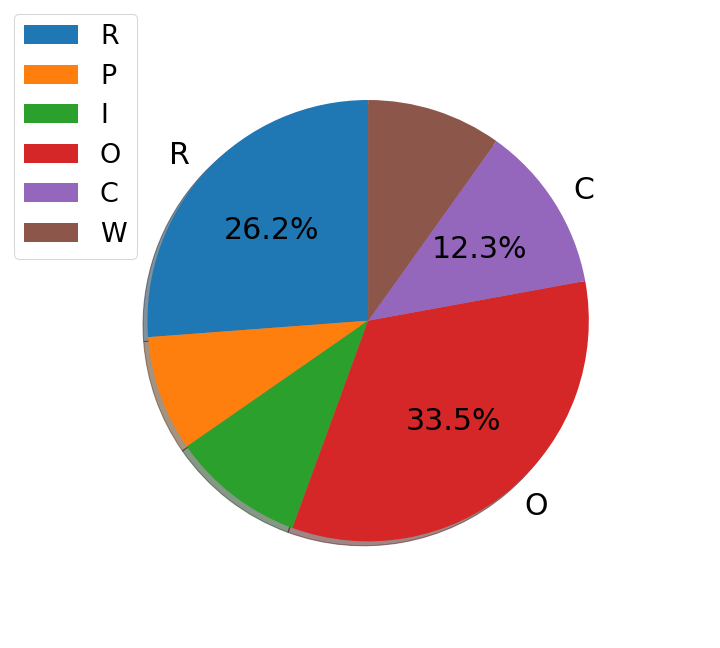}
\includegraphics[width=0.25\linewidth]{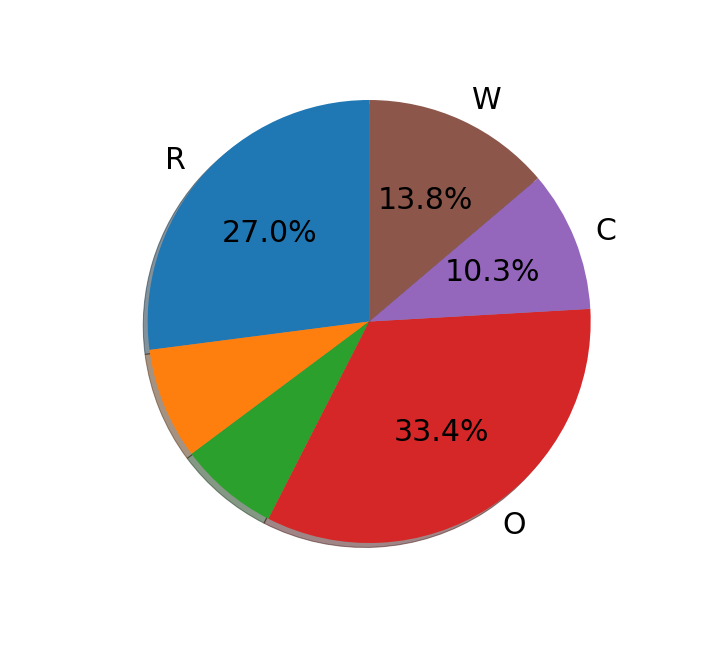}
\includegraphics[width=0.25\linewidth]{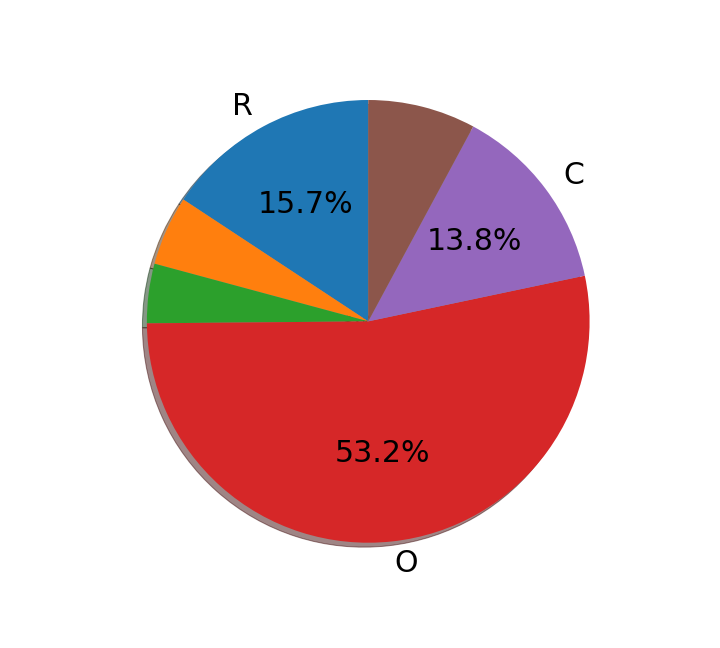}
\includegraphics[width=0.25\linewidth]{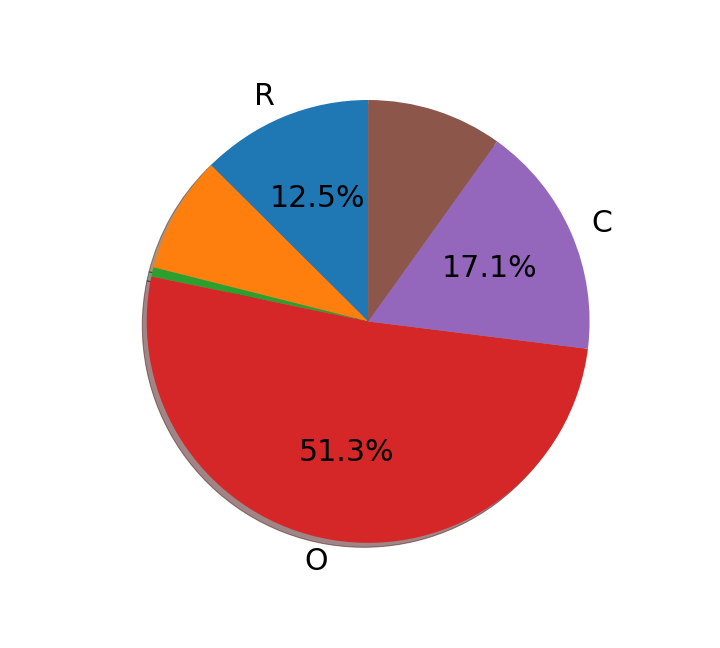}
}
\vspace{-2ex}

\subfloat[\ti{\texttt{CollegeMsg}: three-event \dw vs. three-event \dc vs. four-event \dw vs. four-event \dc}]
{\label{fig:CollegeMsg-pie}\includegraphics[width=0.25\linewidth]{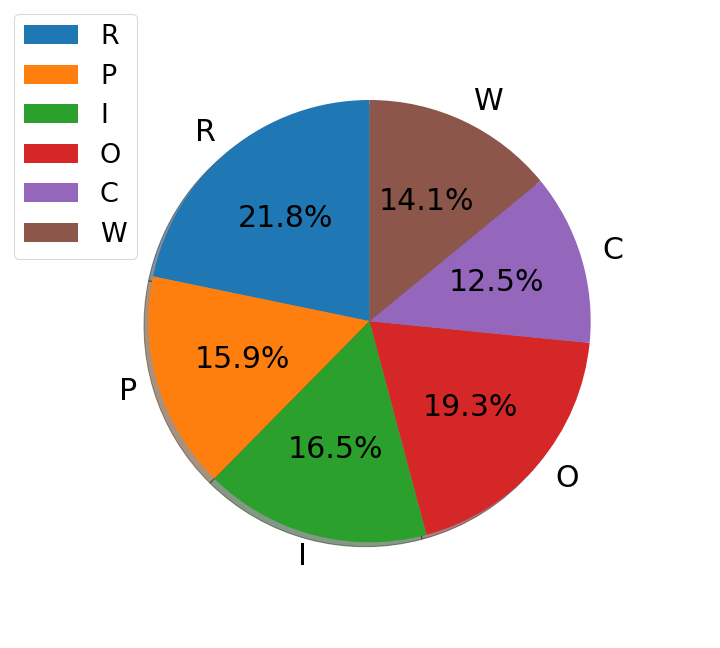}
\includegraphics[width=0.25\linewidth]{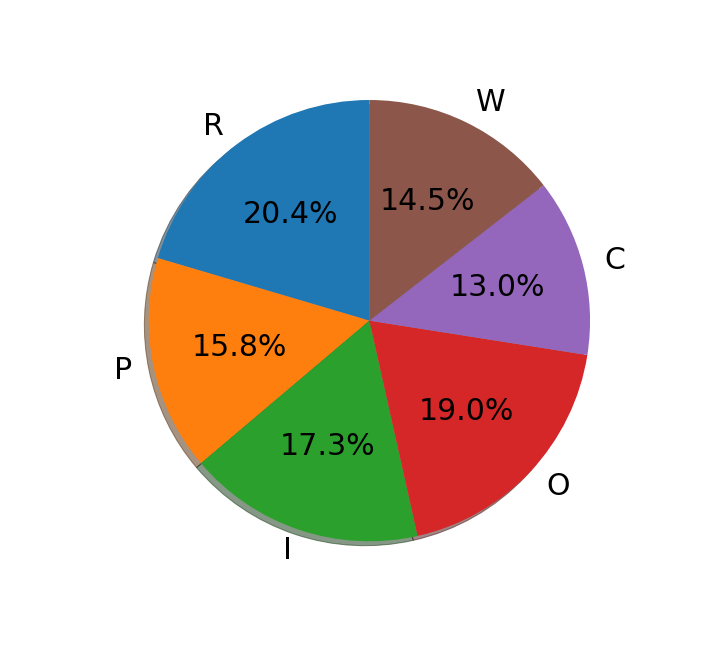}
\includegraphics[width=0.25\linewidth]{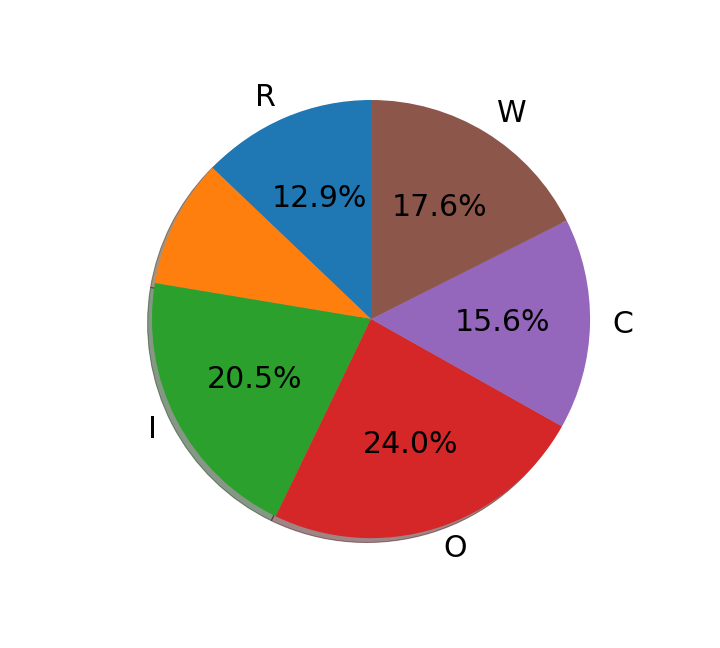}
\includegraphics[width=0.25\linewidth]{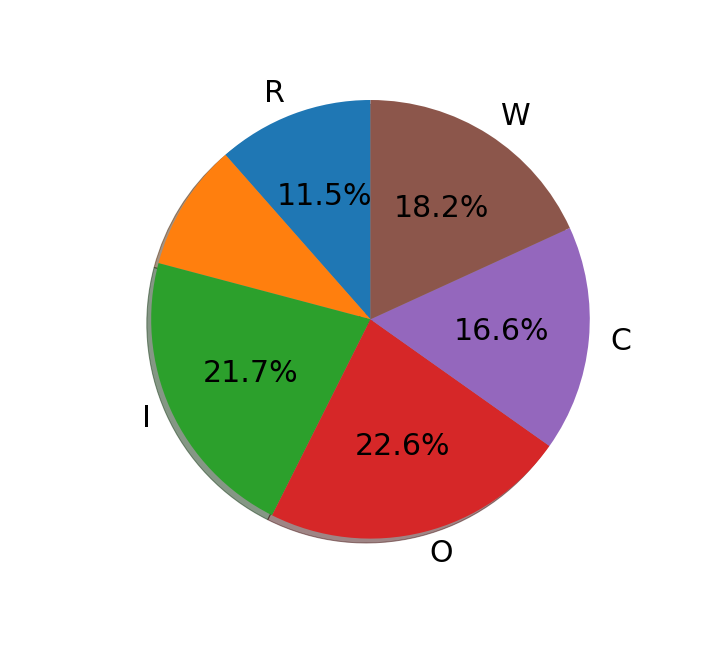}
}
\vspace{-2ex}

\subfloat[\ti{\texttt{Email}: three-event \dw vs. three-event \dc vs. four-event \dw vs. four-event \dc}]
{\label{fig:Email-pie}\includegraphics[width=0.25\linewidth]{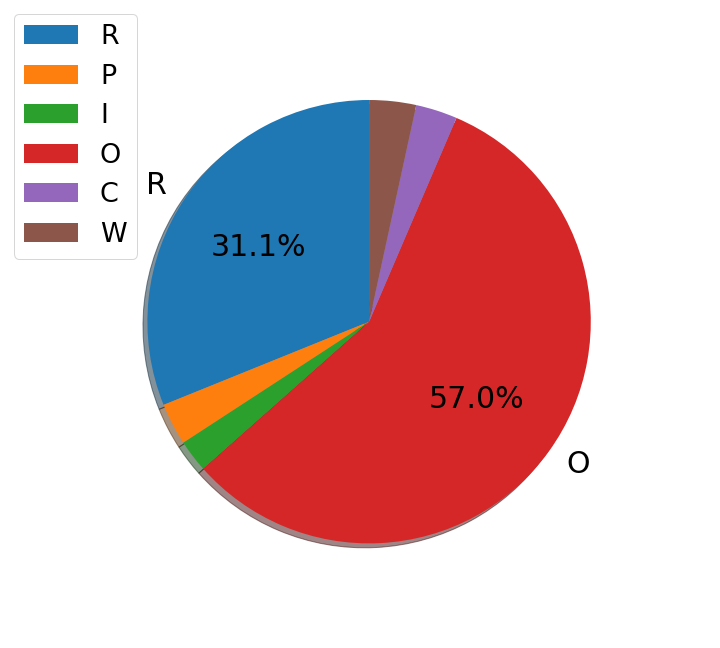}
\includegraphics[width=0.25\linewidth]{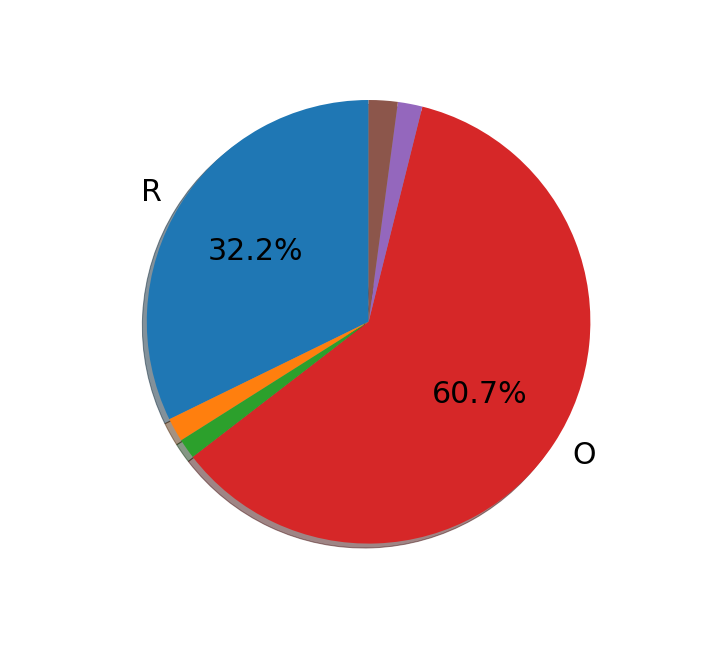}
\includegraphics[width=0.25\linewidth]{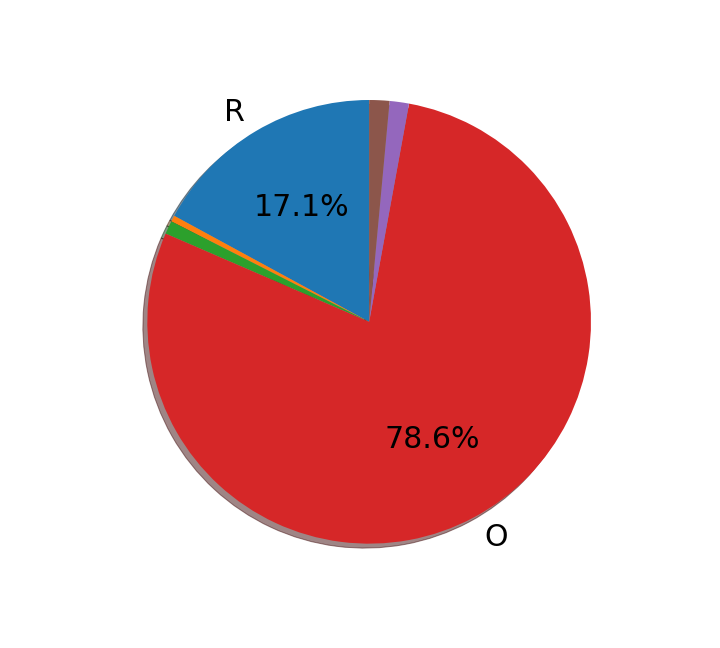}
\includegraphics[width=0.25\linewidth]{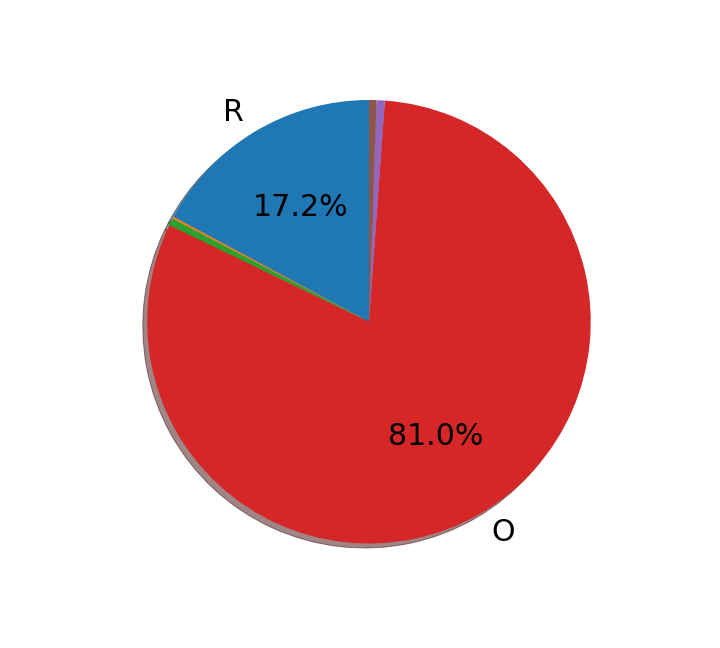}
}
\vspace{-2ex}

\subfloat[\ti{\texttt{FBWall}: three-event \dw vs. three-event \dc vs. four-event \dw vs. four-event \dc}]
{\label{fig:FBWall-pie}\includegraphics[width=0.25\linewidth]{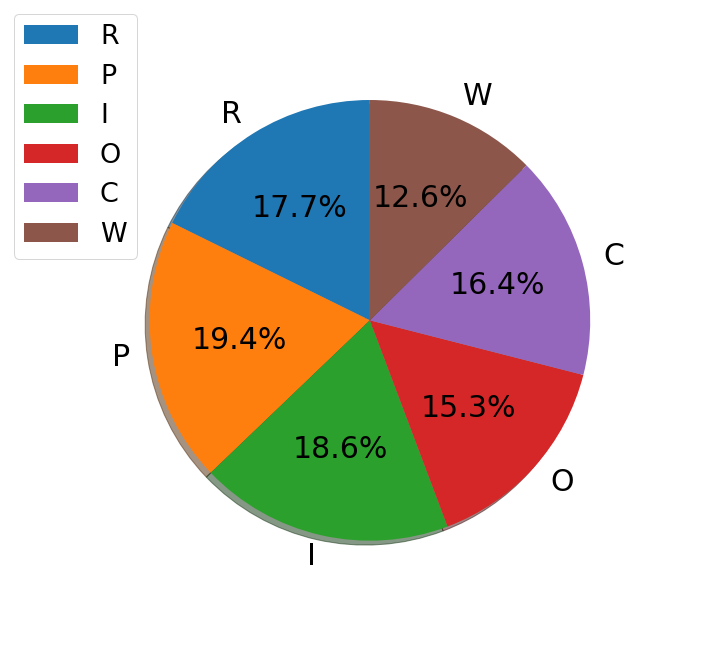}
\includegraphics[width=0.25\linewidth]{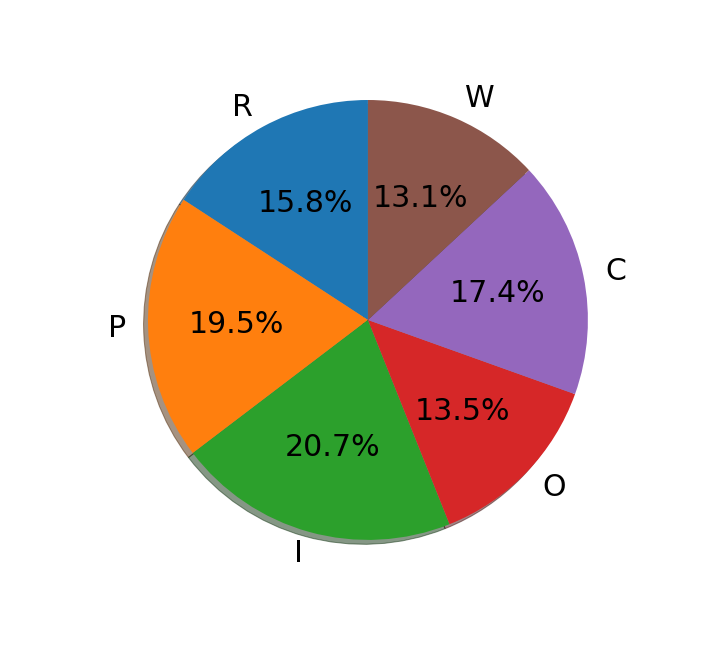}
\includegraphics[width=0.25\linewidth]{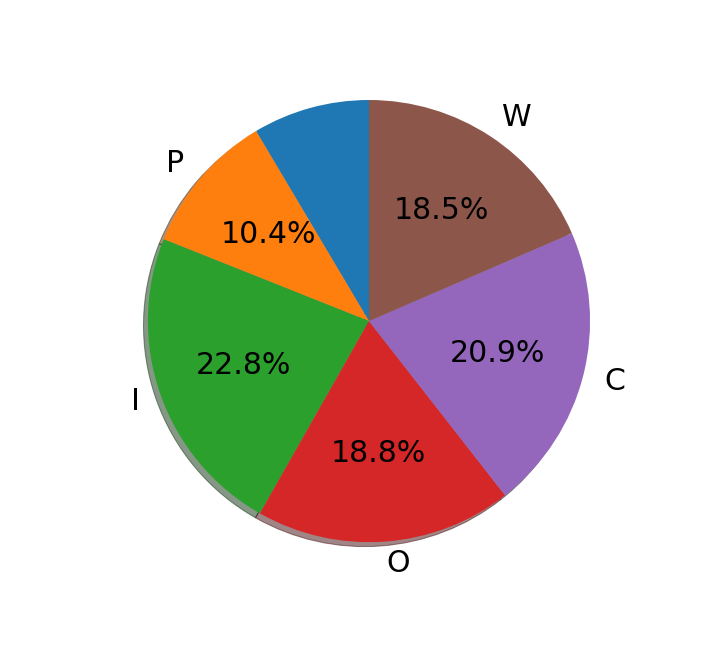}
\includegraphics[width=0.25\linewidth]{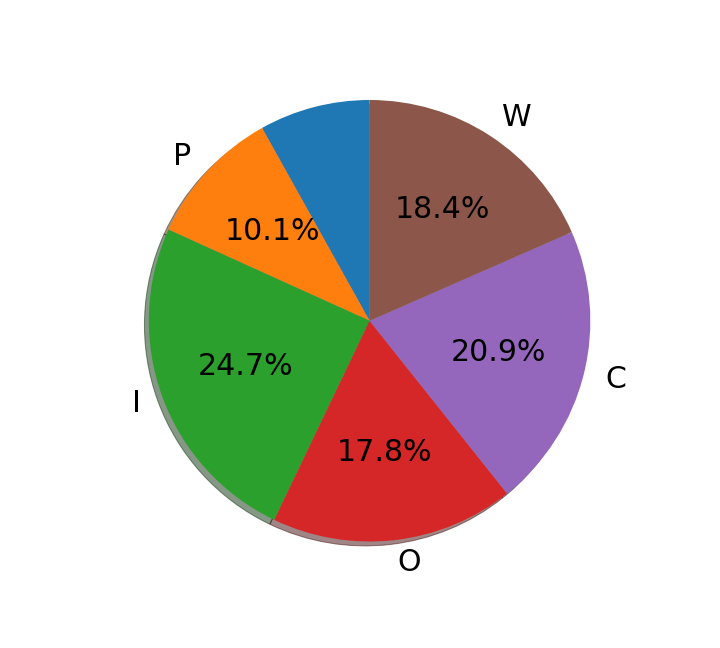}
}
\vspace{-2ex}
\caption{\it Proportion of event pairs in three-event and four-event motifs between \dw and \dc configurations (PART 1). Each figure shows the ratio of event pairs in six categories.}
\vspace{-3ex}
\label{fig:pie-1}
\end{figure*}

\begin{figure*}[!hp]
\centering
\captionsetup[subfigure]{captionskip=-1ex}

\subfloat[\ti{\texttt{Bitcoin-otc}: three-event \dw vs. three-event \dc vs. four-event \dw vs. four-event \dc}]
{\label{fig:Otc-pie}\includegraphics[width=0.25\linewidth]{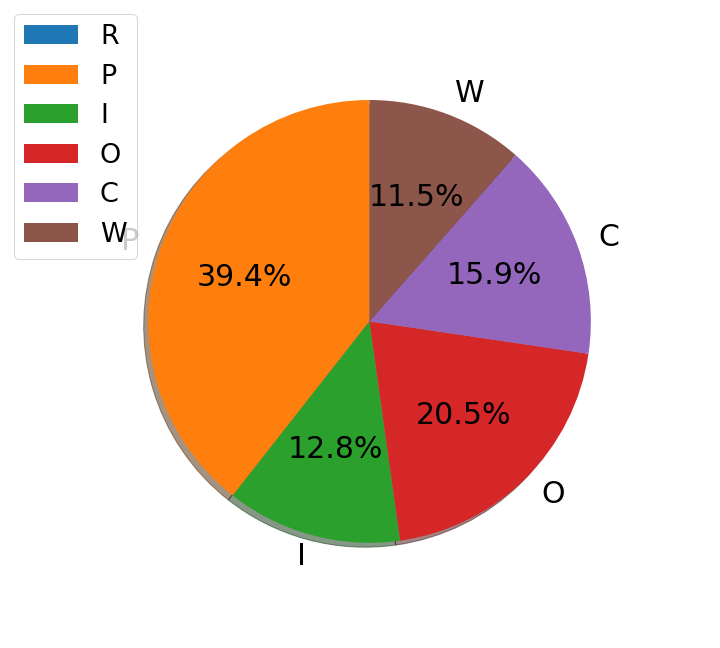}
\includegraphics[width=0.25\linewidth]{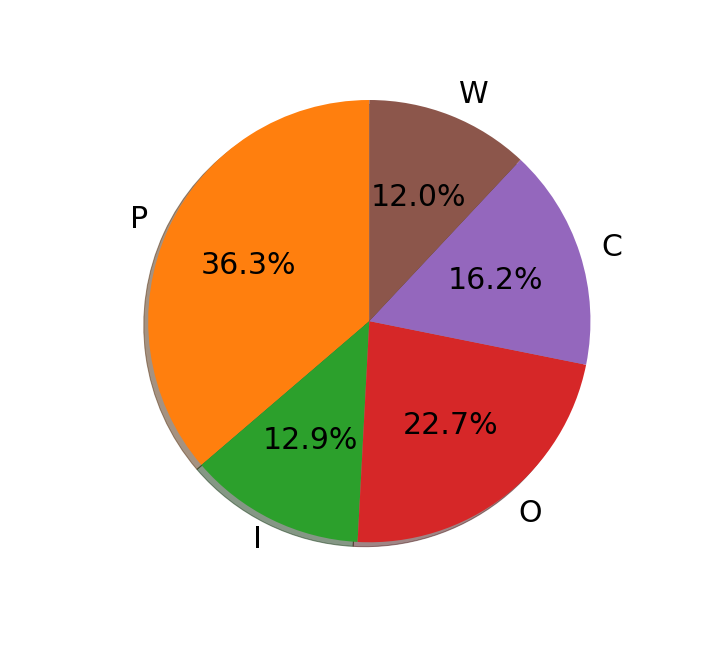}
\includegraphics[width=0.25\linewidth]{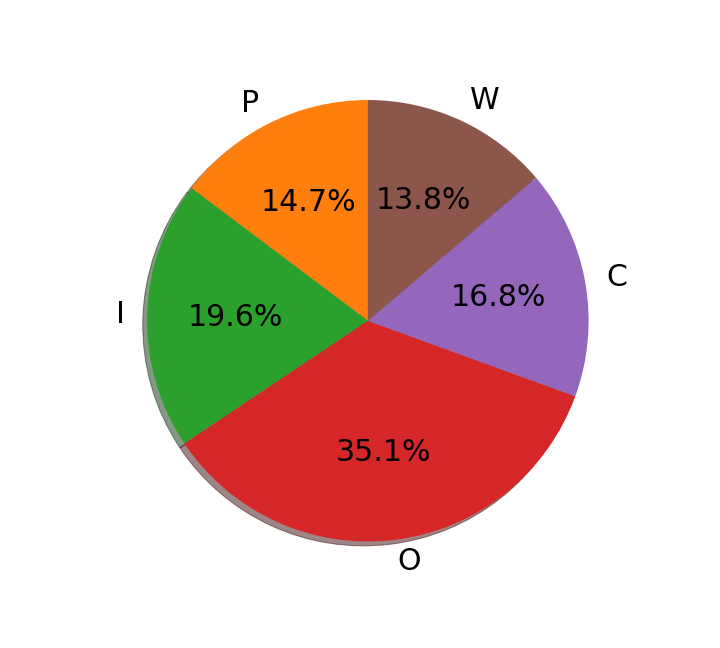}
\includegraphics[width=0.25\linewidth]{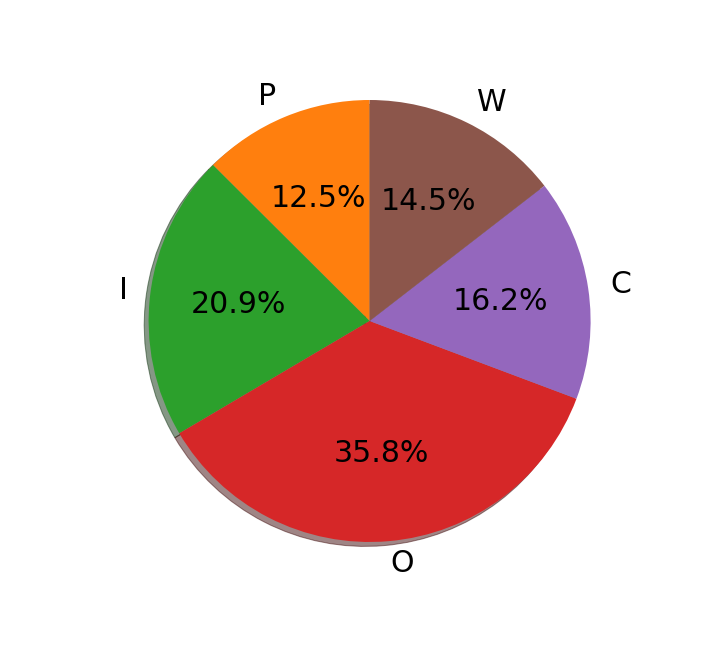}
}
\vspace{-2ex}

\subfloat[\ti{\texttt{SMS-A}: three-event \dw vs. three-event \dc vs. four-event \dw vs. four-event \dc}]
{\label{fig:SMS-A-pie}\includegraphics[width=0.25\linewidth]{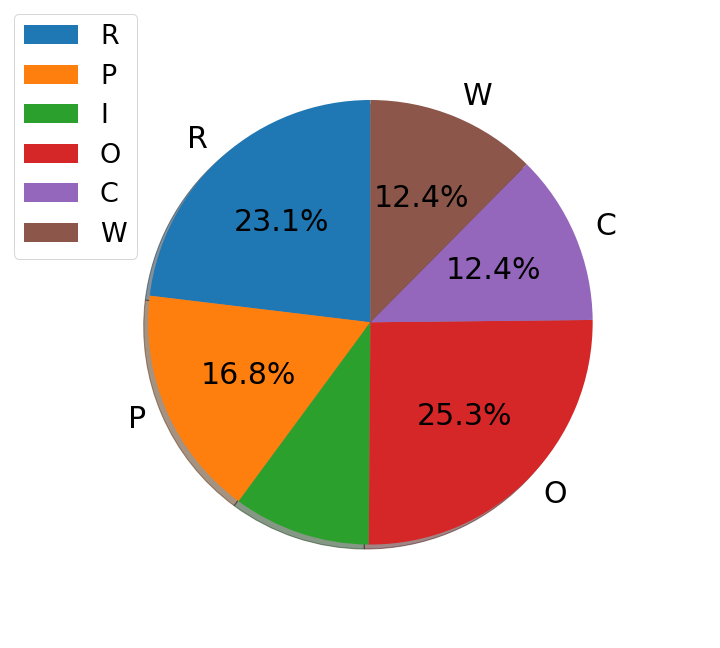}
\includegraphics[width=0.25\linewidth]{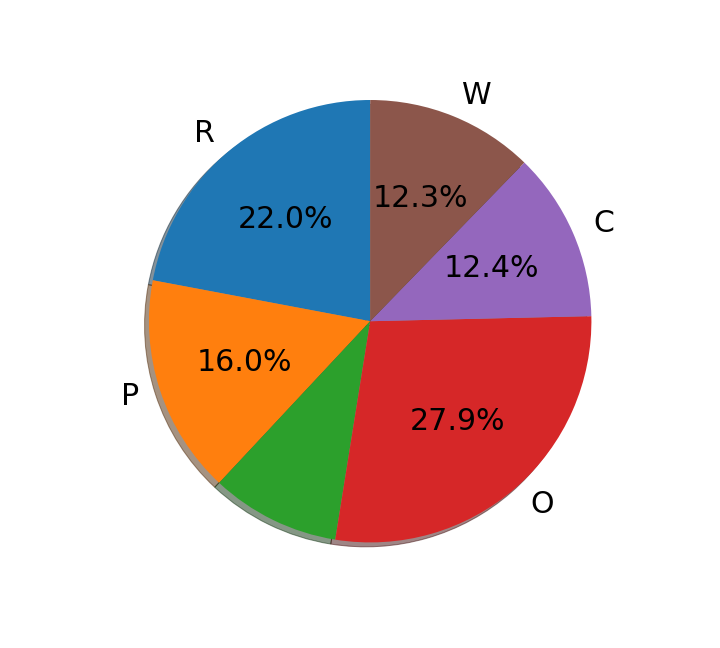}
\includegraphics[width=0.25\linewidth]{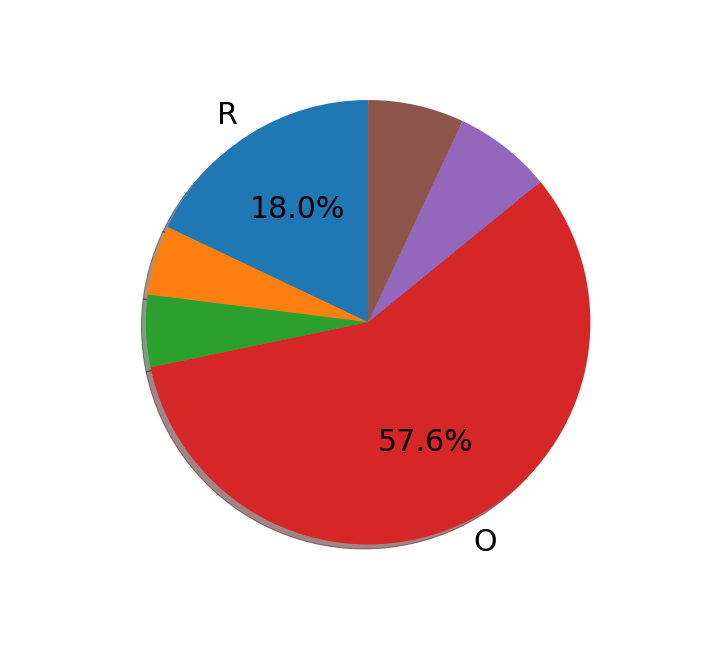}
\includegraphics[width=0.25\linewidth]{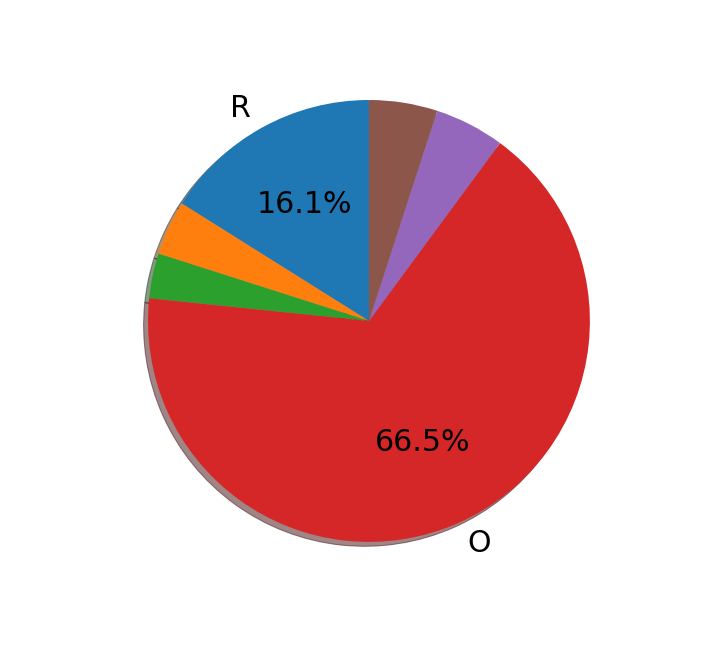}
}
\vspace{-2ex}

\subfloat[\ti{\texttt{SMS-Copenhagen}: three-event \dw vs. three-event \dc vs. four-event \dw vs. four-event \dc}]
{\label{fig:sms-pie}\includegraphics[width=0.25\linewidth]{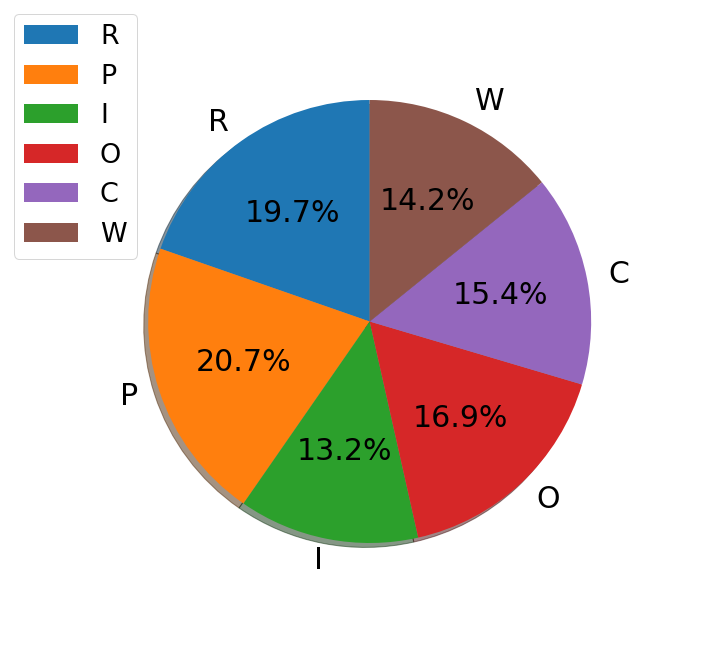}
\includegraphics[width=0.25\linewidth]{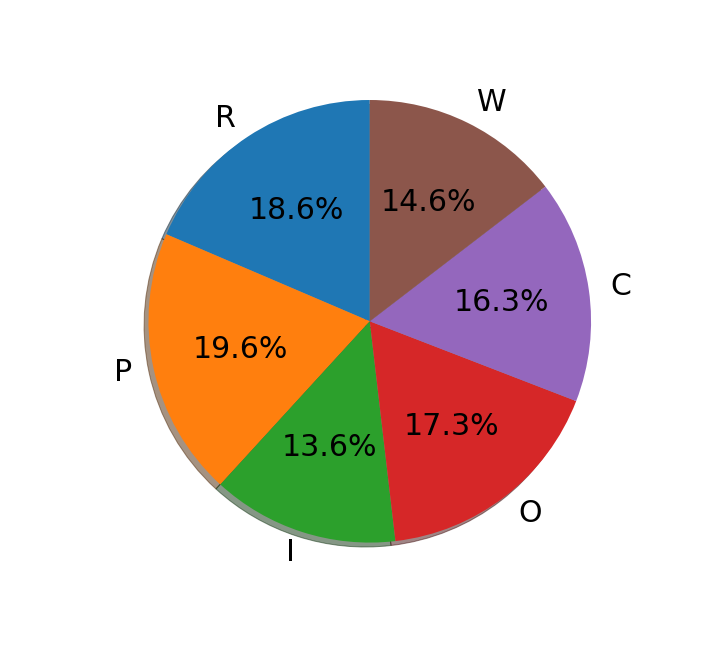}
\includegraphics[width=0.25\linewidth]{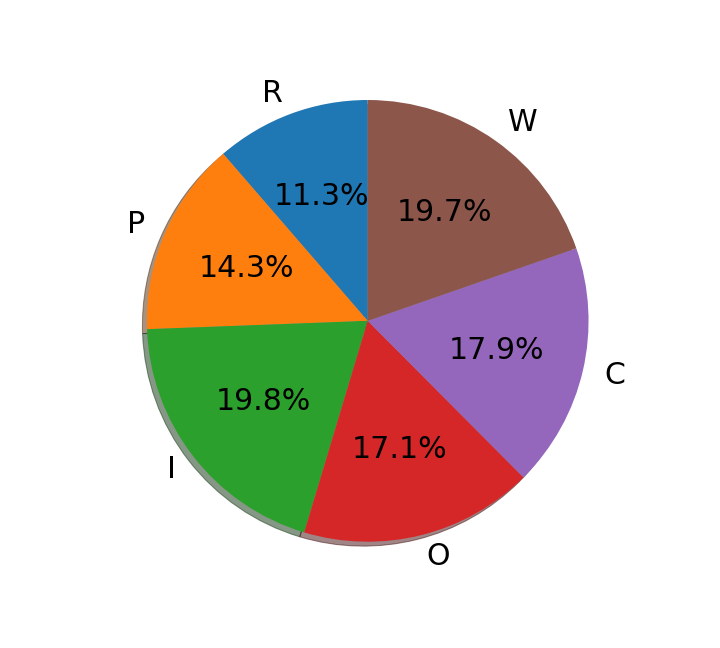}
\includegraphics[width=0.25\linewidth]{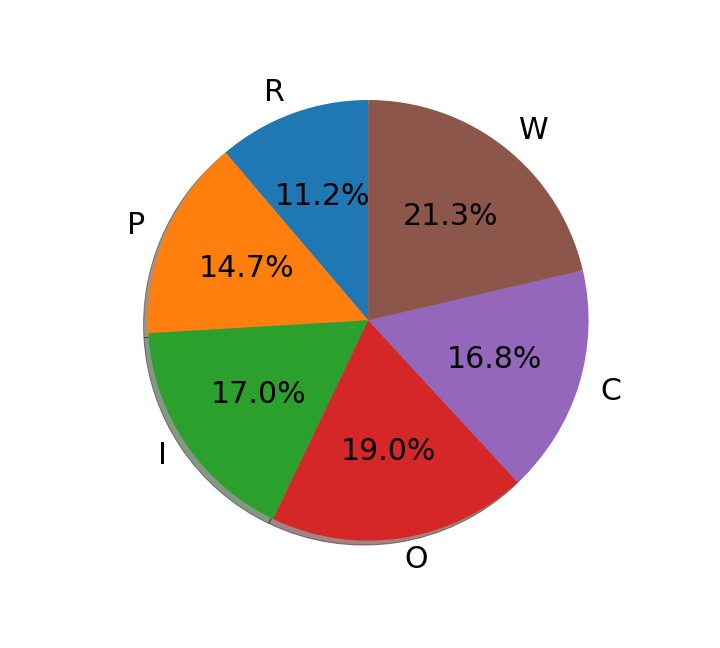}
}
\vspace{-2ex}

\subfloat[\ti{\texttt{StackOverflow}: three-event \dw vs. three-event \dc vs. four-event \dw vs. four-event \dc}]
{\label{fig:stack-pie}\includegraphics[width=0.25\linewidth]{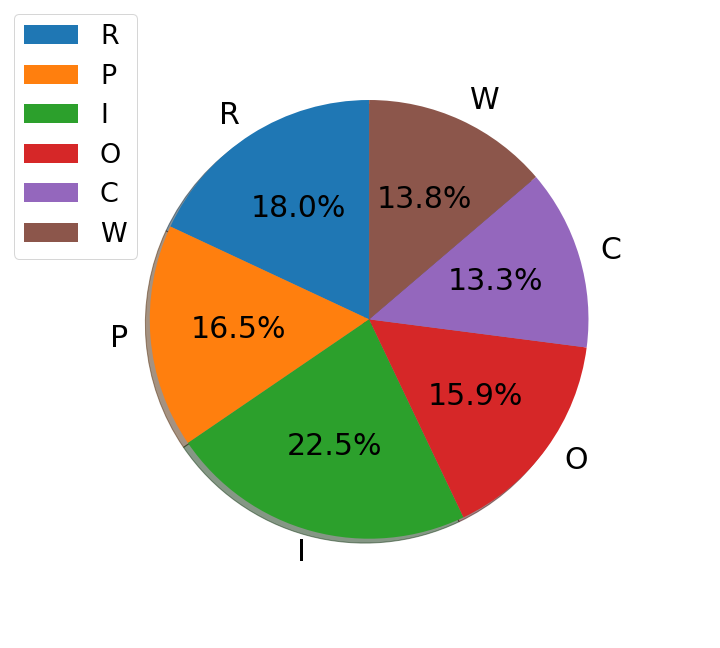}
\includegraphics[width=0.25\linewidth]{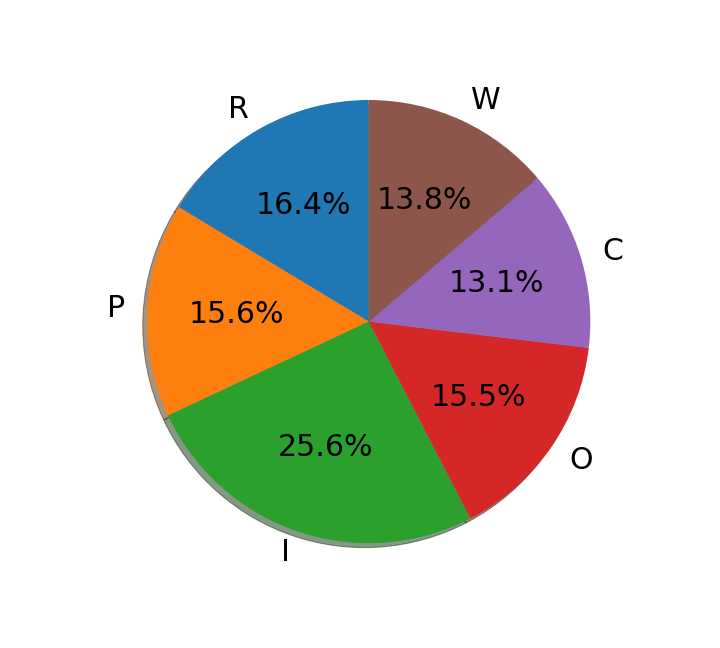}
\includegraphics[width=0.25\linewidth]{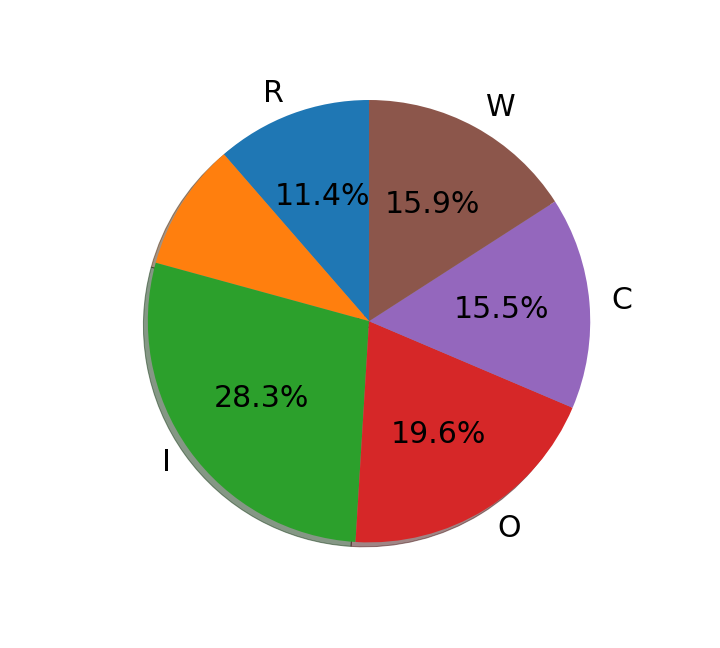}
\includegraphics[width=0.25\linewidth]{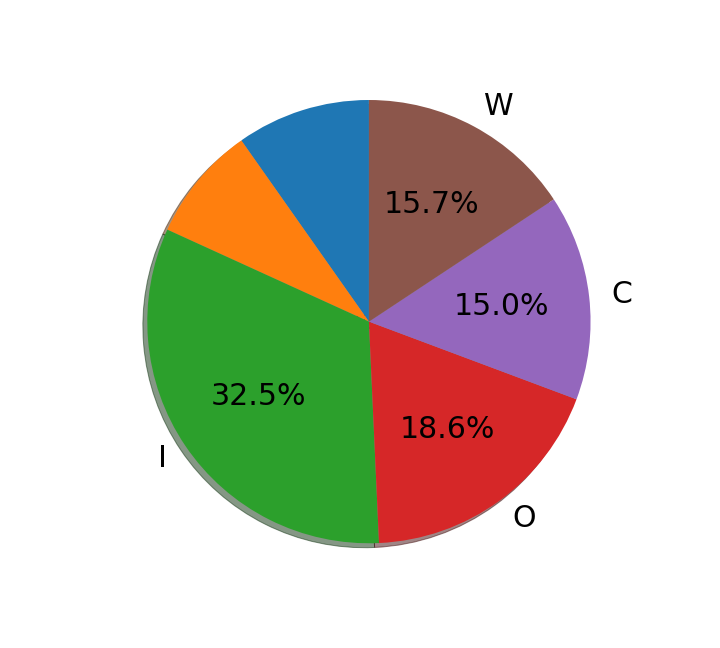}
}
\vspace{-2ex}

\subfloat[\ti{\texttt{SuperUser}: three-event \dw vs. three-event \dc vs. four-event \dw vs. four-event \dc}]
{\label{fig:superuser-pie}\includegraphics[width=0.25\linewidth]{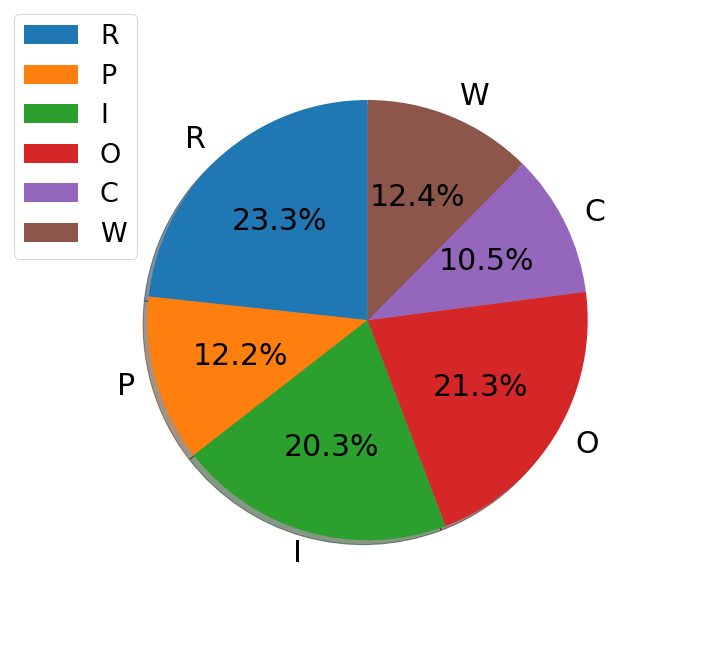}
\includegraphics[width=0.25\linewidth]{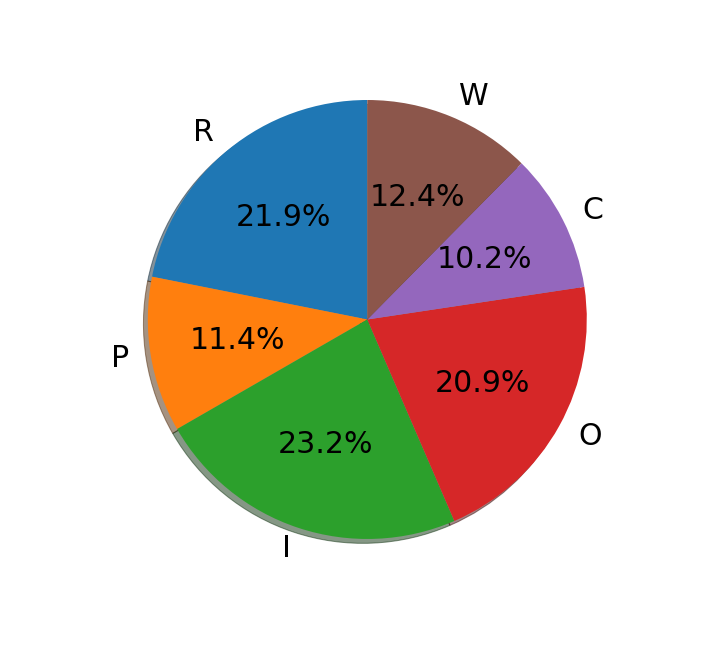}
\includegraphics[width=0.25\linewidth]{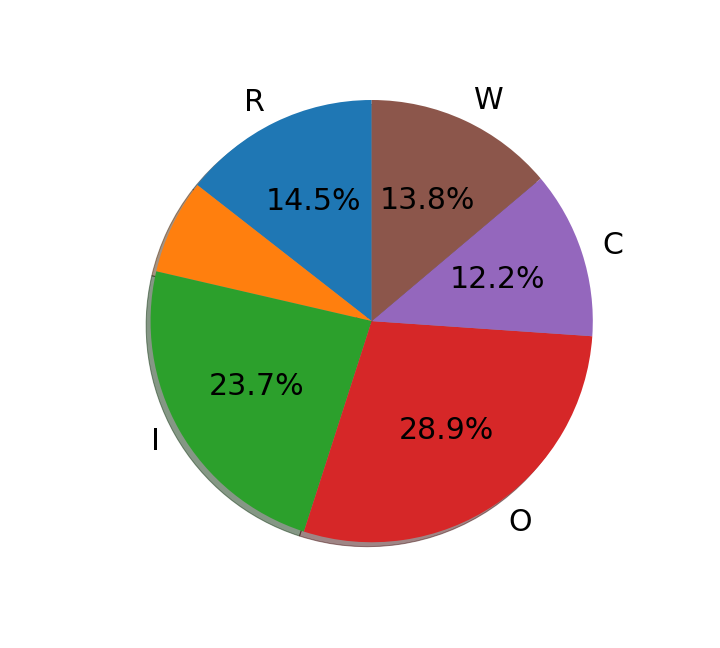}
\includegraphics[width=0.25\linewidth]{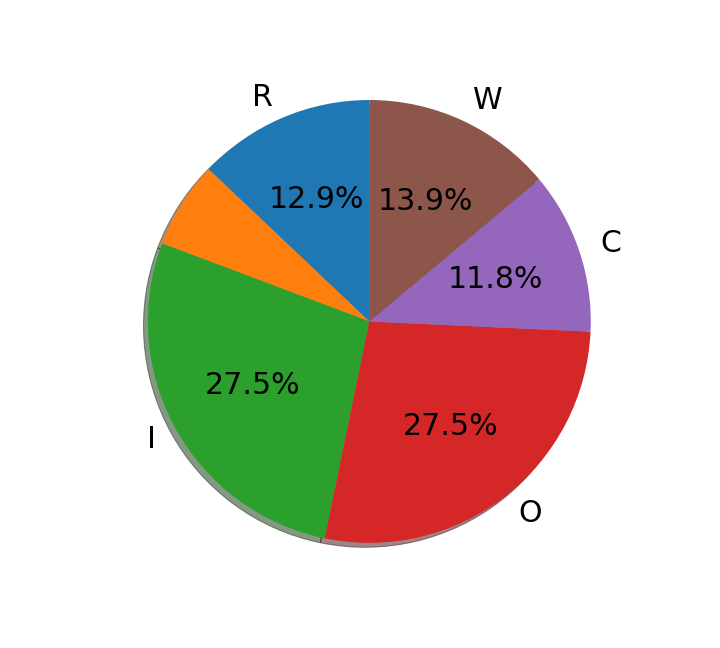}
}

\caption{\it Proportion of event pairs in three-event and four-event motifs between \dw and \dc configurations (PART 2). Each figure shows the ratio of event pairs in six categories.}
\vspace{-3ex}
\label{fig:pie-2}
\end{figure*}

Figure \ref{fig:pie-1} and \ref{fig:pie-2} show the proportions of event pairs for all datasets. For each dataset we show the patterns for three-event and four-event motifs, between \dw and \dc configurations.

\begin{figure*}[!bp]
\centering

\subfloat[\ti{010102 motif for \texttt{Calls-Copenhagen}}] 
{
\label{fig:calls-3-intermediate}
\includegraphics[width=0.30\linewidth]{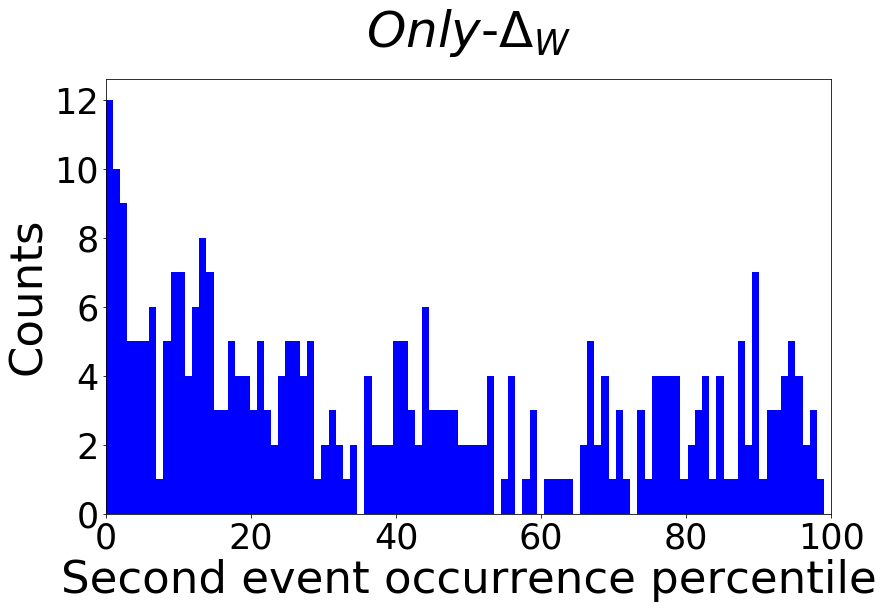}
\includegraphics[width=0.30\linewidth]{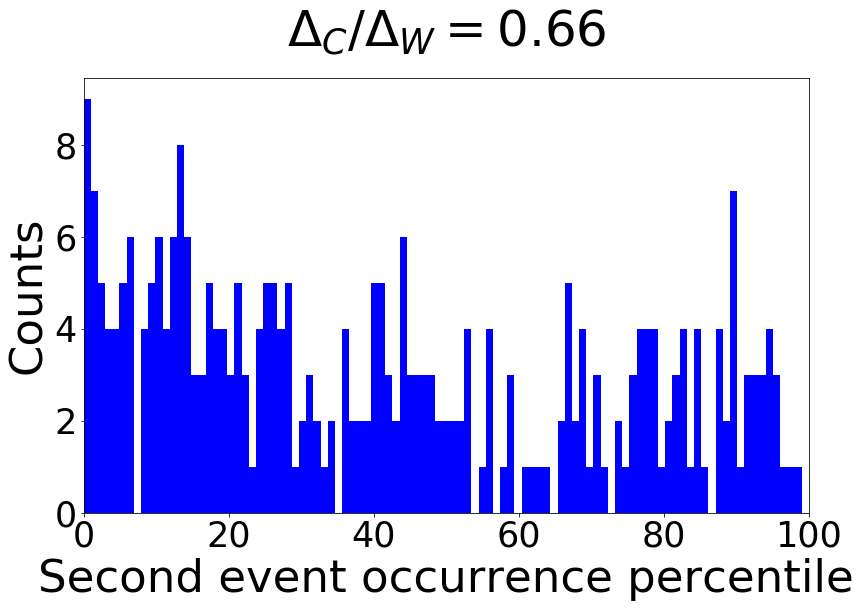}
\includegraphics[width=0.30\linewidth]{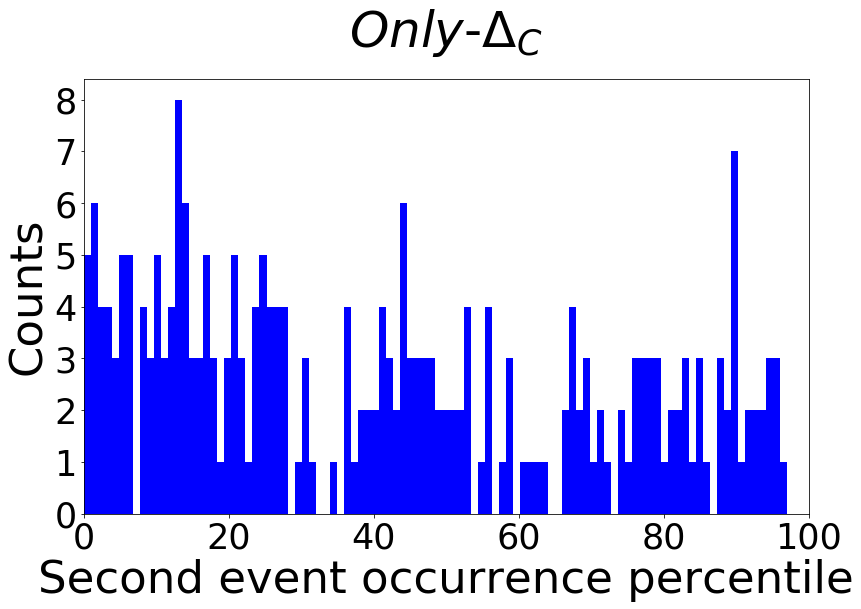}
}
\vspace{-1ex}

\subfloat[\ti{010102 motif for \texttt{Email}}] 
{
\label{fig:email-3-intermediate}
\includegraphics[width=0.30\linewidth]{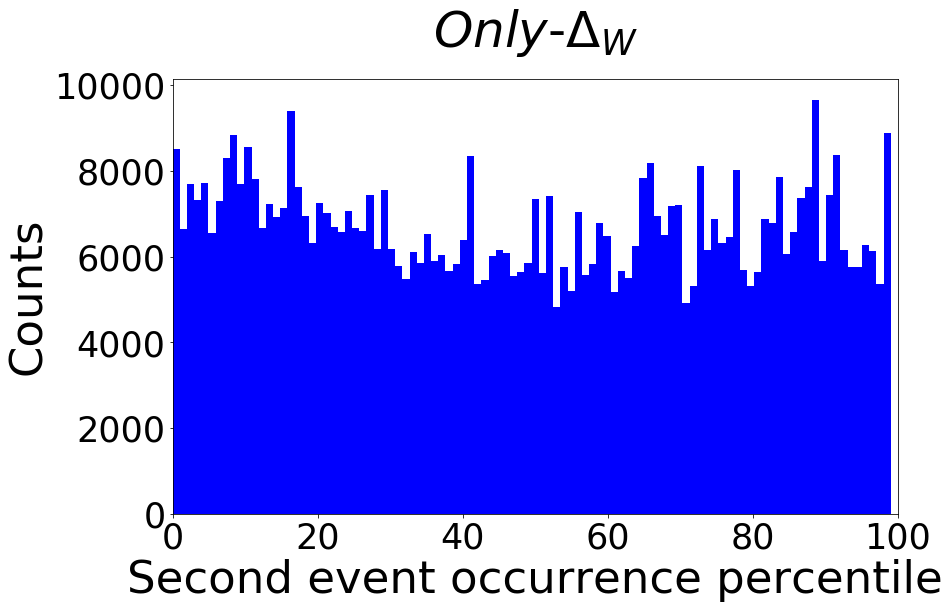}
\includegraphics[width=0.30\linewidth]{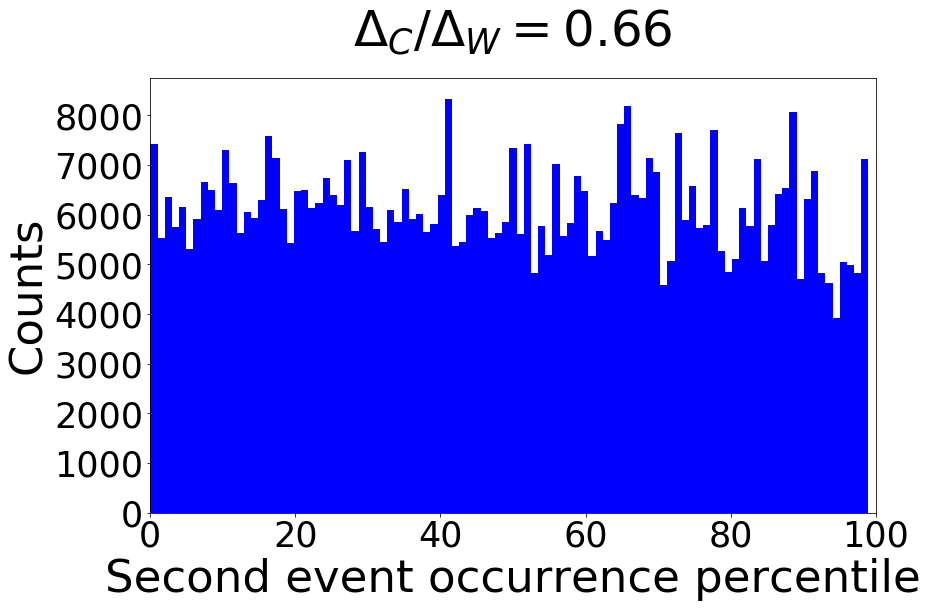}
\includegraphics[width=0.30\linewidth]{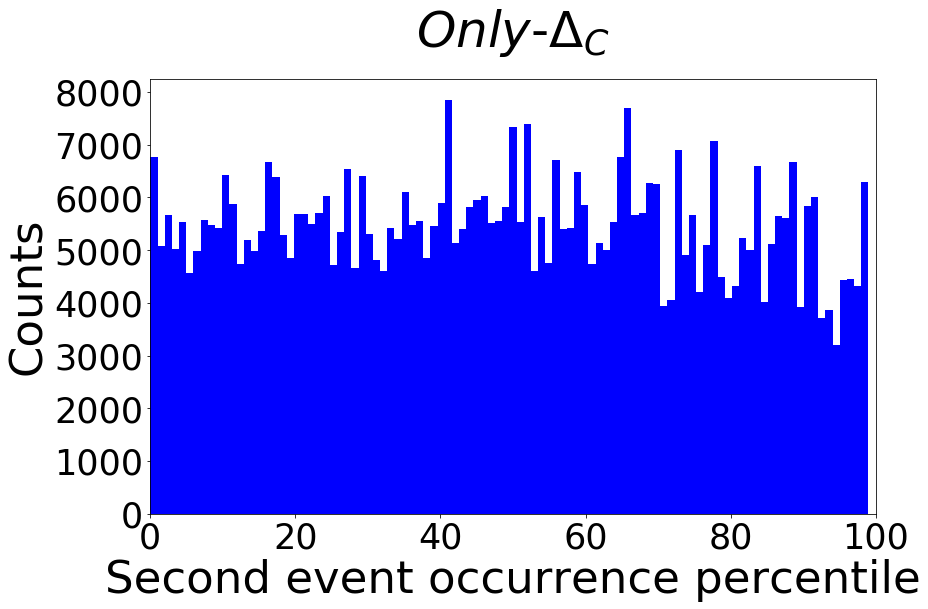}
}
\vspace{-1ex}

\subfloat[\ti{01022123 motif for \texttt{FBWall}}] 
{
\label{fig:fb-4-intermediate}
\includegraphics[width=0.25\linewidth]{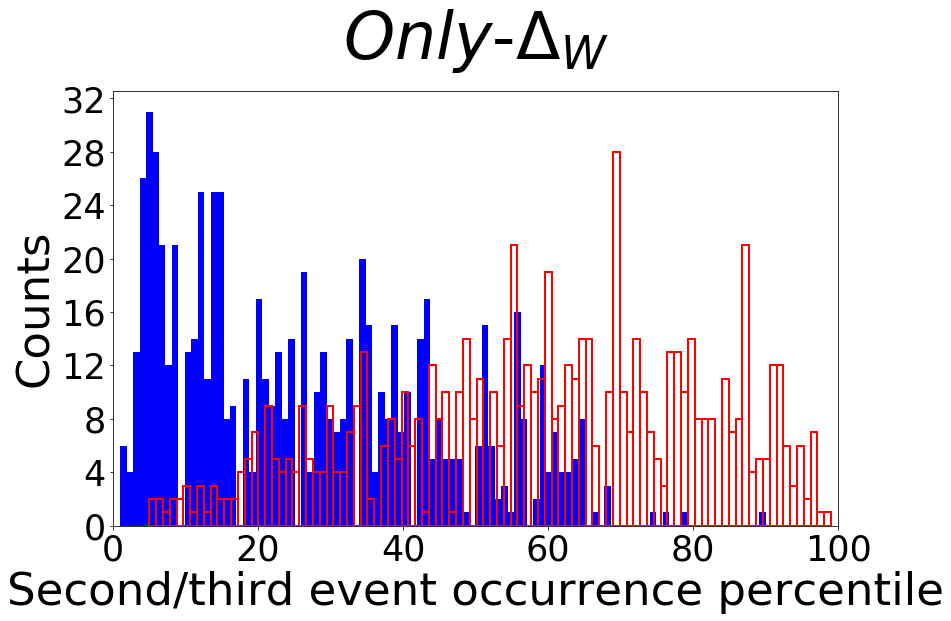}
\includegraphics[width=0.25\linewidth]{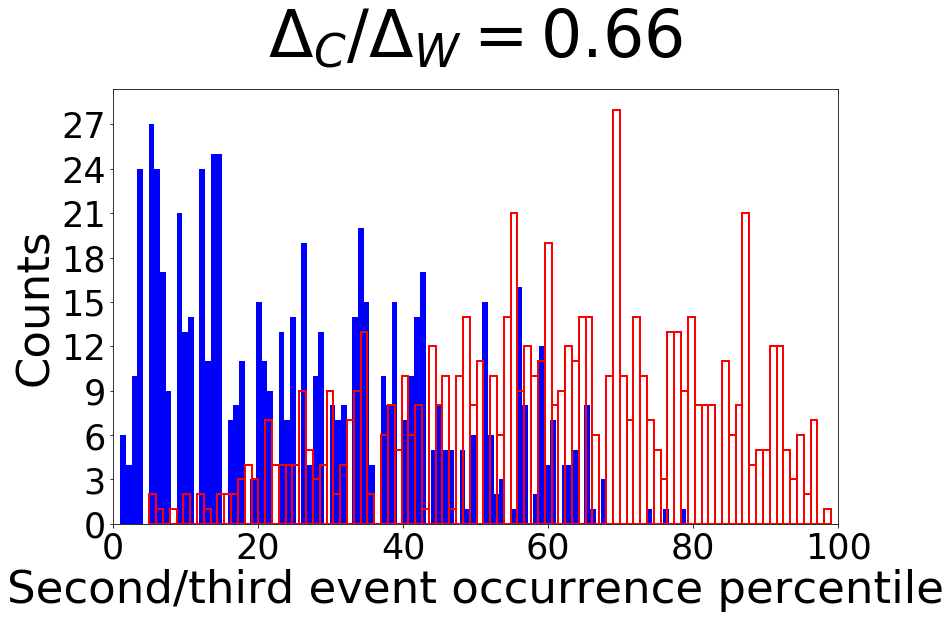}
\includegraphics[width=0.25\linewidth]{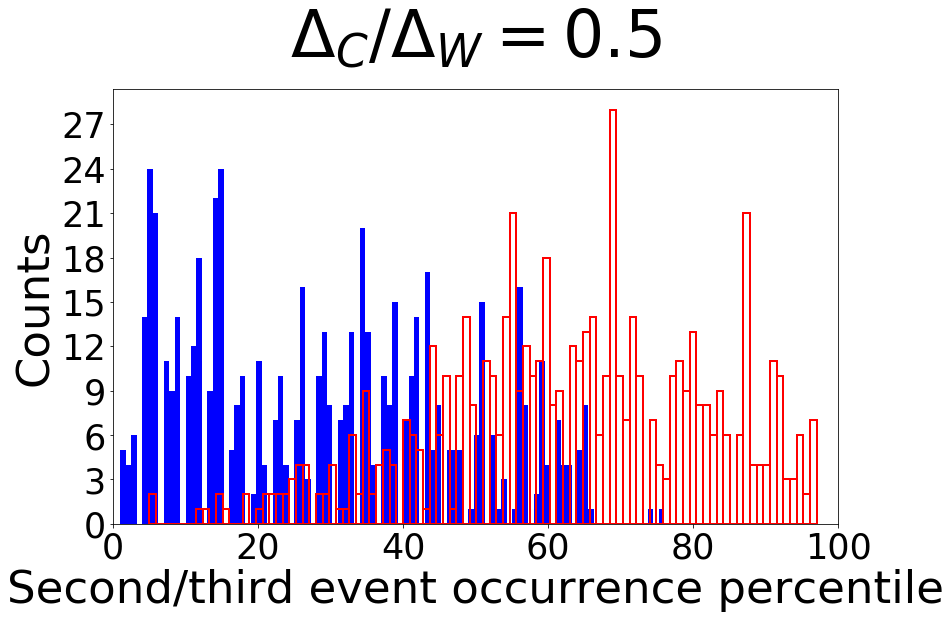}
\includegraphics[width=0.25\linewidth]{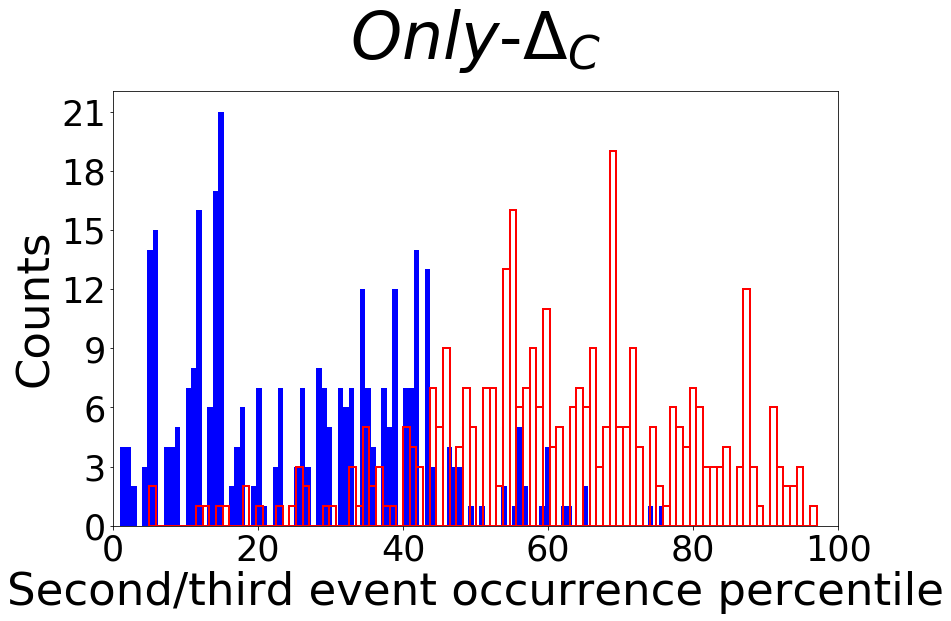}
}
\vspace{-1ex}

\subfloat[\ti{01022123 motif for \texttt{Bitcoin-otc}}] 
{
\label{fig:otc-4-intermediate}
\includegraphics[width=0.25\linewidth]{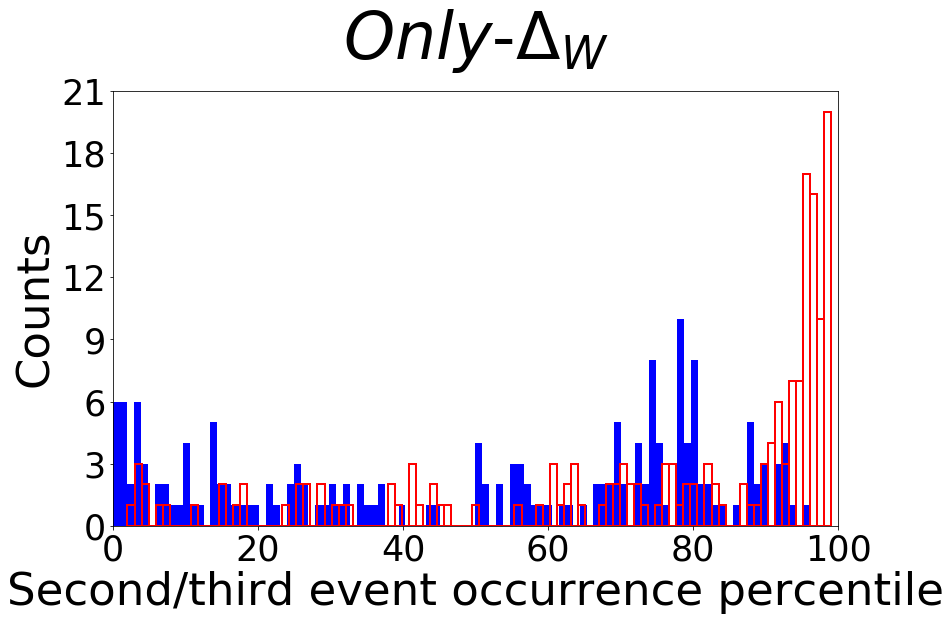}
\includegraphics[width=0.25\linewidth]{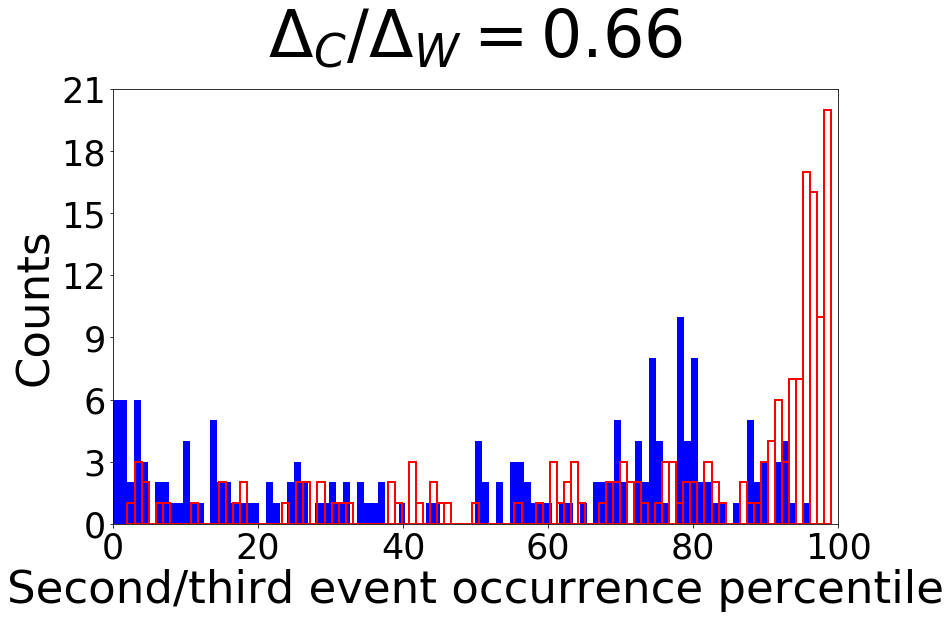}
\includegraphics[width=0.25\linewidth]{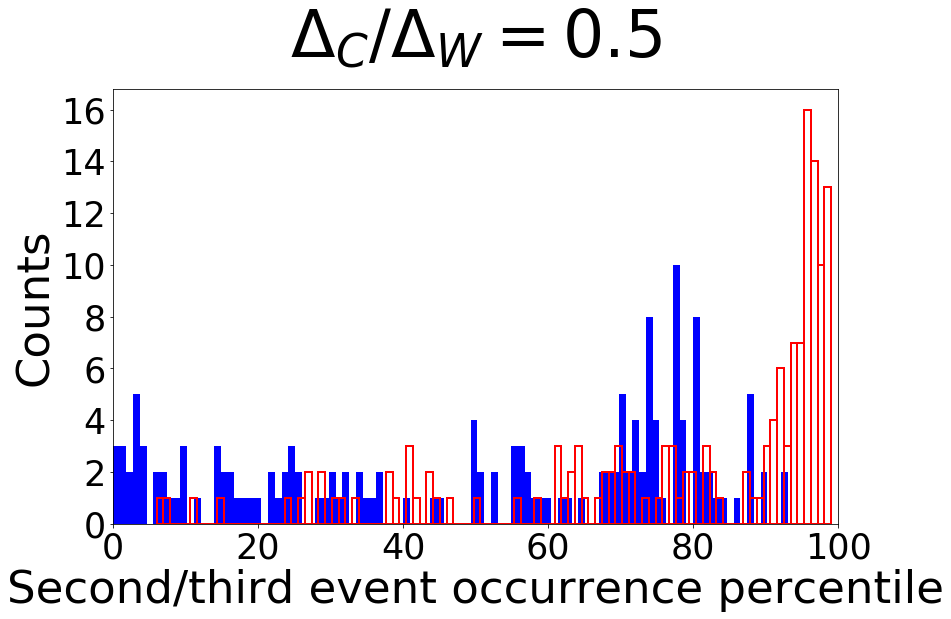}
\includegraphics[width=0.25\linewidth]{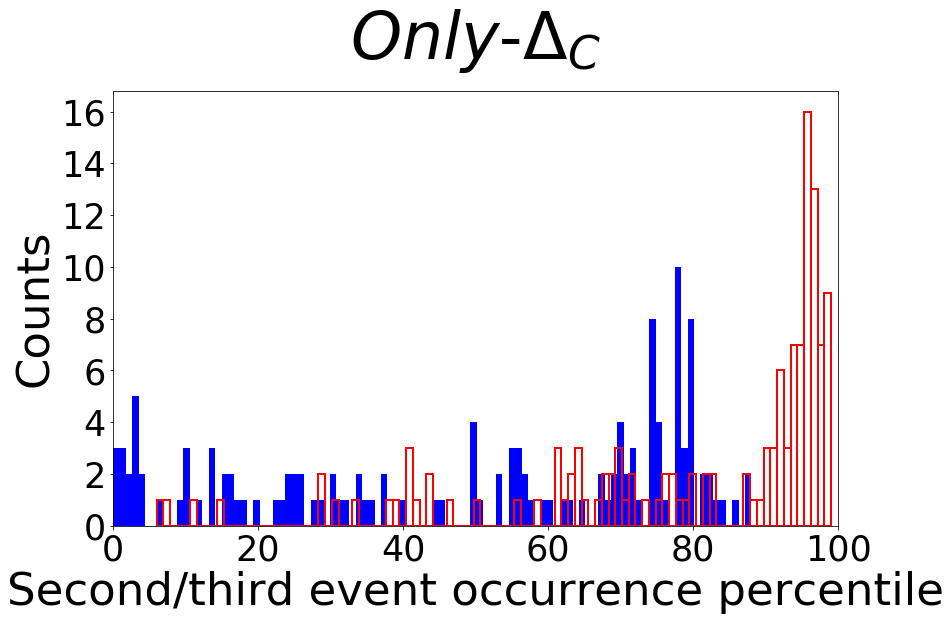}
}
\vspace{-1ex}

\subfloat[\ti{01022123 motif for \texttt{SuperUser}}] 
{
\label{fig:super-4-intermediate}
\includegraphics[width=0.25\linewidth]{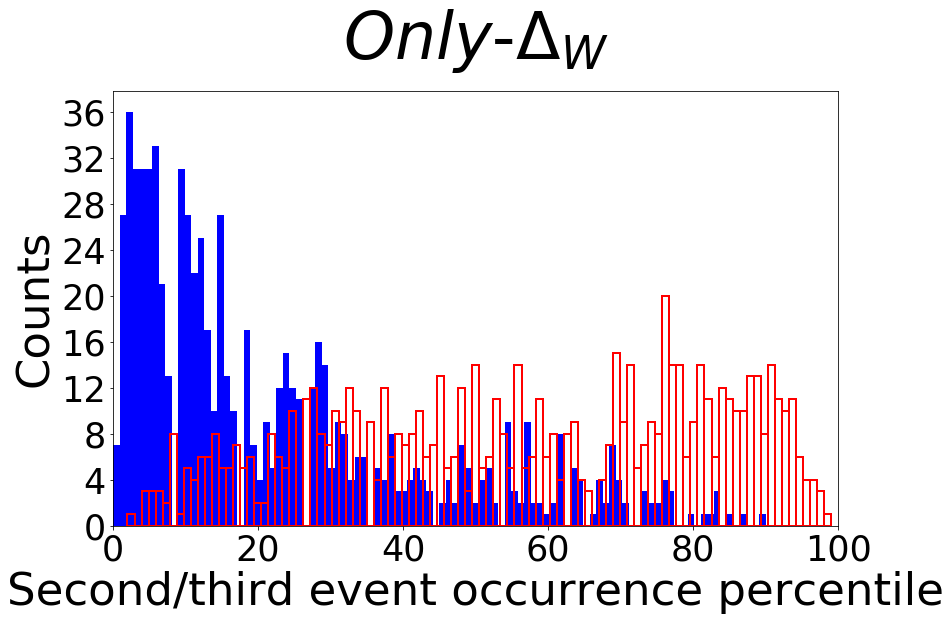}
\includegraphics[width=0.25\linewidth]{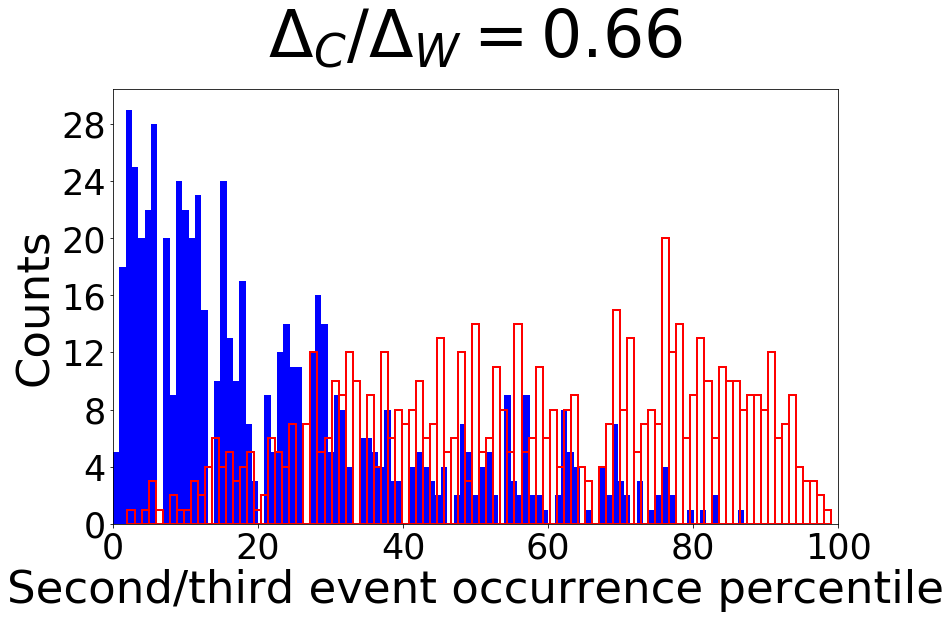}
\includegraphics[width=0.25\linewidth]{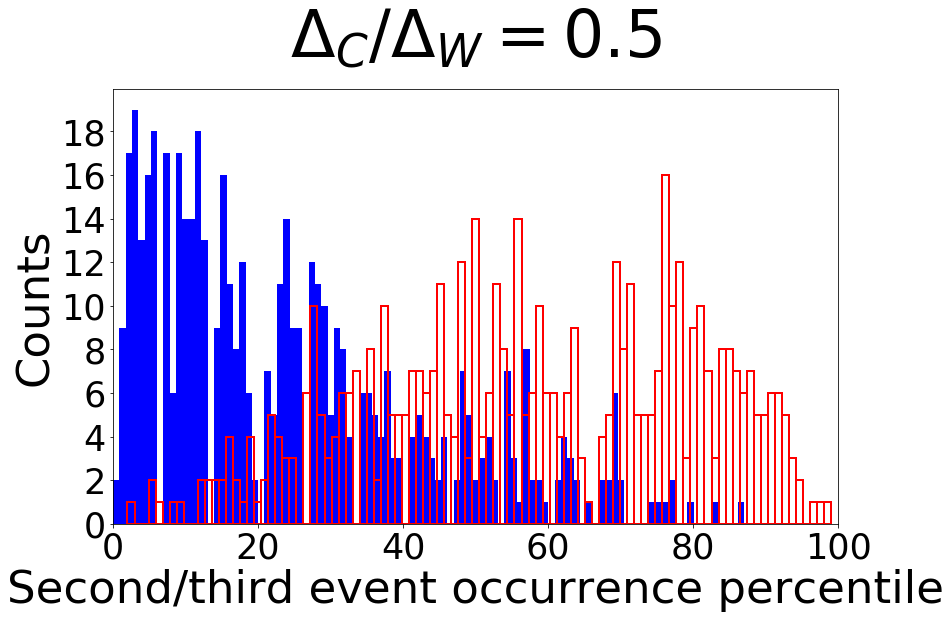}
\includegraphics[width=0.25\linewidth]{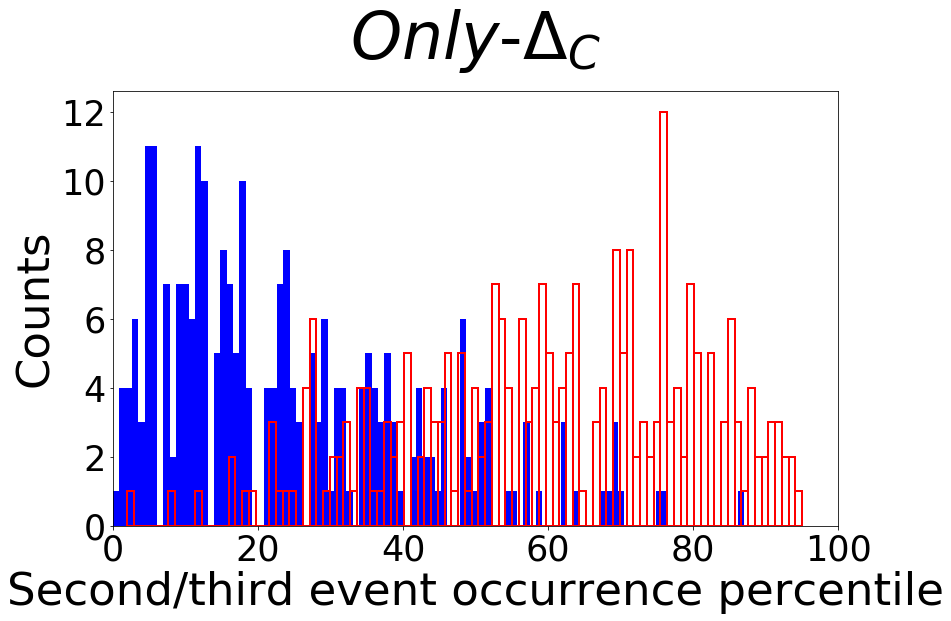}
}

\caption{\it Behavior of intermediate event occurrences. In each figure, the x-axis denotes the occurrence time of intermediate events with respect to the first and last events. 0\% denotes the first event occurrence and 100\% represents the last event occurrence. The y-axis shows the frequency of the intermediate events; second events (blue) in three-event motifs and second \& third events (blue \& red) in four-event motifs). In all cases, enforcing the $\Delta_C$ constraint regularizes the skewness in \dw case.}
\label{fig:intermediate-appdx}
\vspace{-3ex}
\end{figure*}

\begin{figure*}[!bp]
\centering

\subfloat[\ti{010102 motif for \texttt{FBWall}}] 
{
\label{fig:fb-span}
\includegraphics[width=0.30\linewidth]{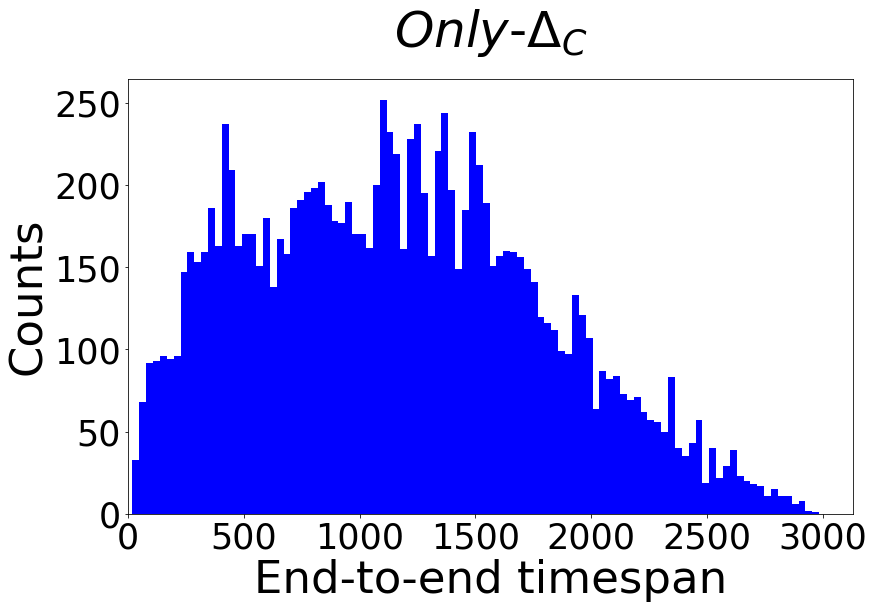}
\includegraphics[width=0.30\linewidth]{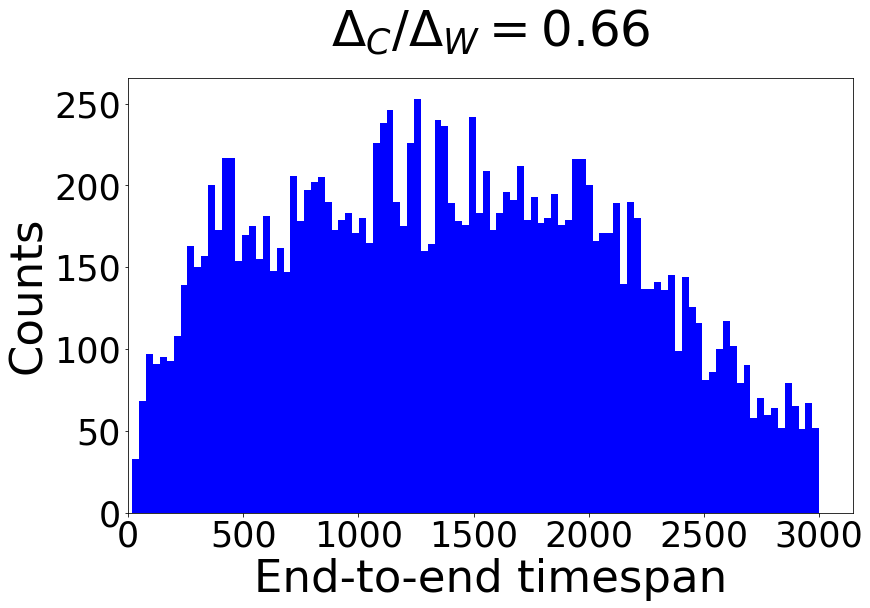}
\includegraphics[width=0.30\linewidth]{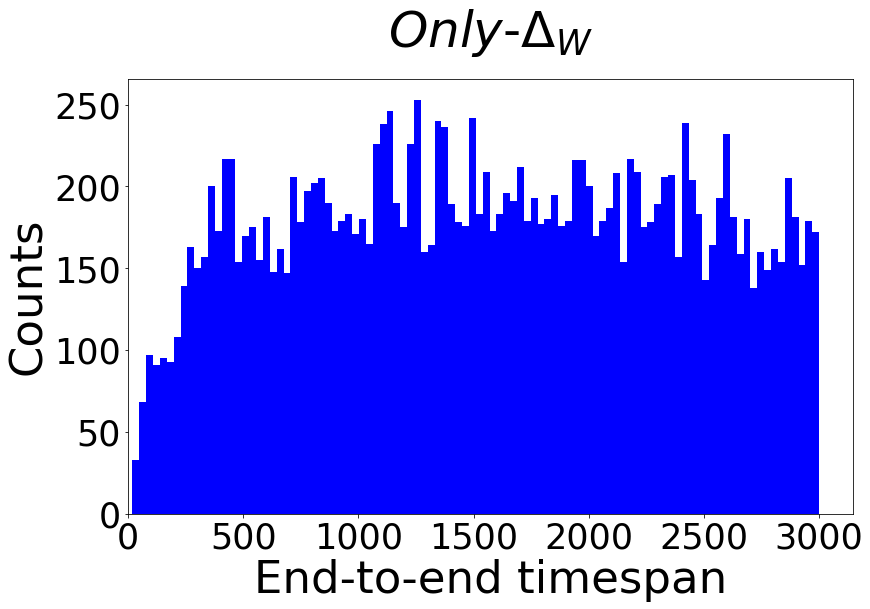}
}
\vspace{-2ex}

\subfloat[\ti{010102 motif for \texttt{SMS-Copenhagen}}] 
{
\label{fig:sms-span}
\includegraphics[width=0.30\linewidth]{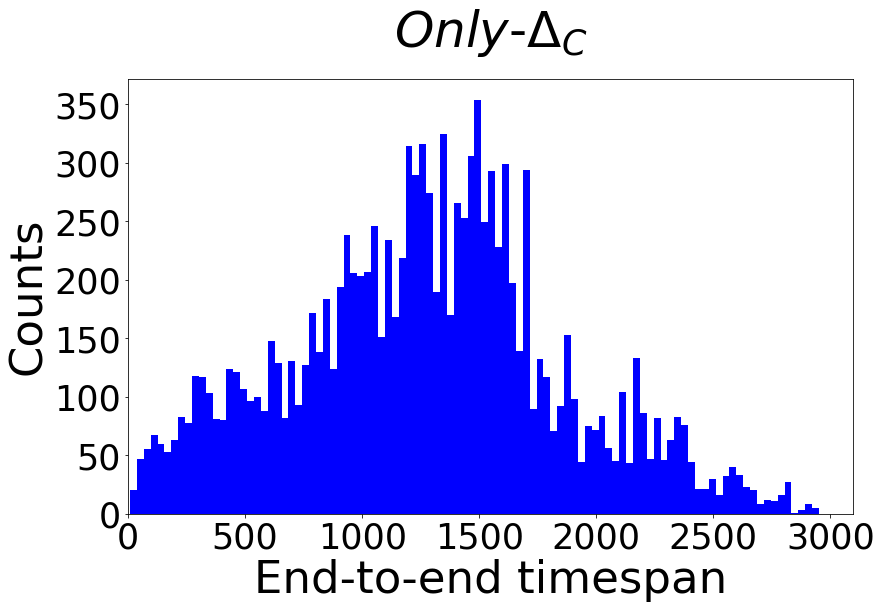}
\includegraphics[width=0.30\linewidth]{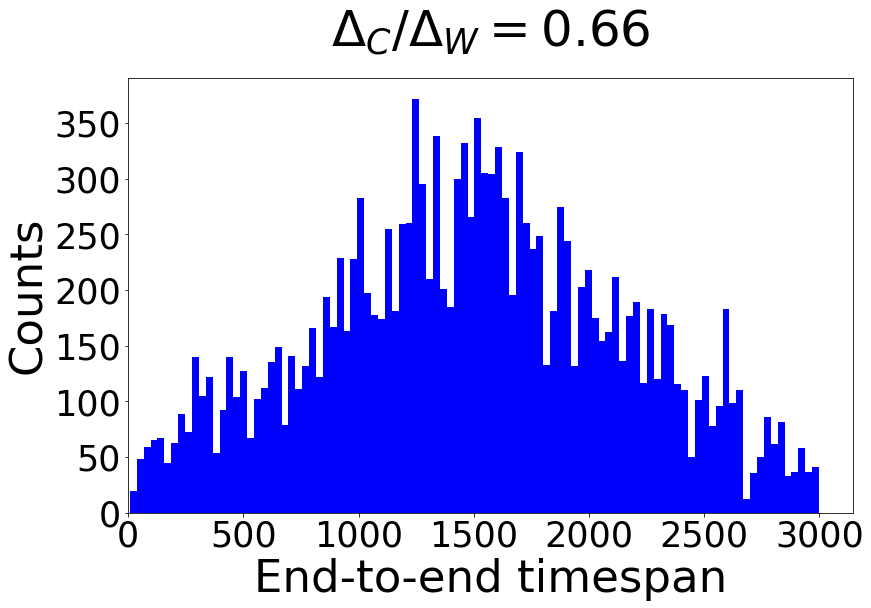}
\includegraphics[width=0.30\linewidth]{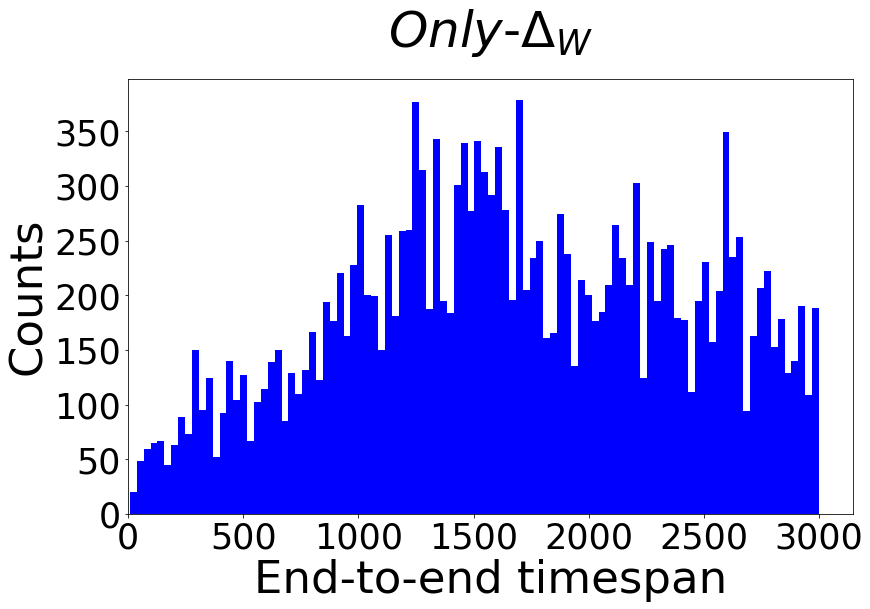}
}
\vspace{-2ex}

\subfloat[\ti{010102 motif for \texttt{SuperUser}}] 
{
\label{fig:super-span}
\includegraphics[width=0.30\linewidth]{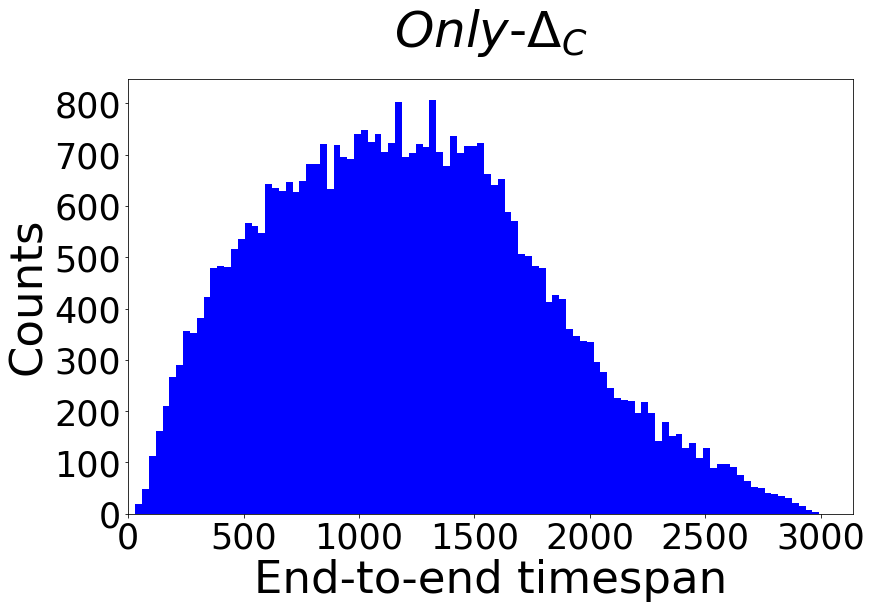}
\includegraphics[width=0.30\linewidth]{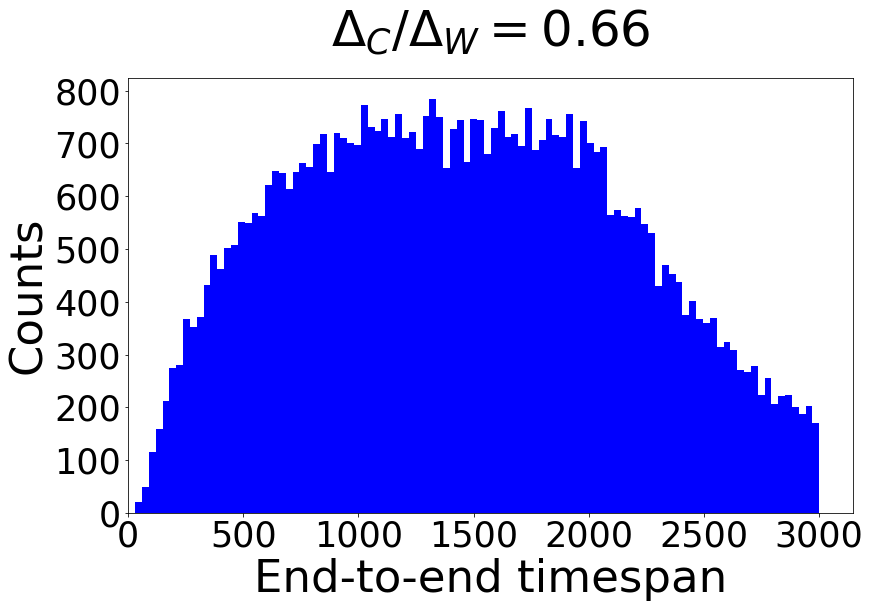}
\includegraphics[width=0.30\linewidth]{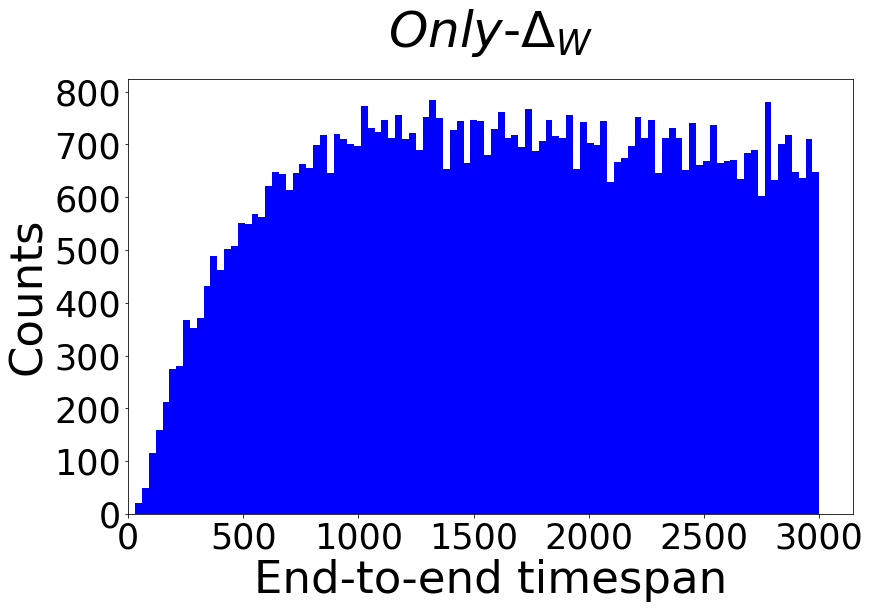}
}
\vspace{-2ex}

\subfloat[\ti{010102 motif for \texttt{Calls-Copenhagen}}] 
{
\label{fig:calls-span}
\includegraphics[width=0.30\linewidth]{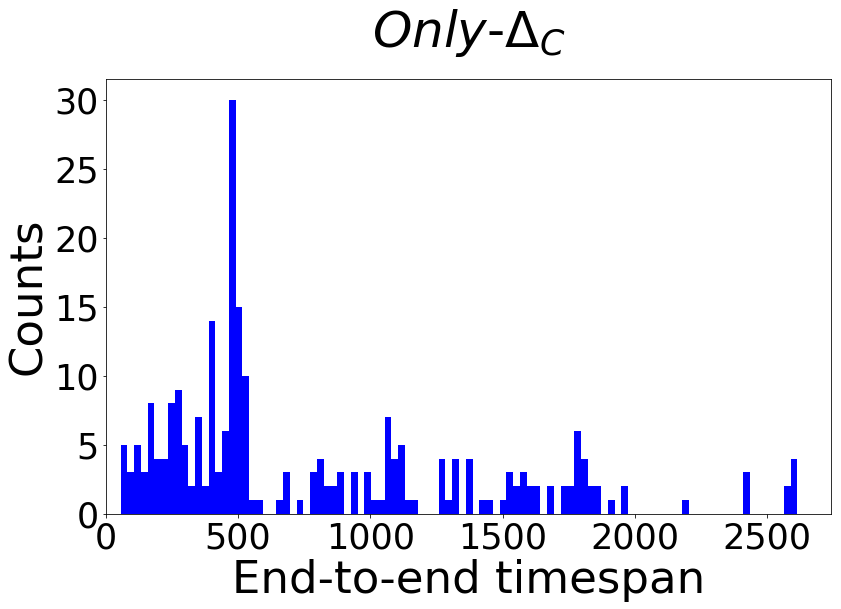}
\includegraphics[width=0.30\linewidth]{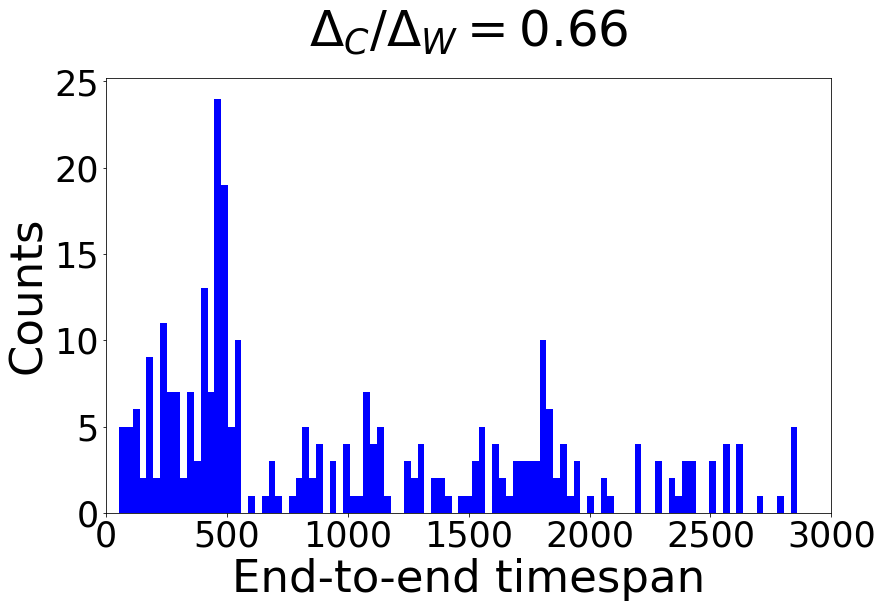}
\includegraphics[width=0.30\linewidth]{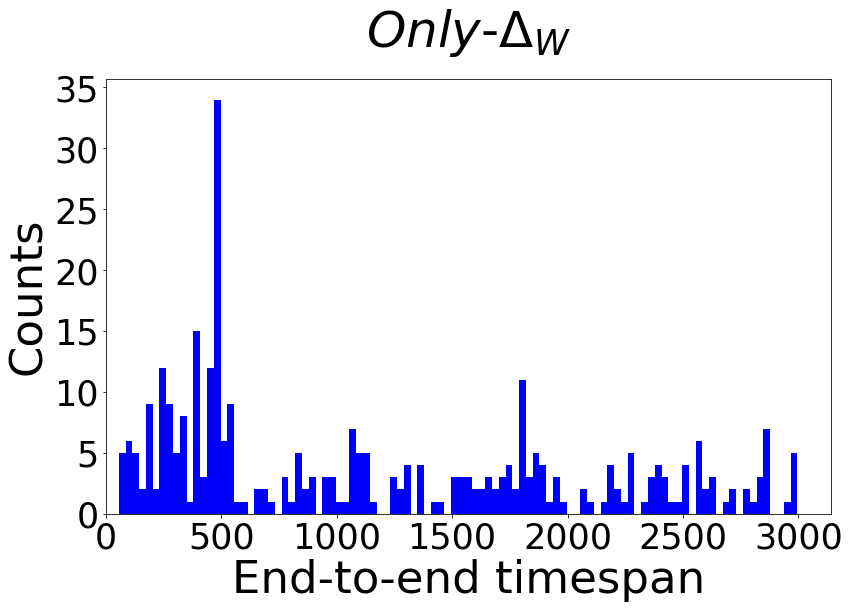}
}
\vspace{-2ex}

\subfloat[\ti{011012 motif for \texttt{Bitcoin-otc}}] 
{
\label{fig:otc-span}
\includegraphics[width=0.30\linewidth]{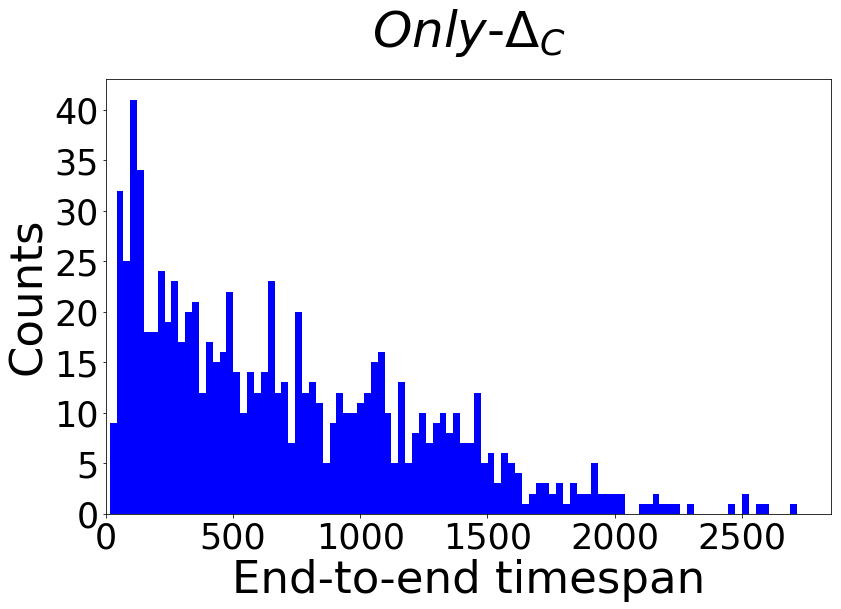}
\includegraphics[width=0.30\linewidth]{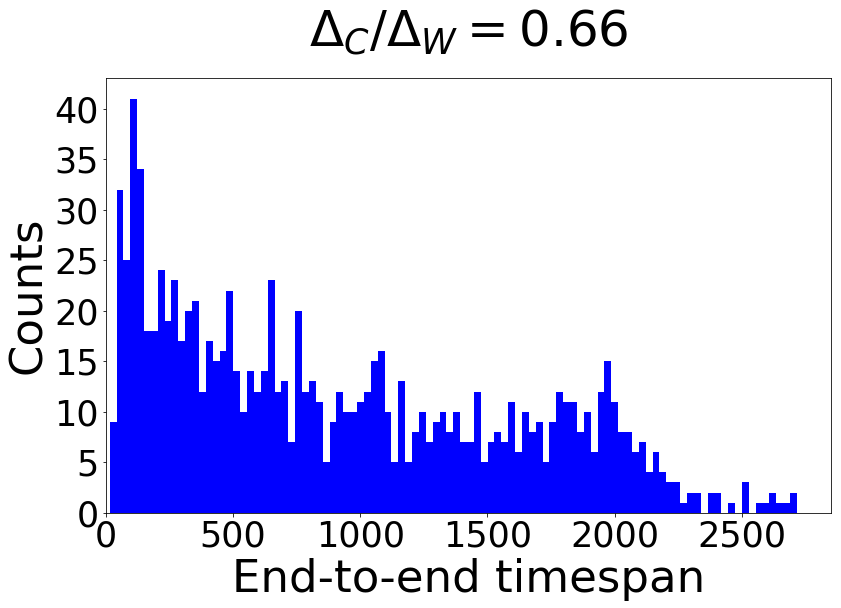}
\includegraphics[width=0.30\linewidth]{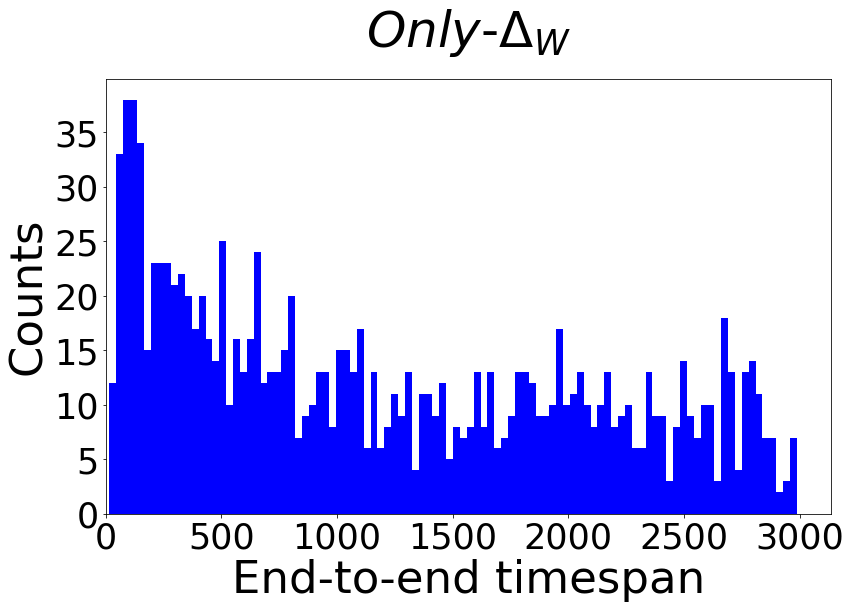}
}
\vspace{-2ex}

\caption{\it Distribution of the motif timespans. The x-axis denotes the timespan of the motif and the y-axis shows the count of such motifs. The distributions are more regularized when going from \dc to \dw configuration.}
\label{fig:span-appdx}
\vspace{-3ex}
\end{figure*}

\subsection{Intermediate event behaviors}
\cref{fig:intermediate-appdx} shows the intermediate event behaviors for more datasets.
In both three-event and four-event motifs, adding the $\Delta_C$ constraint mitigates the skewness in the intermediate event occurrences.

\subsection{Motif timespan distributions}
\cref{fig:span-appdx} gives the distributions of motif timepspans for more datasets.

\section{Temporal inducedness constraints}
Here we present the supplementary tables for the impacts of temporal inducedness constraints.
\cref{tab:consecutive-appdx} shows the impacts of consecutive event restriction regarding the motif ranking. \cref{tab:dgc-appdx} illustrate the proportion changes from vanilla temporal motifs to constrained dynamic graphlets.

\begin{table*}[hp!]
\setlength\tabcolsep{1.5pt}
\captionsetup{justification=centering}
\caption{\it \\The ranking changes of all three-node three-motifs after applying the consecutive event restriction, where $\Delta_C=1500s$. Positive values denote ascensions, and negative values indicate descending ranks. }
\small
\centering
\begin{tabular}{|l|r|r|r|r|r|r|r|r|r|}
\hline
Motif  & \texttt{Calls-Copenhagen} & \texttt{CollegeMsg} & \texttt{SMS-Copenhagen} & \texttt{SMS-A} & \texttt{Email} & \texttt{FBWall}  & \texttt{Bitcoin-otc} & \texttt{StackOverflow} & \texttt{SuperUser} \\ \hline
\texttt{010102} & 0     & -4      & -6  & 1    & -1    & -1  & 2   & 0     & 0     \\ \hline
\texttt{010112} & -9    & -8      & -14 & -12  & -1    & -9  & -5  & 0     & -8    \\ \hline
\texttt{010120} & 0     & -5      & -12 & -5   & -6    & -2  & -3  & -1    & 1     \\ \hline
\texttt{010121} & -4    & -13     & -15 & 11   & 0     & 5   & 1   & 0     & -1    \\ \hline
\texttt{010201} & -3    & -4      & -1  & -9   & 0     & 0   & 3   & -6    & 0     \\ \hline
\texttt{010202} & 0     & -4      & -8  & -1   & 1     & -3  & 4   & -8    & -2    \\ \hline
\texttt{010210} & 7     & 18      & 16  & 11   & 1     & 14  & 0   & -11   & 4     \\ \hline
\texttt{010212} & 6     & -1      & 2   & 1    & 2     & 0   & -3  & 4     & -2    \\ \hline
\texttt{010220} & 4     & 8       & 8   & -4   & -3    & -6  & -2  & -20   & -11   \\ \hline
\texttt{010221} & 0     & -4      & 2   & 2    & -2    & 4   & -2  & 2     & 1     \\ \hline
\texttt{011002}& 1     & 2       & -4  & -14  & 3     & -4  & 1   & -9    & 0     \\ \hline
\texttt{011012} & -2    & 8       & -2  & -7   & 2     & -1  & 1   & -18   & -8    \\ \hline
\texttt{011020} & 0     & 0       & -2  & -8   & -2    & -1  & 0   & -8    & -2    \\ \hline
\texttt{011021} & 4     & -2      & -6  & -3   & -3    & 1   & -1  & 0     & 1     \\ \hline
\texttt{011201} & 1     & -3      & -4  & 5    & -2    & 2   & 1   & 7     & 5     \\ \hline
\texttt{011202} & 0     & 0       & 3   & 2    & 1     & -2  & 2   & 16    & 1     \\ \hline
\texttt{011210} & -9    & 23      & 18  & 11   & 4     & 10  & 0   & 8     & 4     \\ \hline
\texttt{011212} & -1    & -6      & -2  & -12  & -4    & -16 & -6  & -11   & -14   \\ \hline
\texttt{011220} & 0     & 0       & -1  & 1    & 0     & 2   & -1  & 1     & 0     \\ \hline
\texttt{011221} & 2     & -4      & 2   & 5    & -4    & -3  & 0   & -3    & -1    \\ \hline
\texttt{012001} & -2    & -2      & 0   & 8    & -1    & 2   & 3   & 17    & 5     \\ \hline
\texttt{012002} & 8     & 8       & 4   & 2    & 4     & -1  & 2   & 9     & 0     \\ \hline
\texttt{012010} & 0     & 10      & 14  & 2    & 4     & 13  & 2   & -5    & -4    \\ \hline
\texttt{012012} & 0     & 0       & -1  & -2   & 0     & 2   & 0   & -2    & 0     \\ \hline
\texttt{012020} & -1    & -11     & -3  & 4    & -1    & -5  & -2  & -8    & -4    \\ \hline
\texttt{012021} & 2     & 3       & -3  & 11   & -2    & -2  & 0   & 13    & 4     \\ \hline
\texttt{012101} & -3    & -7      & 0   & -2   & 2     & 1   & 1   & 3     & 4     \\ \hline
\texttt{012102} & 0     & -1      & -2  & -5   & 1     & -2  & -1  & 19    & 10    \\ \hline
\texttt{012110} & 0     & 16      & 17  & 6    & 4     & 2   & 3   & 0     & 5     \\ \hline
\texttt{012112} & 10    & -5      & -3  & 4    & 1     & -7  & -1  & -9    & -1    \\ \hline
\texttt{012120} & 0     & 3       & 0   & -1   & 1     & 4   & 0   & 20    & 14    \\ \hline
\texttt{012121} & -11   & -15     & 3   & -2   & 1     & 3   & 1   & 0     & -1   \\ \hline
\end{tabular}
\label{tab:consecutive-appdx}
\end{table*}

\begin{table*}[hp!]
\setlength\tabcolsep{1.5pt}
\captionsetup{justification=centering}
\caption{\it \\The proportion changes (percentage) of all three-node three-event motifs when going from vanilla temporal motifs to constrained dynamic graphlets, where the resolution of all datasets is degraded to 300s.}
\small
\centering
\begin{tabular}{|l|r|r|r|r|r|r|r|r|}
\hline
Motif  & \texttt{Calls-Copenhagen} & \texttt{CollegeMsg} & \texttt{SMS-Copenhagen} & \texttt{SMS-A} & \texttt{Email} & \texttt{FBWall} & \texttt{StackOverflow} & \texttt{SuperUser} \\ \hline
\texttt{010102} & 0.22\%    & 3.31\%      & 2.37\%  & 4.09\%  & -9.63\%   & 1.06\%  & 0.26\%    & 0.63\%    \\ \hline
\texttt{010112} & 0.52\%    & 2.03\%      & 2.34\%  & 1.44\%   & 2.50\%    & 1.09\%  & 0.24\%    & 0.23\%    \\ \hline
\texttt{010120} & 0.51\%    & 2.16\%      & 2.36\%  & 1.34\%   & 2.76\%    & 1.38\%  & 0.20\%    & 0.15\%    \\ \hline
\texttt{010121} & 0.13\%    & 2.25\%      & 2.61\%  & 0.94\%   & 1.74\%    & 1.84\%  & -0.07\%   & -0.14\%   \\ \hline
\texttt{010201} & -5.60\%   & -2.12\%     & -0.99\% & -1.93\%  & -18.00\%  & -0.78\% & -0.09\%   & -0.14\%   \\ \hline
\texttt{010202} & -3.45\%   & 4.36\%      & 3.23\%  & 4.98\%   & -10.05\%  & 1.09\%  & 0.27\%    & 0.65\%    \\ \hline
\texttt{010210} & 0.20\%    & -1.04\%     & -0.69\% & -0.88\%  & 1.86\%    & -0.83\% & -0.06\%   & -0.08\%   \\ \hline
\texttt{010212} & -0.09\%   & 0.03\%      & 0.01\%  & -0.02\%  & 0.09\%    & -0.12\% & -0.12\%   & -0.09\%   \\ \hline
\texttt{010220} & 1.03\%    & -1.64\%     & -2.54\% & -1.76\%  & 1.82\%    & -0.67\% & 0.32\%    & 0.27\%    \\ \hline
\texttt{010221} & -0.36\%   & 0.02\%      & 0.00\%  & -0.03\%  & 0.07\%    & -0.12\% & -0.07\%   & -0.06\%   \\ \hline
\texttt{011002} & 0.40\%    & -1.03\%     & -1.85\% & -1.10\%  & 2.20\%    & -0.71\% & 0.22\%    & 0.15\%    \\ \hline
\texttt{011012} & 0.25\%    & -1.31\%     & -2.21\% & -1.39\%  & 2.07\%    & -0.94\% & -0.15\%   & -0.13\%   \\ \hline
\texttt{011020} & 0.06\%    & -1.33\%     & -1.42\% & -0.73\%  & 1.50\%    & -0.39\% & 0.00\%    & -0.06\%   \\ \hline
\texttt{011021} & 0.83\%    & -1.58\%     & -1.68\% & -0.94\%  & 1.39\%    & -0.76\% & 0.16\%    & 0.01\%    \\ \hline
\texttt{011201} & -0.20\%   & -0.95\%     & -0.41\% & -0.42\%  & 1.63\%    & -0.58\% & -0.09\%   & -0.12\%   \\ \hline
\texttt{011202} & -0.42\%   & 0.00\%      & 0.01\%  & -0.03\%  & 0.19\%    & -0.10\% & -0.10\%   & -0.08\%   \\ \hline
\texttt{011210} & 0.63\%    & -1.31\%     & -1.33\% & -0.90\%  & 0.91\%    & -0.77\% & -0.27\%   & -0.18\%   \\ \hline
\texttt{011212} & 1.29\%    & 2.12\%      & 2.87\%  & 1.25\%   & 2.49\%    & 2.98\%  & 0.19\%    & 0.21\%    \\ \hline
\texttt{011220} & 0.01\%    & 0.00\%      & 0.05\%  & -0.02\%  & 0.22\%    & -0.08\% & -0.02\%   & -0.01\%   \\ \hline
\texttt{011221} & 0.61\%    & -1.86\%     & -1.93\% & -1.05\%  & 1.24\%    & -0.95\% & -0.04\%   & -0.06\%   \\ \hline
\texttt{012001} & 0.81\%    & -0.84\%     & -0.64\% & -0.19\%  & 1.57\%    & -0.39\% & -0.08\%   & -0.13\%   \\ \hline
\texttt{012002} & -0.24\%   & -1.68\%     & -3.03\% & -2.21\%  & 1.75\%    & -1.15\% & 0.07\%    & 0.01\%    \\ \hline
\texttt{012010} & 0.33\%    & -1.32\%     & -0.52\% & -0.55\%  & 1.23\%    & -1.04\% & -0.31\%   & -0.29\%   \\ \hline
\texttt{012012} & 0.00\%    & 0.01\%      & -0.02\% & -0.03\%  & 0.28\%    & -0.11\% & -0.01\%   & 0.00\%    \\ \hline
\texttt{012020} & 1.39\%    & 3.76\%      & 3.26\%  & 1.54\%   & 2.88\%    & 0.75\%  & 0.26\%    & 0.24\%    \\ \hline
\texttt{012021} & -0.20\%   & 0.01\%      & -0.02\% & -0.04\%  & 0.08\%    & -0.13\% & -0.09\%   & -0.10\%   \\ \hline
\texttt{012101} & -0.25\%   & -1.67\%     & 0.09\%  & -0.29\%  & 0.96\%    & -0.20\% & -0.48\%   & -0.57\%   \\ \hline
\texttt{012102} & -0.05\%   & 0.02\%      & 0.04\%  & -0.03\%  & 0.20\%    & -0.17\% & -0.01\%   & -0.04\%   \\ \hline
\texttt{012110} & 0.06\%    & -1.41\%     & -1.28\% & -0.59\%  & 0.99\%    & -0.23\% & -0.29\%   & -0.24\%   \\ \hline
\texttt{012112} & 0.23\%    & -1.69\%     & -2.27\% & -1.22\%  & 1.25\%    & -0.91\% & 0.30\%    & 0.11\%    \\ \hline
\texttt{012120} & 0.33\%    & 0.01\%      & 0.10\%  & -0.04\%  & -0.01\%   & -0.17\% & -0.10\%   & -0.10\%   \\ \hline
\texttt{012121} & 1.01\%    & 2.69\%      & 3.49\%  & 0.84\%   & 1.83\%    & 2.11\%  & -0.08\%   & -0.04\%  \\ \hline
\end{tabular}

\label{tab:dgc-appdx}
\end{table*}

\section{Ordered sequences of event pairs}
\begin{figure*}[h]
\centering
\subfloat[\ti{\texttt{CollegeMsg}}]{\label{fig:heat-CollegeMsg}\includegraphics[width=0.33\linewidth]{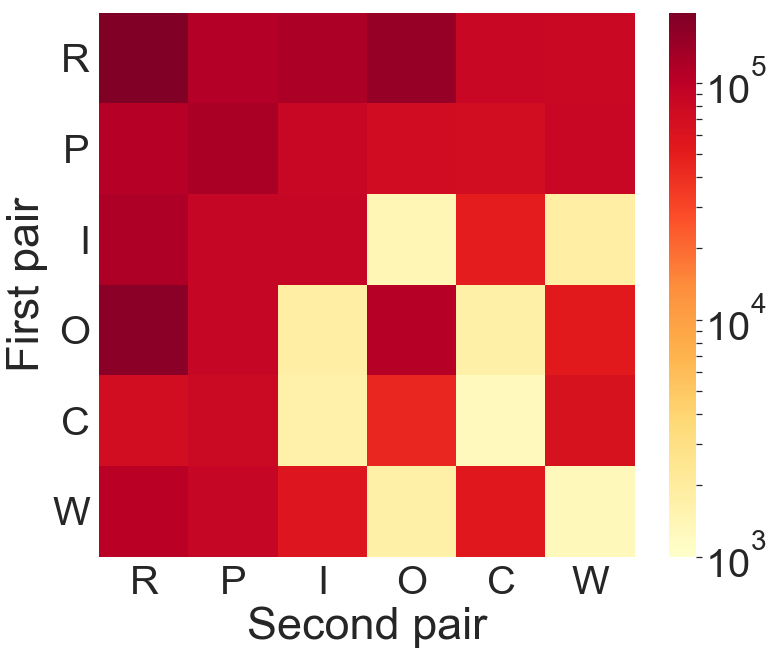}}
\subfloat[\ti{\texttt{FBWall}}]{\label{fig:heat-FBWall}\includegraphics[width=0.33\linewidth]{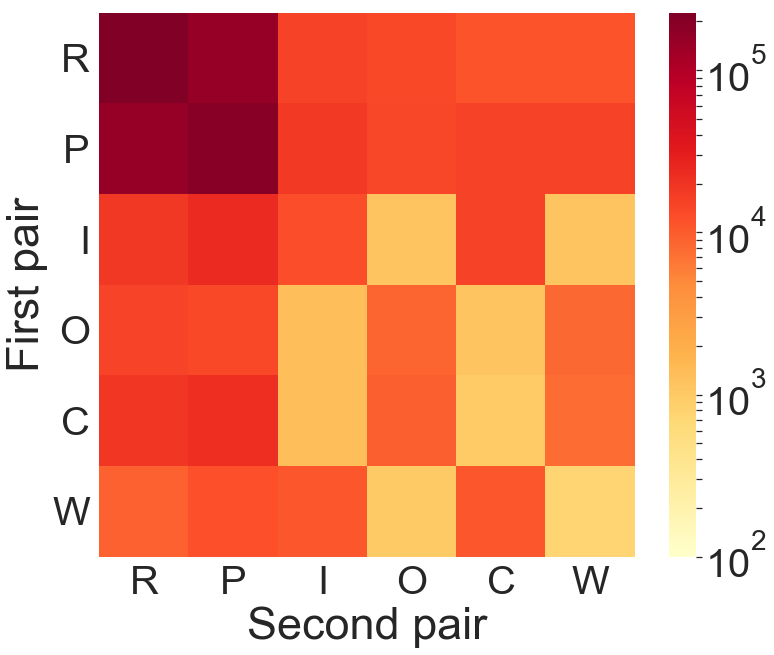}}
\subfloat[\ti{\texttt{StackOverflow}}]{\label{fig:heat-stack}\includegraphics[width=0.33\linewidth]{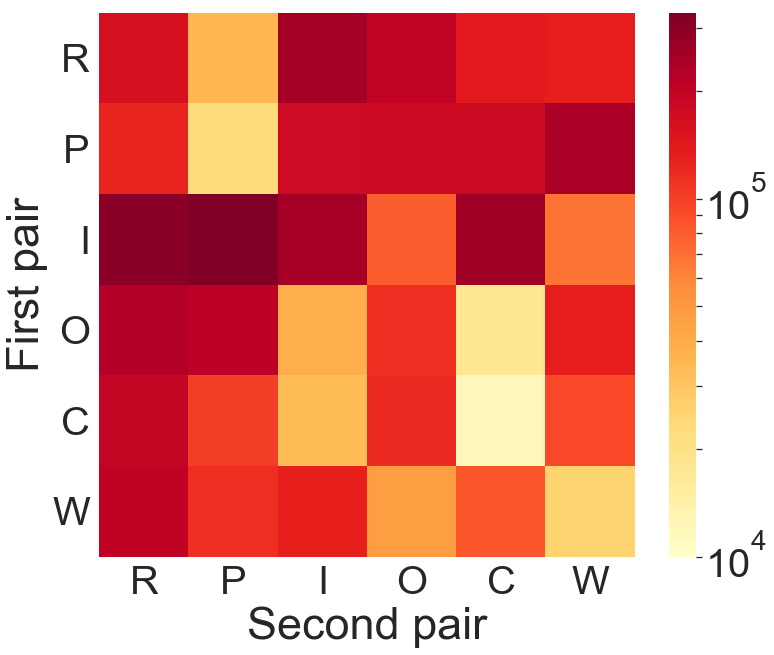}}

\subfloat[\ti{\texttt{SuperUser}}]{\label{fig:heat-superuser}\includegraphics[width=0.33\linewidth]{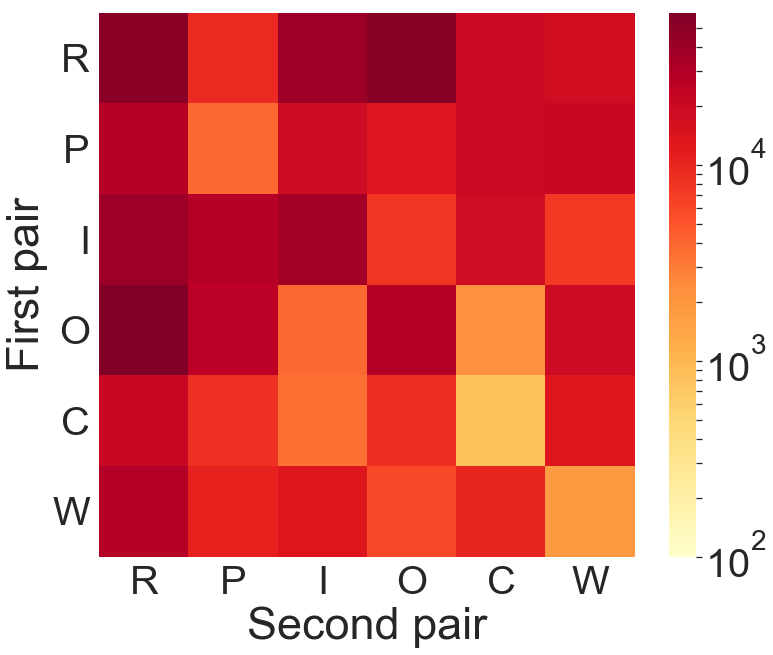}}
\subfloat[\ti{\texttt{Bitcoin-otc}}]{\label{fig:heat-otc}\includegraphics[width=0.33\linewidth]{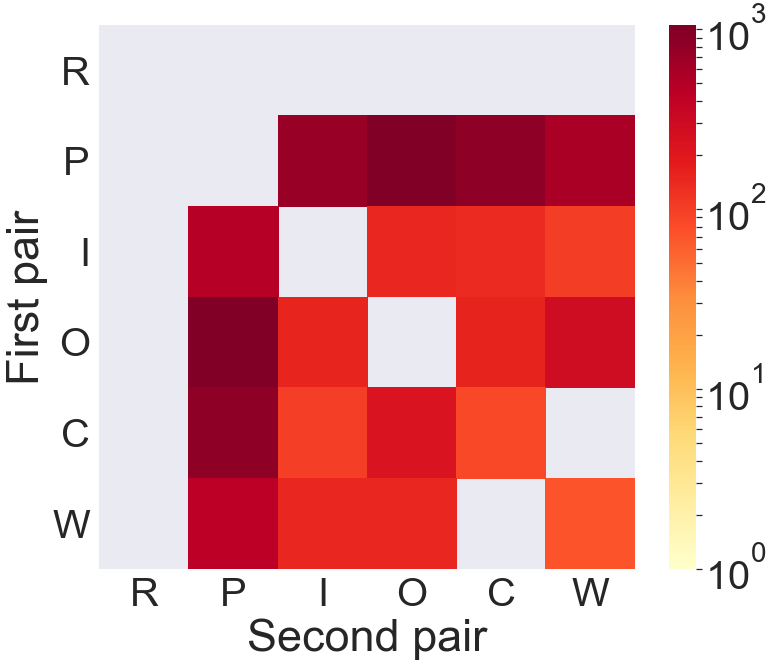}}

\caption{\it Ordered sequences of event pairs for three-event motifs. Each block denotes a type of three-event motif, where y-axis shows the first pair of events (first and second event) and the x-axis shows the second pair of events (second and third event). The color indicates the motif counts in log scale and calculated with respect to the minimum and maximum counts in each dataset. 
}
\vspace{-3ex}
\label{fig:heat-appdx}
\end{figure*}

\cref{fig:heat-appdx} shows the heat map of event pairs in more datasets.

